\documentclass[notitlepage,a4paper,aps,prd,tightenlines,preprintnumbers,nofootinbib,showkeys,superscriptaddress,11pt]{revtex4-2}
\pdfoutput=1

\usepackage{XCharter}
\usepackage[T1]{fontenc}
\usepackage{mathptmx}

\usepackage{fullpage}
\usepackage{amsfonts}
\usepackage{amsmath}
\usepackage{slashed}
\usepackage{amssymb}
\usepackage{graphicx}
\usepackage{array}
\usepackage{epic}
\usepackage{eepic}
\usepackage{epsfig}
\usepackage{latexsym}
\usepackage[dvipsnames]{xcolor}
\usepackage[export]{adjustbox}
\usepackage{float}
\usepackage{multirow}
\usepackage[linktocpage]{hyperref}
\usepackage{enumitem}
\hypersetup{colorlinks=true,citecolor=red,linkcolor=purple,urlcolor=NavyBlue}
\usepackage[caption=false]{subfig}
 
\usepackage{natbib}
\usepackage{relsize}
\usepackage[left=2.4cm,right=2.4cm,top=2.0cm,bottom=2.5cm]{geometry}
\definecolor{OliveGreen}{rgb}{0,0.6,0}

\usepackage{shorthand}

\newcommand{\bulletsubsec}[1]{\noindent\underline{$\blacksquare$~#1:}}

\begin{document}

\title{Precise limits on the charge-$2/3$ $U_1$ vector leptoquark}

\author{Arvind Bhaskar}
\email{arvind.bhaskar@research.iiit.ac.in}
\affiliation{Center for Computational Natural Sciences and Bioinformatics, International Institute of Information Technology, Hyderabad 500~032, India}

\author{Diganta Das}
\email{diganta.das@iiit.ac.in}
\affiliation{Center for Computational Natural Sciences and Bioinformatics, International Institute of Information Technology, Hyderabad 500~032, India}
\affiliation{Department of Physics and Astrophysics, University of Delhi, Delhi 110~007, India}

\author{Tanumoy Mandal}
\email{tanumoy@iisertvm.ac.in}
\affiliation{Indian Institute of Science Education and Research Thiruvananthapuram, Vithura, Kerala, 695~551, India}

\author{Subhadip Mitra}
\email{subhadip.mitra@iiit.ac.in}
\affiliation{Center for Computational Natural Sciences and Bioinformatics, International Institute of Information Technology, Hyderabad 500~032, India}

\author{Cyrin Neeraj}
\email{cyrin.neeraj@research.iiit.ac.in}
\affiliation{Center for Computational Natural Sciences and Bioinformatics, International Institute of Information Technology, Hyderabad 500~032, India}

\date{\today}

\begin{abstract}
The $U_1$ leptoquark is known to be a suitable candidate for explaining the semileptonic $B$-decay anomalies. We derive precise limits on its parameter space relevant for the anomalies from the current LHC high-$p_{\rm T}$ dilepton data. We consider an exhaustive list of possible $B$-anomalies-motivated simple scenarios with one or two new couplings that can also be used as templates for obtaining bounds on more complicated scenarios. 
To obtain precise limits, we systematically consider all possible $U_1$ production processes that can contribute to the dilepton searches, including the resonant pair and single productions, nonresonant $t$-channel $U_1$ exchange, as well as its large interference with the Standard Model background. We demonstrate how the inclusion of resonant production  contributions in the dilepton signal can lead to  appreciably improved exclusion limits.  We point out new search channels of $U_1$ that can act as unique tests of the flavour-motivated models. The template scenarios can also be used for future $U_1$ searches at the LHC. 
We compare the LHC limits with other relevant flavour bounds and find that a TeV-scale $U_1$ can accommodate both  $R_{D^{(*)}}$ and $R_{K^{(*)}}$ anomalies while satisfying all the bounds. 
\end{abstract}


\maketitle
\noindent\rule{\textwidth}{0.5pt}\vspace{-55pt}
{\linespread{0.95}\relscale{0.9}\tableofcontents}
\makeatletter
\let\toc@pre\relax
\let\toc@post\relax
\makeatother 
\noindent\rule{\textwidth}{0.5pt}


\hypertarget{sec:intro}{\section{Introduction}
\label{sec:intro}}
\noindent
The concept of lepton flavour universality, a key prediction of the Standard Model (SM), seems to be in tension with the present experimental measurements of some 
semileptonic $B$-meson decays~\cite{Lees:2012xj,Lees:2013uzd,
Aaij:2014ora,Aaij:2017vbb,Aaij:2015yra,Aaij:2017uff,Aaij:2017deq,Huschle:2015rga,Sato:2016svk,Hirose:2016wfn,Hirose:2017dxl}. Differences between theoretical predictions and experimental measurements, hinting towards the existence of some physics beyond the SM (BSM), have been observed in the 
$R_{D^{(*)}}$ and $R_{K^{(*)}}$ observables:
\begin{align}
R_{D^{(*)}}  =  \dfrac{{\mc B}(B\rightarrow D^{(*)}\tau\bar\nu)}{{\mc B}(B\rightarrow D^{(*)}\hat{\ell}\bar\nu)} \quad \mbox{and}\quad  R_{K^{(*)}} = \dfrac{{\mc B}(B\rightarrow K^{(*)}\mu^+\mu^-)}{{\mc B}(B\rightarrow K^{(*)}e^+e^-)}\ .\label{eq:anomalies}
\end{align}
We use $\hat{\ell}$ to denote the light charged leptons, $e$ or $\mu$ and ${\mc B}(x\to y)$ for the $x\to y$ decay branching ratio (BR). The experimental values of $R_D$ and $R_{D^*}$ exceed their SM predictions by $1.4\sigma$ and $3.1\sigma$, respectively~\cite{Bigi:2016mdz,Bernlochner:2017jka,Bigi:2017jbd,Jaiswal:2017rve} (combined excess of $3.1\sigma$ in $R_{D^{(*)}}$, according to the $2019$ world averages~\cite{Amhis:2016xyh}), whereas, 
the $R_K$ and $R_{K^*}$ measurements \cite{Aaij:2019wad, Aaij:2021vac} are smaller than the theoretical predictions by about $3.1\sigma$~\cite{Hiller:2003js,Bordone:2016gaq}.

A TeV-scale vector leptoquark (vLQ), a color-triplet vector boson with nonzero lepton and baryon numbers, is considered to be a suitable candidate to address these anomalies in the literature~\cite{Alonso:2015sja,Calibbi:2015kma,Fajfer:2015ycq,Barbieri:2015yvd,Becirevic:2016oho,Sahoo:2016pet,Bhattacharya:2016mcc,Duraisamy:2016gsd,Buttazzo:2017ixm,Assad:2017iib,Calibbi:2017qbu,Blanke:2018sro,Greljo:2018tuh,Sahoo:2018ffv,Kumar:2018kmr,Crivellin:2018yvo,Angelescu:2018tyl,Aebischer:2018acj,Chauhan:2018lnq,Fornal:2018dqn,Baker:2019sli,Hati:2019ufv,Cornella:2019hct,DaRold:2019fiw,Cheung:2020sbq,Dev:2020qet,Kumbhakar:2020okw,Iguro:2020keo,Hati:2020cyn,Alda:2020okk}.\footnote{See Refs.~\cite{Mandal:2015lca,Das:2016vkr,Bandyopadhyay:2016oif,Dey:2017ede,Bandyopadhyay:2018syt,Aydemir:2018cbb,Bansal:2018eha,Biswas:2018snp,Mandal:2018czf,Biswas:2018iak,Roy:2018nwc,Alves:2018krf,Aydemir:2019ynb,Chandak:2019iwj,Hou:2019wiu,Bordone:2019uzc,Padhan:2019dcp,Bhaskar:2020kdr,Bandyopadhyay:2020klr,Buonocore:2020erb,Bordone:2020lnb,Greljo:2020tgv,Haisch:2020xjd,Bandyopadhyay:2020jez,Crivellin:2021egp} and the references therein for other recent phenomenological studies on LQs.} It is shown in~\cite{Angelescu:2018tyl} that a charge-$2/3$ weak-singlet vLQ, $U_1\equiv(\mathbf{3},\mathbf{1},2/3)$, can resolve both $R_{D^{(*)}}$ and $R_{K^{(*)}}$ anomalies simultaneously.
If the vLQ is really responsible for these anomalies, it is then  essential to scrutinise its parameter space that can address the anomalies simultaneously while satisfying all relevant experimental bounds. In the literature, various flavour and collider data have already been used in this context. However, we find that even though a lot of emphasis has been put on obtaining regions of parameter space that are either ruled out or favoured by the observed anomalies and other flavour data, relatively less attention has been paid to obtain \emph{precise} bounds from the Large Hadron Collider (LHC) data.

It is known that the regions of parameter spaces favoured by the flavour anomalies in various leptoquark (LQ) models are already in tension with the high-$p_{\rm T}$ dilepton data~\cite{Greljo:2015mma,Faroughy:2016osc,Raj:2016aky,Buttazzo:2017ixm,Dorsner:2017ufx,Becirevic:2018afm,Mandal:2018kau,Baker:2019sli}. In this paper, we specifically investigate the case of $U_1$ and argue that the bounds from the LHC data might actually be underestimated and, in some regions of the parameter space, the data could be more constraining than what has been considered so far if one systematically computes all relevant processes and considers the latest direct  search limits. 
As we see, different production processes of $U_1$ contribute to the dilepton or monolepton plus missing energy (MET) signals affecting various kinematic distributions. When incorporated in the statistical analysis, they can give strong bounds on the unknown LQ-$q$-$\ell$ (where $\ell$ can be any charged lepton) couplings together. However, while most of these processes contribute constructively to the signal, a significant contribution (in fact, the most dominant one, in some cases) comes from the nonresonant $t$-channel $U_1$ exchange process that interferes destructively with the SM background. Hence, there is a competition among the $U_1$ production processes, which are highly sensitive to the $U_1$ parameters. Usually, the contribution of the resonant production processes (i.e., pair and single productions) to the  $\ell\ell$ or $\ell+\slashed E_{\rm T}$ signals are ignored assuming that it would give only minor corrections. However, we find, especially in the lower mass region, that the resonant productions' effect on the exclusion could be significant. In this paper, we systematically put together all the sources of resonant and nonresonant dilepton events in our analysis and obtain robust and precise limits on the $U_1$ parameters to date.

\begin{figure}
\captionsetup[subfigure]{labelformat=empty}
\subfloat[(a)]{\includegraphics[width=0.45\textwidth]{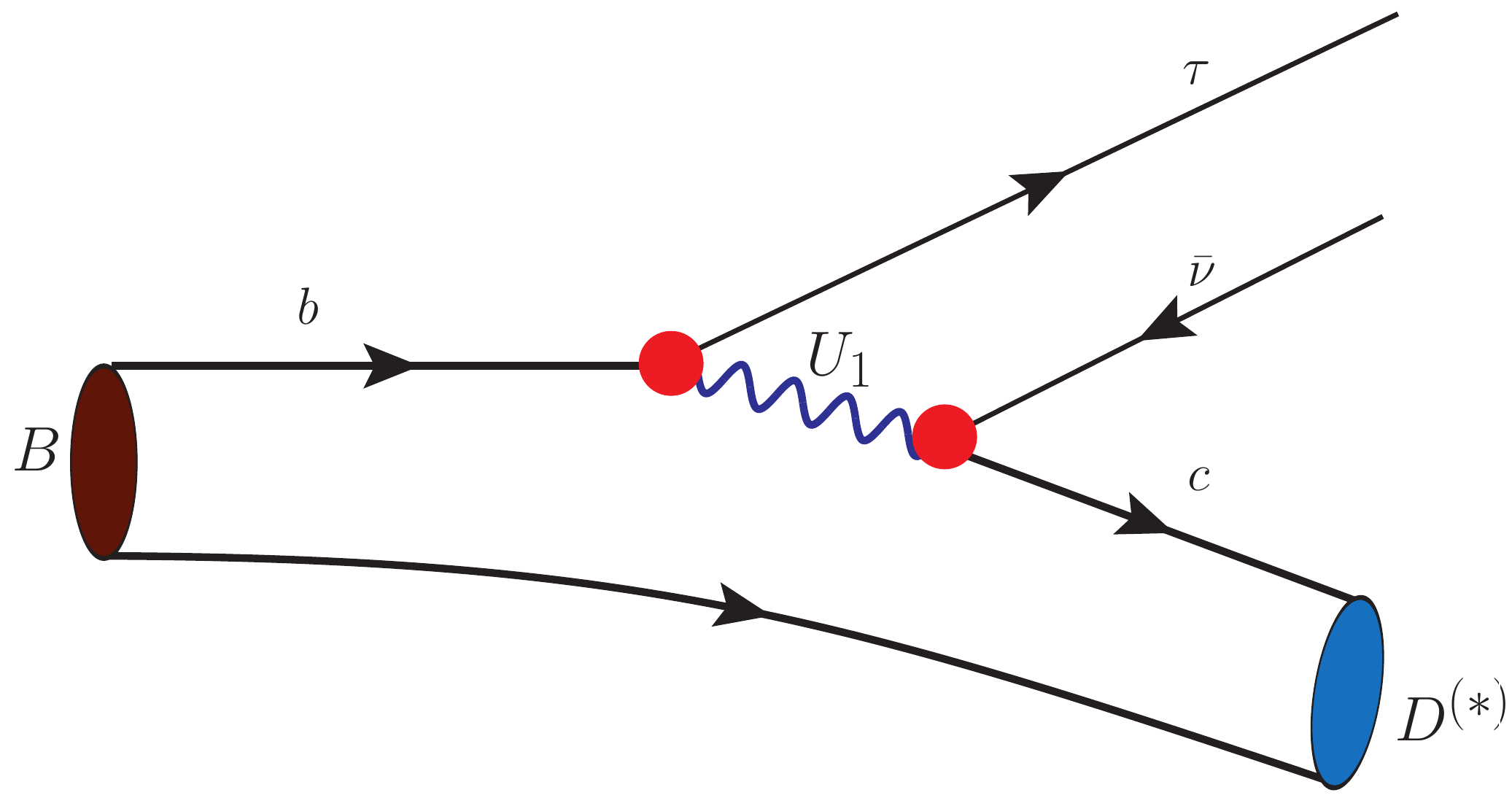}\label{fig:BtoDb}}\quad
\subfloat[(b)]{\includegraphics[width=0.45\textwidth]{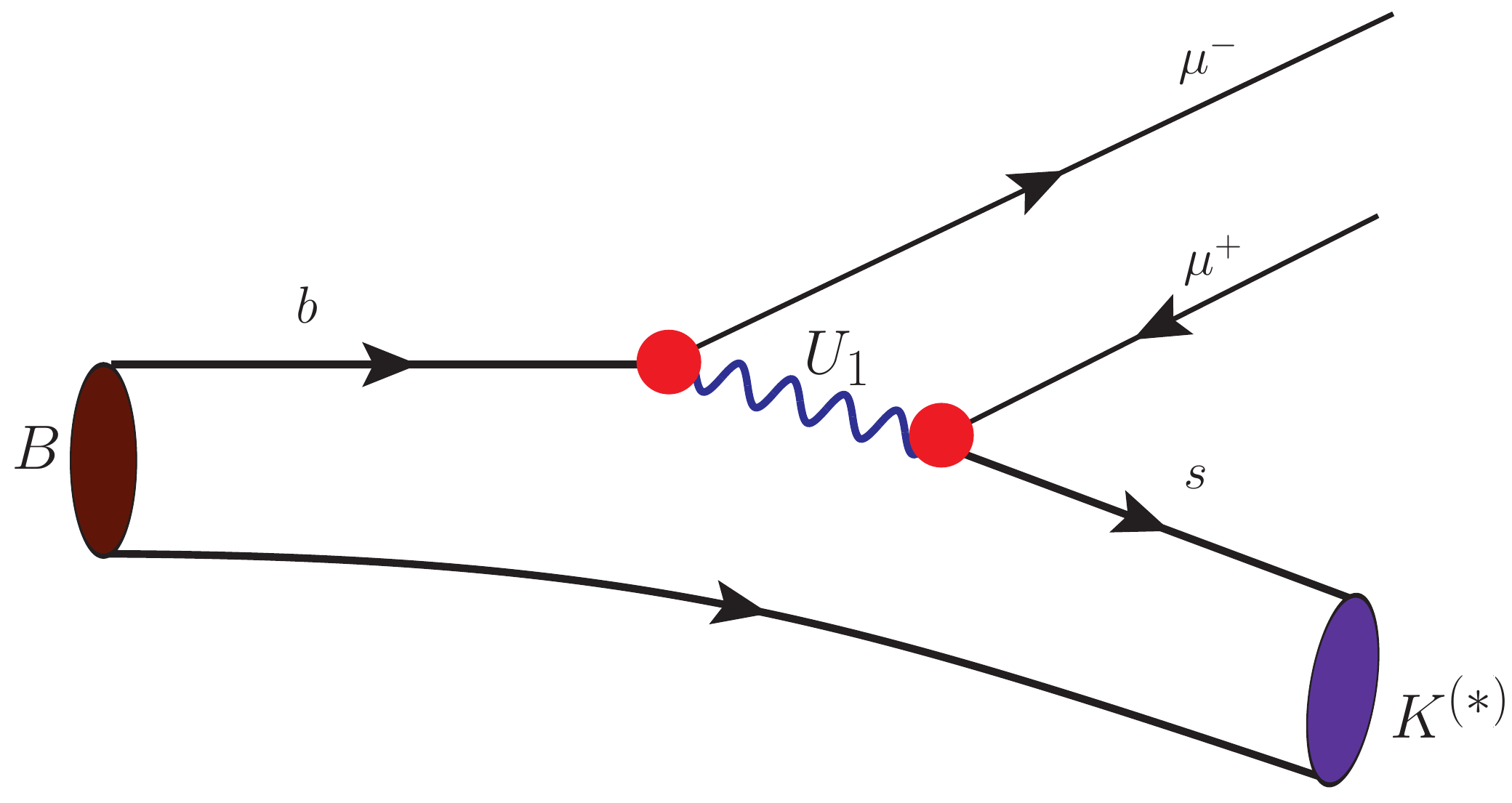}\label{fig:BtoKb}}\\
\subfloat[(c)]{\includegraphics[width=0.45\textwidth]{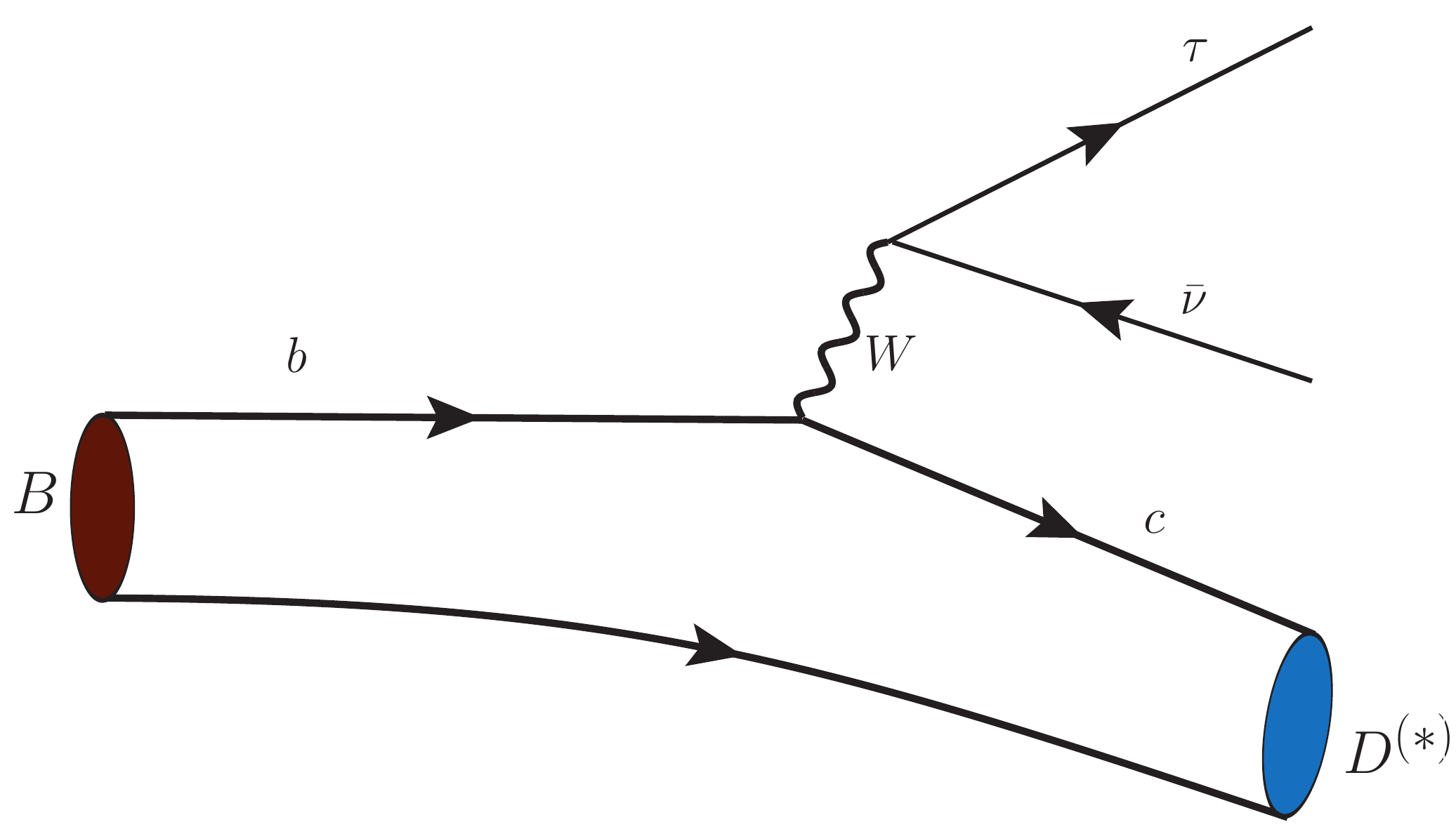}\label{fig:BtoDa}}\quad
\subfloat[(d)]{\includegraphics[width=0.45\textwidth]{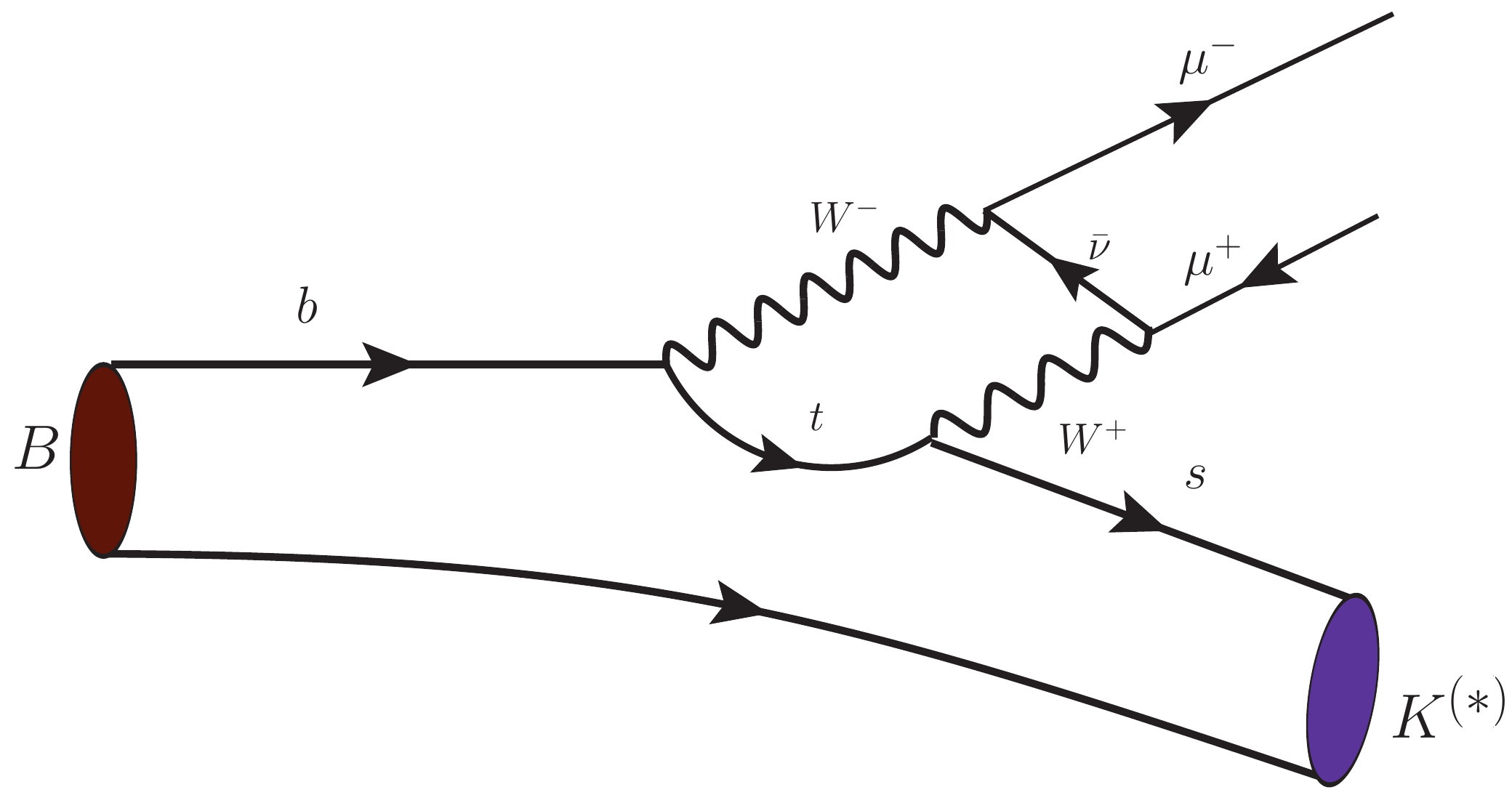}\label{fig:BtoKa}}\\
\caption{Representative leading order diagrams showing the $B\to {D^{(*)}}\tau\bar{\nu}$ and $B\to {K^{(*)}}\m^+\m^-$ decays: the tree level $U_1$ contribution to (a) $B\to {D^{(*)}}\tau\bar{\nu}$ 
and (b) $B\to {K^{(*)}}\m^+\m^-$, and (c) and (d) the corresponding SM processes, respectively.}\label{fig:fynd}
\end{figure} 

To contribute to $R_{D^{(*)}}$, a $U_1$ must couple to the third-generation lepton(s) and, second and third-generation quarks [see Fig.~\ref{fig:BtoDb}, assuming that it does not alter the denominators in Eq.~\eqref{eq:anomalies}] and to contribute to the $R_{K^{(*)}}$ observables, it should  couple to the second-generation leptons [see Fig.~\ref{fig:BtoKb}]. Within the SM,  the $b\to c\tau\bar{\nu}$ decay is mediated by a tree-level charge current interaction, and the neutral current $b\to s\mu^+\mu^-$ decay occurs through a loop. However, the 
$U_1$ LQ can mediate both the flavour-changing transitions, $b\to c\tau\bar{\nu}$ and $b\to s\mu^+\mu^-$ at the tree level, as shown in Fig.~\ref{fig:fynd}. Here, we adopt a bottom-up approach and construct all possible minimal or next-to-minimal scenarios within the $U_1$ LQ model with one or two new couplings at a time that can accommodate either the $R_{D^{(*)}}$ or the $R_{K^{(*)}}$ anomalies. These scenarios can be used as templates to obtain bounds on more complicated scenarios (as explained in Ref.~\cite{Mandal:2018kau} for the $S_1$ LQ). 

There is another motivation for considering various minimal scenarios with different coupling combinations.
An effective field theory suitable for describing the outcomes of low-energy experiments is not well suited for high-energy collider experiments where some of the heavy degrees of freedom are directly accessible. The SM-like Wilson operator, $\mc{O}_{v_L}=\lt[\bar{c}\gm^\mu P_L b\rt]\lt[\bar{\tau}\gm_\mu P_L\nu\rt]$ plays the most important role in the $R_{D^{(*)}}$ observables. However, by looking only at this operator, it is not obvious that the $\ell\ell$ data would lead to strong bounds and the interference between the new physics and the SM background processes would play the prominent role in determining the bounds. Scenarios with very different LHC signatures can lead to the same effective operator (we discuss such an example later). Hence, even though these two scenarios would look similar in low-energy experiments, the limits from LHC would be different.

In the case of the scalar LQ $S_1$, we have seen the dilepton data putting stronger bounds than the monolepton plus MET data~\cite{Mandal:2018kau}. Hence, in this paper, we consider only the dilepton  ($\tau\tau$ and $\mu\mu$~\cite{Aad:2020zxo,Sirunyan:2021khd}) data to put bounds on the regions of $U_1$ parameter space relevant for the $R_{D^{(*)}}$ and $R_{K^{(*)}}$ observables. Unlike the existing bounds on LQ masses from their pair production searches at the LHC, the bounds thus obtained are 
model dependent (i.e., they depend on unknown couplings). However, for large new couplings
they become more restrictive than the pair production ones. We obtain the LHC bounds for various scenarios with different coupling structures and show that they are competitive and complimentary to other flavour bounds.
Also, these bounds are independent of other known theoretical constraints on the
$U_1$ parameter space. Obtaining them requires a systematic consideration of different LQ signal processes at the 
LHC (including their interference with the SM backgrounds which plays the dominant role in determining the bounds). Here, for systematics, we largely follow the analysis of 
Ref.~\cite{Mandal:2018kau} (where a similar analysis was done for a $S_1$-type scalar LQ that can alleviate the $R_{D^{(*)}}$ anomalies).

Before we proceed further, we review the direct detection bounds on LQs that couple with second- and third-generation fermions. Assuming the extra gluon-$U_1$ coupling $\kappa=0$ (we follow the same convention as~\cite{Bhaskar:2020gkk}), a recent LQ pair production search at the CMS detector has excluded vLQs with masses 
below $1460$ GeV for ${\mc B}({\rm LQ}\to t \nu)=1$~\cite{Sirunyan:2018kzh}. For a vLQ decaying to a light quark and a neutrino with $100\%$ BR, the mass exclusion limit is at $1410$ GeV. In the case where it decays to a bottom quark and a neutrino, the limit goes to $1475$ GeV. If the vLQ decays to a top quark $+$ a neutrino and a bottom quark $+$ $\ta$ with equal BRs, then the mass points below $1115$ GeV are excluded. For $\kappa=1$, the limits go up ~\cite{Sirunyan:2018kzh}. 
Pair produced scalar LQs decaying to a light quark and a neutrino with branching ratio unity can be excluded up to $980$ GeV. A scalar LQ decaying to a $b$-quark and a neutrino with a $100\%$ branching ratio can be excluded up to $1100$ GeV ~\cite{Sirunyan:2018kzh}. The ATLAS experiment searched for scalar LQs decaying to the following final states, $\m c$, $\m~+$ a light quark, and $\m b$~\cite{Aad:2020iuy}. The exclusion limits from these channels and the above are summarised in Table~\ref{tab:lhcdirect}. 

\begin{table}[t!]
\caption{Summary of LQ mass exclusion limits from recent direct searches by the CMS (ATLAS) Collaboration. We recast some of the recent scalar searches (marked with ``$*$'') for better limits on $U_1$ than the ones for vLQs.}
\begin{center}
{\linespread{1.3}\footnotesize
\begin{tabular*}{\textwidth}{l@{\extracolsep{\fill}}cccc}
\hline
&Integrated & Scalar LQ & Vector LQ, $\kappa=0$ & Vector LQ, $\kappa=1$ \\  
&Luminosity [fb$^{-1}$]& Mass~[GeV] & Mass~[GeV] & Mass~[GeV] \\
\hline\hline
LQ $ \rightarrow t\nu~(\mc B=1.0)$~\cite{Sirunyan:2018kzh, Aaboud:2019bye} &$35.9$ ($36.1$)& $1020~(992)$ & $1460$ & $1780$ \\ 
 LQ $ \rightarrow q\nu~(\mc B=1.0)$~\cite{Sirunyan:2018kzh} &$35.9$& $980$ & $1410$ & $1790$ \\
 LQ $ \rightarrow b\nu~(\mc B=1.0)$~\cite{Sirunyan:2018kzh, Aaboud:2019bye} &$35.9$ ($36.1$)& $1100~(968)$ & $1475$ & $1810$ \\
 LQ $ \rightarrow b\tau~/t\nu(\mc B=0.5)$~\cite{Sirunyan:2020zbk} & $137$&$950$ & $1290$ & $1650$ \\ 
 LQ $ \rightarrow b\tau~(\mc B=1.0)$~\cite{Aaboud:2019bye}~$*$ &($36.1$)& $(1000)$ & $-$ & $-$ \\
 LQ $ \rightarrow $ $\mu j$ $(\mc B=1.0)$~\cite{Aad:2020iuy}~$*$ &($139$)& $(1733)$ & $-$ & $-$ \\
 LQ $ \rightarrow $ $\mu c$ $(\mc B=1.0)$~\cite{Aad:2020iuy} &($139$)& $(1680)$ & $-$ & $-$ \\
 LQ $ \rightarrow $ $\mu b$ $(\mc B=1.0)$~\cite{Aad:2020iuy}~$*$ &($139$)& $(1721)$ & $-$ & $-$ \\
\hline
\end{tabular*}}
\label{tab:lhcdirect}
\end{center}
\end{table}

The paper is organized as follows. In the next section, we introduce the $U_1$ LQ model and the relevant scenarios. In Section~\ref{sec:pheno}, we describe its LHC phenomenology. In Section~\ref{sec:recast}, we discuss the dilepton search and their recasts. In Section~\ref{sec:results}, we present the numerical results, and finally, in Section~\ref{sec:conclu}, we conclude.

\section{The $U_1$ leptoquark model}
\label{sec:model}

\noindent
The interaction between $U_1$ and the SM quarks and leptons can be expressed as~\cite{Buchmuller:1986zs,Blumlein:1994qd,Blumlein:1996qp,Dorsner:2016wpm},
\begin{equation}
\label{eq:GenLagU1}
\mathcal{L} \supset
x^{LL}_{1~ij}~\bar{Q}^{i}\gamma_{\mu}U^{\mu}_{1}P_{L}L^{j} + x_{1~ij}^{RR}~\bar{d}^i_{R}\gamma_{\mu}U^{\mu}_{1}P_{R}\ell^j_{R} + \textrm{H.c.},
\end{equation}
if we ignore the diquark interactions which are severely constrained by the proton decay bounds. Here, $Q_i$ and $L_j$ denote the SM left-handed quark and lepton doublets, respectively and $d_{R}^{i}$ and $\ell_{R}^{j}$ are the down-type right-handed quarks and leptons, respectively. The indices $i,j=\{1,\ 2,\ 3\}$ stand for quark and lepton generations; i.e., $x_{1~ij}^{LL}$ and $x_{1~ij}^{RR}$ are $3\times 3$ matrices in flavour space. In general, these matrices are complex. We, however, simply assume them to be real since the LHC would be mostly insensitive to their complex natures. Global fits to experimental data with complex couplings are similar to the fits obtained with real couplings, albeit with slightly greater significance~\cite{Alda:2018mfy, Alok:2017jgr}. Hence, predictions for flavor observables with complex couplings are expected to be similar to the ones obtained with purely real couplings. Moreover, since we are interested in only those $U_1$ scenarios that can accommodate the $R_{D^{(*)}}$ and $R_{K^{(*)}}$ anomalies, we further simplify the $x_{1~ij}^{LL}$ and $x_{1~ij}^{RR}$ matrices by setting all the components that do not participate directly in these decays to zero. We refer to any type of neutrinos simply as $\n$, i.e., without any flavour index as this would not affect our LHC analysis.
As the $b\to c\tau\bar{\nu}$ and $b\to s\mu^+\mu^-$ decays involve independent couplings, we analyse the $R_{D^{(*)}}$- and $R_{K^{(*)}}$-anomalies-motivated scenarios separately.\footnote{From here onwards, we 
refer to the $R_{D^{(*)}}$- and $R_{K^{(*)}}$-anomalies-motivated scenarios simply as $R_{D^{(*)}}$ and $R_{K^{(*)}}$ scenarios for brevity.}
\bigskip

\subsection*{$R_{D^{(*)}}$ scenarios}\label{sec:rd}
\noindent 
In the SM, the $b\to c\tau\bar{\nu}$ transition is a tree-level charged-current-mediated process and the Lagrangian responsible for it can be written as
\begin{align}
\mc{L}_{\rm SM} = -\frac{4G_F}{\sqrt{2}}V_{cb}~\mc{O}_{V_L} =-\frac{4G_F}{\sqrt{2}}V_{cb}~\lt[\bar{c}\gm^\mu P_L b\rt]\lt[\bar{\tau}\gm_\mu P_L\nu_\tau\rt]\, .
\end{align}
New physics can generate additional contributions to the $b\to c\tau\bar{\nu}$ transition in the form of four-fermion operators. The most general form of the Lagrangian can be written as~\cite{Tanaka:2012nw}
\begin{align}
\mc{L} \supset -\frac{4G_F}{\sqrt{2}}V_{cb}&\lt[  \lt(1+\mc{C}_{V_L}\rt)\mc{O}_{V_L} + \mc{C}_{V_R}\mc{O}_{V_R} 
+ \mc{C}_{S_L}\mc{O}_{S_L} 
+ \mc{C}_{S_R}\mc{O}_{S_R} + \mc{C}_{T_R}\mc{O}_{T_R}\rt]\, ,
\end{align}
where the Wilson coefficient corresponding to an operator $\mc{O}_i$ is denoted as 
$\mc{C}_i$. The operators have three different Lorentz structures:

\begin{itemize}
\item  Vector: 
$\left[\begin{array}{ccc}
\mc{O}_{V_L} &=& \lt[\bar{c}\gm^\mu P_L b\rt]\lt[\bar{\tau}\gm_\mu P_L\nu\rt]\\~\\
\mc{O}_{V_R} &=& \lt[\bar{c}\gm^\mu P_R b\rt]\lt[\bar{\tau}\gm_\mu P_L\nu\rt]
\end{array}\right.$

\item Scalar: 
$\left[\begin{array}{ccc}
\mc{O}_{S_L} &=& \lt[\bar{c}P_L b\rt]\lt[\bar{\tau}P_L\nu\rt]\\~\\
\mc{O}_{S_R} &=& \lt[\bar{c}P_R b\rt]\lt[\bar{\tau}P_L\nu\rt]
\end{array}\right.$

\item Tensor:~~
$\begin{array}{ccc}
\mc{O}_{T_L} &=& \lt[\bar{c}\sg^{\mu\nu} P_L b\rt]\lt[\bar{\tau}\sg_{\mu\nu} P_L\nu\rt].
\end{array}$
\end{itemize}
\vspace{0.1cm}

From Fig.~\ref{fig:fynd} we see that the $\bar c\n U_1$ and $\bar b\ta U_1$ couplings have to be nonzero for $U_1$ to contribute in the $b\to c\tau\bar{\n}$ process. We make the following flavour Ansatz for simplicity:
\begin{equation}
x^{LL}_1 = 
\begin{pmatrix}
0 & 0 & 0 \\
0 & 0 & \lm^L_{23} \\
0 & 0 & \lm^L_{33}
\end{pmatrix},~~
x^{RR}_1 = 
\begin{pmatrix}
0 & 0 & 0 \\
0 & 0 & 0 \\
0 & 0 & \lm^R_{33}
\end{pmatrix}.\label{eq:rdst_couplings}
\end{equation}
Given the Ans\"atze of the five operators listed above,  
only $\mc{O}_{V_L}$ and $\mc{O}_{S_L}$ can be generated by $U_1$,  i.e., $\mc{C}_{V_R}^{U_1}=\mc{C}_{S_R}^{U_1}=\mc C_{T_R}^{U_1}=0$. Note that the simplified assumption of several zeros in the coupling matrices are purely phenomenological. This may not be strictly valid in some specific models, e.g., in the models in Refs.~\cite{Bordone:2019uzc,Bordone:2020lnb} where the LQ induced flavour structures are parametrised by Froggatt-Nielsen charges.

\begin{figure}
\captionsetup[subfigure]{labelformat=empty}
\subfloat[(a)]{\raisebox{22pt}{\includegraphics[width=0.44\textwidth,valign=t]{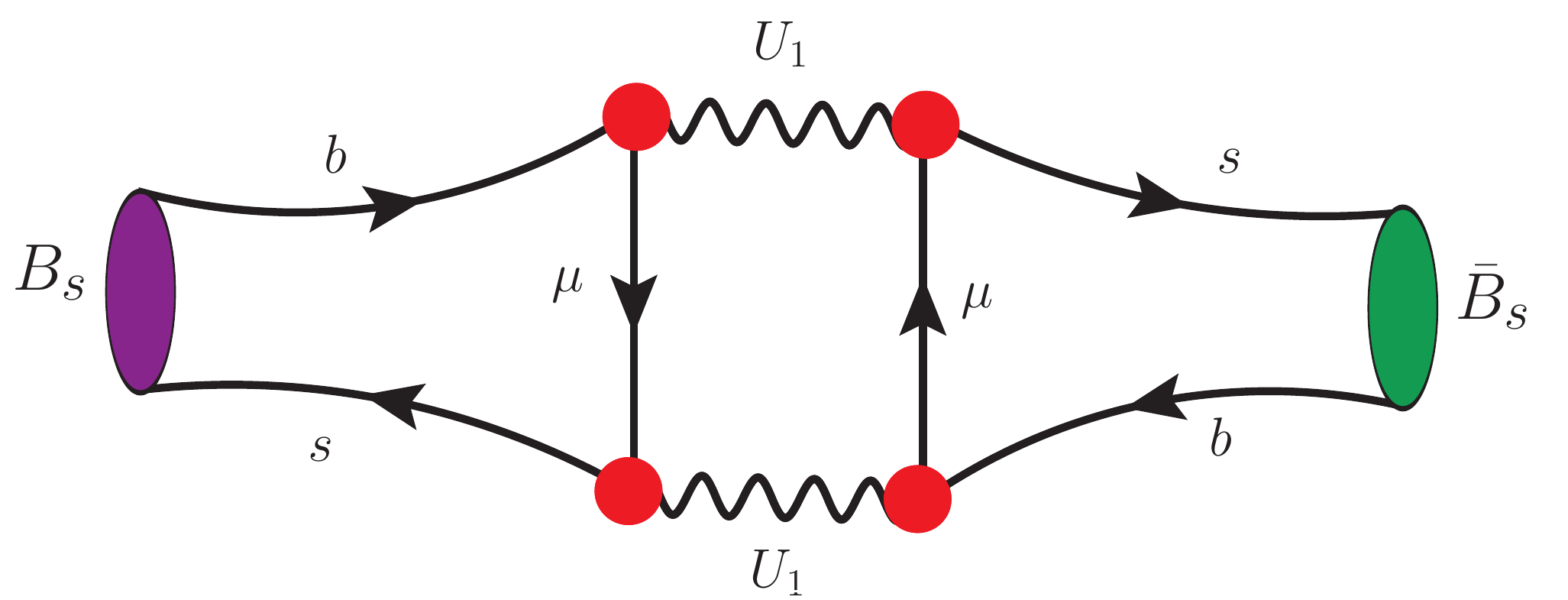}}\label{fig:BSBSbar}
\vphantom{\includegraphics[width=0.45\textwidth,valign=c]{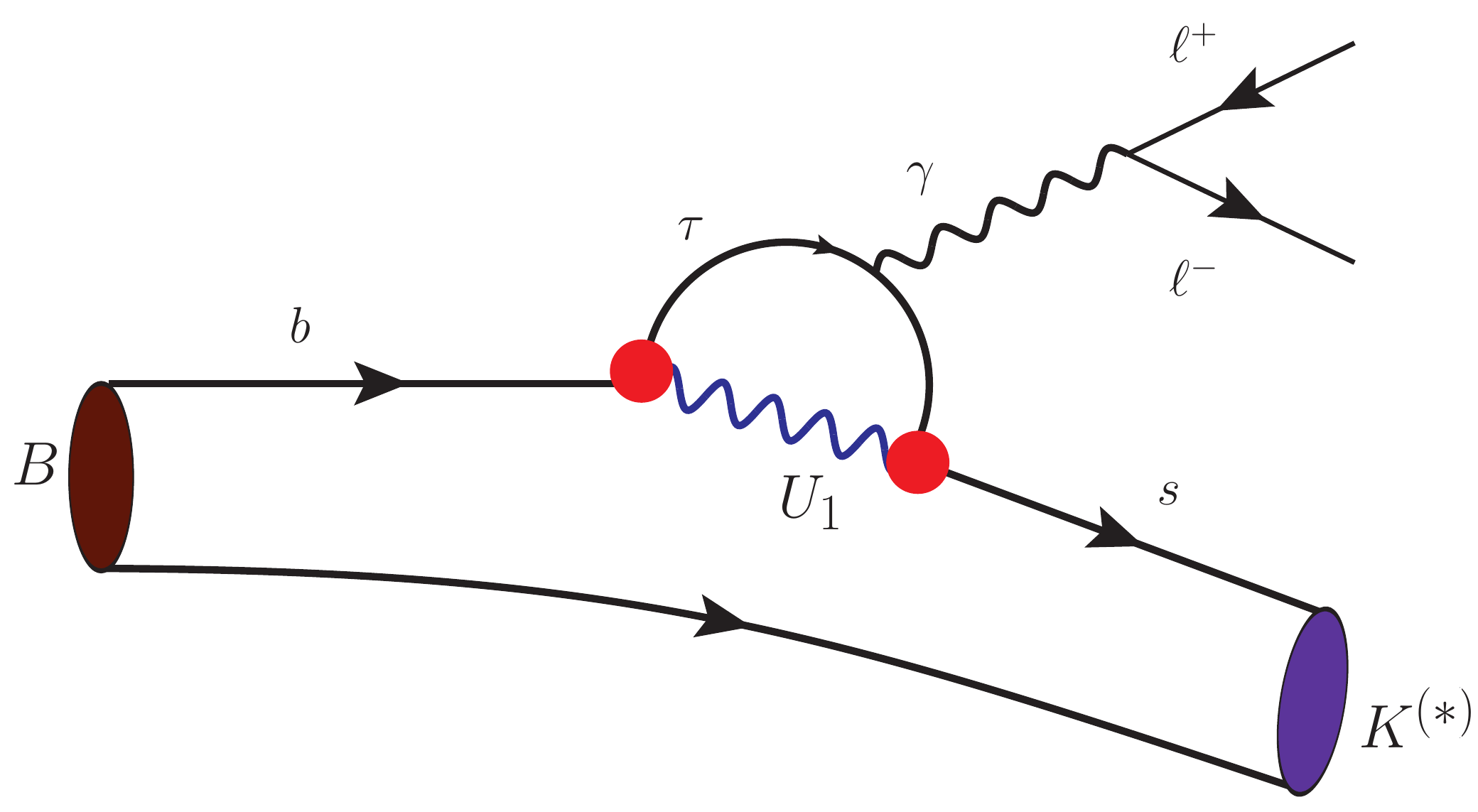}}
}\quad
\subfloat[(b)]{\includegraphics[width=0.45\textwidth,valign=c]{Fig2}\label{fig:BStoll}}
\caption{(a) A representative diagram showing the $U_1$ contribution to $B_s$-$\bar B_s$ mixing  and (b) a $U_1$-mediated  photon penguin diagram contributing to $b\to s\ell^+\ell^-$.}\label{fig:fynd2}
\end{figure}

The nonzero coefficients, $\mc{C}_{V_L}$ and $\mc{C}_{S_L}$ can be written in terms of the $\bar c\n U_1$ and $\bar b\ta U_1$ couplings,
\begin{eqnarray}
\left.\begin{array}{lcl}
\mc{C}_{V_L}^{U_1} &=&  \displaystyle\frac{1}{2\sqrt{2}G_FV_{cb}}\frac{\lm_{c\n}^{L}\left(\lm_{b\ta}^L\right)^*}{M_{U_1}^2} \\
\mc{C}_{S_L}^{U_1} &=& \displaystyle-\frac{1}{2\sqrt{2}G_FV_{cb}}\frac{2\lm_{c\n}^{L}\lt(\lm_{b\ta}^R\rt)^*}{M_{U_1}^2}
\end{array}\right\}\label{eq:cvlcslctl}.
\end{eqnarray}
The actual relationship of $\lm_{c\n}^{L}$ and $\lm_{b\ta}^{L/R}$ with $\lm^L_{23}$ and $\lm^{L/R}_{33}$, defined in Eq.~\eqref{eq:rdst_couplings}, varies from scenario to scenario.
We can express the ratios, $r_{D^{(*)}} = R_{D^{(*)}}/R_{D^{(*)}}^{\mathrm{SM}}$ in terms of the nonzero Wilson 
coefficients as~\cite{Iguro:2018vqb},
\begin{align}
\label{eq:rdrdst}
r_D \equiv \frac{R_{D}}{R_{D}^{\mathrm{SM}}}\approx&\ \left|1+\mc{C}_{V_L}^{U_1}\right|^2 + 1.02\ \lt|\mc{C}_{S_L}^{U_1}\rt|^2 
+ 1.49\ \textrm{Re}\lt[(1+\mc{C}_{V_L}^{U_1})\mc{C}_{S_L}^{U_1*}\rt],\\
r_{D^{*}} \equiv \frac{R_{D^{*}}}{R_{D^*}^{\mathrm{SM}}} \approx&\ \lt|1+\mc{C}_{V_L}^{U_1}\rt|^2 + 0.04\ \lt|\mc{C}_{S_L}^{U_1}\rt|^2 - 0.11\ \textrm{Re}\lt[(1+\mc{C}_{V_L}^{U_1})\mc{C}_{S_L}^{U_1*}\rt].
\end{align}
There are two other observables where nonzero $\mc{C}_{V_L}^{U_1}$ and $\mc{C}_{S_L}^{U_1}$ would contribute to -- the longitudinal $D^*$ polarization $F_L(D^*)$ and the longitudinal $\tau$ polarization asymmetry $P_{\tau}(D^*)$.
They have been measured by the Belle Collaboration~\cite{Adamczyk:2019wyt,Hirose:2016wfn,Hirose:2017dxl}. 
For our purpose, we can express $F_L(D^*)$ and $P_{\tau}(D^*)$ as~\cite{Iguro:2018vqb},
\begin{align}
f_L(D^*) \equiv& \frac{F_{L}(D^*)}{F_{L}^{\textrm{SM}}(D^*)} \approx\ \frac{1}{r_{D^{*}}}\Big\{|1+\mc{C}_{V_L}^{U_1}|^2 + 0.08\  |\mc{C}_{S_L}^{U_1}|^2 
- 0.24\ \textrm{Re}\lt[(1+\mc{C}_{V_L}^{U_1})\mc{C}_{S_L}^{U_1*}\rt]\Big\},\\ 
p_{\tau}(D^*) \equiv& \frac{P_{\tau}(D^*)}{P_{\tau}^{\textrm{SM}}(D^*)} \approx\ \frac{1}{r_{D^{*}}}\Big\{|1+\mc{C}_{V_L}^{U_1}|^2 - 0.07\ |\mc{C}_{S_L}^{U_1}|^2 + 0.22\ \textrm{Re}\lt[(1+\mc{C}_{V_L}^{U_1})\mc{C}_{S_L}^{U_1*}\rt]\Big\}.
\end{align}
\begin{table}[t!]
\caption{Bounds on the $R_{D^{(*)}}$ scenarios.}
\begin{center}
{\linespread{1.3}\footnotesize
\begin{tabular*}{\textwidth}{l@{\extracolsep{\fill}}cclc}
\hline
$\vphantom{\Big|}$
Observable & Experimentally Allowed Range& SM Expectation & Ratio & Value \\
\hline\hline
$R_{D}$ 				& $0.340\pm 0.027\pm 0.013$~\cite{Amhis:2016xyh}						& $0.299\pm 0.003$~\cite{Bigi:2016mdz}		& $r_{D}$ & $1.137\pm 0.101$\\
$R_{D^{*}}$ 		& $0.295 \pm 0.011\pm 0.008$~\cite{Amhis:2016xyh} 						& $0.258 \pm 0.005$~\cite{Amhis:2016xyh} 	& $r_{D^{*}}$ & $1.144\pm 0.057$\\
$F_L(D^*)$ 		& $0.60\pm 0.08 \pm 0.035$~\cite{Hirose:2016wfn,Hirose:2017dxl}	& $0.46  \pm 0.04$~\cite{Bhattacharya:2018kig} & $f_L(D^*)$ & $1.313\pm 0.198$\\
$P_\tau(D^*)$ 	& $-0.38 \pm 0.51^{+0.21}_{-0.16}$~\cite{Adamczyk:2019wyt} 		& $-0.497\pm0.013$ ~\cite{Tanaka:2012nw}	& $p_\tau(D^*)$ & $0.766\pm 1.093$ \\
$\mathcal{B}(B\to\tau\nu)$&$<(1.09\pm 0.24)\times 10^{-4}$~\cite{Tanabashi:2018oca}&$(0.812\pm 0.054)\times 10^{-4}$~\cite{Bona:2017cxr} \\
$\mathcal{B}(B_c\to \tau\nu)$&$<10\%$~\cite{Akeroyd:2017mhr}\\
\hline
\end{tabular*}}
\label{tab:obsval}
\end{center}
\end{table}
\noindent
A nonzero $\mc C_{V_L}^{U_1}$ and $\mc C_{S_L}^{U_1}$ would also contribute to leptonic decays $B_c\to\ta\n$ and $B\to\ta\n$ as,
\begin{eqnarray}
\label{eq:b2taunu}
\mathcal{B}(B_c\to \tau\nu) &=& \frac{\tau_{B_c} m_{B_c} f^2_{B_c} G_F^2 |V_{cb}|^2 }{8\pi} m_\tau^2 \bigg( 1 - \frac{m_\tau^2}{m_{B_c}^2} \bigg)^2 
\left|1+\mc{C}_{V_L}^{U_1}+\frac{m_{B_c}^2}{m_{\ta}(m_b+m_c)}\mc C_{S_L}^{U_1}\right|^2,\\
\mathcal{B}(B\to\tau\nu) &=& \mathcal{B}(B\to \tau\nu)_{\rm SM} \left|1+\mc{C}_{V_L}^{U_1}+\frac{m_{B}^2}{m_{\ta}(m_b+m_u)}\mc C_{S_L}^{U_1}\right|^2 
\end{eqnarray}
where $\tau_{B_c}$ is the lifetime of the $B_c$ meson, $f_{B_c}$ is its decay constant, and $\mathcal{B}(B\to \tau\nu)_{\rm SM} $ is the branching ratio within the SM. The LEP data put a constraint on the $B_c\to \tau\nu$ branching ratio~\cite{Akeroyd:2017mhr} as,
$\mathcal{B}(B_c\to\tau\nu)<10\%$.
The experimental upper bound on the $B\to\tau\nu$ decay is given as~\cite{Tanabashi:2018oca} 
$\mathcal{B}(B\to\tau\nu)<(1.09\pm 0.24)\times 10^{-4}$,
and the corresponding SM branching ratio is estimated to be~\cite{Bona:2017cxr} 
$\mathcal{B}(B\to\tau\nu)_{\rm SM} = (0.812\pm 0.054)\times 10^{-4}$.
The current bounds on these observables are summarized in Table~\ref{tab:obsval}.
Wherever applicable, we also consider constraints from $B_s$-$\bar{B}_s$ mixing [see Fig.~\ref{fig:BSBSbar}] through the effective Hamiltonian, 
\begin{equation}
H_{\rm eff} = (\mc C^{\rm SM}_{box} + \mc C^{U_1}_{box})(\bar{s}_L\gamma^\alpha b_L)(\bar{s}_L\gamma_\al b_L)
\end{equation}
where the SM contribution, $\mc C^{\rm SM}_{box}$, and the $U_1$ contribution, $\mc C^{U_1}_{box}$, which generically depends on new coupling(s) as $\sim \lambda^4$, are given as
\begin{eqnarray}
\mc C^{\rm SM}_{box} &=& \frac{G_F^2}{4\pi^2}(V_{tb}V_{ts})^2M_W^2 S_0(x_t),\\
\label{eq:NPmixing}
\mc C^{U_1}_{box} &=& \frac{\lambda^4}{8\pi^2 M_{U_1}^2 }.
\end{eqnarray}
In Eq.~\eqref{eq:NPmixing}, the generation indices and possible Cabibbo-Kobayashi-Maskawa (CKM) elements have been omitted as they depend on the scenario that we are interested in. The loop function is the Inami-Lim function~\cite{Inami:1980fz}, $S_0(x_t\equiv m_t^2/m_W^2)\sim 2.37$~\cite{DiLuzio:2017fdq}. The UT$fit$ Collaboration gives the following bounds on the ratio $\mc C^{U_1}_{box}/\mc C^{\rm SM}_{box}$~\cite{Bona:2017cxr}:
\begin{align}
0.94 < \left|1 + \frac{\mc C^{U_1}_{box}}{\mc C^{\rm SM}_{box}}\right| < 1.29.
\end{align}
Additionally, whenever $\lambda^L_{b\ta}$ and $\lambda^L_{s\ta}$ are simultaneously nonzero, they  contribute to another lepton-flavour-universal operator in a log-enhanced manner through an off-shell photon penguin diagram
as [see Fig.~\ref{fig:BStoll}]
\begin{equation}
\mc L \supset  -\frac{4G_F}{\sqrt2}\left(V_{tb} V_{ts}^\ast\right) \mc C_{9}^{\rm univ}\mc O_{9}^{\rm univ}\ \label{eq:deltac9}
\end{equation}
where
\begin{eqnarray}
\mc O_{9}^{\rm univ} = \displaystyle\frac\al{4\pi}\left(\bar s_L\gm_\al b_L\right)\left(\bar \ell\gm^\al \ell\right)~~~~\textrm{and}~~~~\mc C_{9}^{\rm univ} = -\frac{1}{V_{tb} V_{ts}^\ast} \frac{\lambda^L_{s\ta}\lt(\lambda^L_{b\ta}\rt)^*}{3\sqrt{2} G_F M_{U_1}^2}  \log(m_b^2/M_{U_1}^2).\label{eq:deltac9a}
\end{eqnarray}
We consider the $2\sigma$ limits from the global fits to the $b\to s\mu^+\mu^-$  data \cite{Alguero:2019ptt,Aebischer:2019mlg,Alguero:2021anc} as  $-1.27\leq\mc{C}_9^{\rm univ}\leq-0.51$.

\newsavebox{\myimage}
\bigskip

We now consider different scenarios with different combinations of the three couplings $\lm^L_{23}$, $\lm^L_{33}$ and $\lm^R_{33}$. As indicated in the \hyperlink{sec:intro}{Introduction}, these scenarios may not always appear very different from each other if we look at them only from the perspective of effective operators but their LHC phenomenology are different. As a result, the bounds from the LHC data differ within the scenarios. We elaborate this point further shortly.

\vspace{0.2cm}
\hypertarget{sce:rd1a}{\bulletsubsec{~Scenario RD1A}} 
In this scenario, only $\lm_{23}^L$ is assumed to be nonzero. This directly generates the following two couplings: $\bar c\n U_1$ and $\bar s\ta U_1$. We assume that the $U_1$ interaction is aligned with the physical basis of the up-type quarks. The interactions with the physical down-type quarks are then obtained by rotating them with the CKM matrix (i.e., by considering mixing among the down-type quarks)~\cite{Mandal:2018kau}. This way, an effective $\bar b\tau U_1$ coupling of strength $V^*_{cb}\lm_{23}^L$ is generated. The interaction Lagrangian now reads as
\begin{align}
\mathcal{L} &\supset \lambda_{23}^{L}[\bar{c}_L\gamma_{\mu}\nu_{L} + \bar{s}_L \gamma_{\mu}\tau_{L})] U_{1}^{\mu}\, ,\nn\\
&= \lambda_{23}^{L}[\bar{c}_L\gamma_{\mu}\nu_{L} + (V^*_{cd}\bar{d}_L + V^*_{cs}\bar{s}_L + V^*_{cb}\bar{b}_L) \gamma_{\mu}\tau_{L})] U_{1}^{\mu}
\end{align}
giving
\begin{equation}
\mc{C}_{V_L}^{RD1A} =\frac{1}{2\sqrt{2}G_F }\frac{\left(\lm_{23}^{L}\right)^2}{M_{U_1}^2} \,,\quad\mc{C}^{RD1A}_{S_L} = 0.
\end{equation}
This implies the observables, $R_{D^{(*)}}$, $F_L(D^*)$, $P_\tau(D^*)$, and $\mathcal{B}(B_{(c)}\to \tau\nu)$ would receive contributions from $U_1$. Due to the off-shell photon-penguin diagram shown in Fig.~\ref{fig:fynd2}, there will be a log-enhanced lepton-universal contribution 
to the $b\to s\ell^+\ell^-$ transition \cite{Crivellin:2018yvo}:
\begin{equation}
\mc{C}_9^{\rm univ} = -\frac{V_{cb} V_{cs}^\ast}{V_{tb} V_{ts}^\ast} \frac{\lt(\lambda^L_{23}\rt)^2}{3\sqrt{2} G_F M_{U_1}^2}  \log(m_b^2/M_{U_1}^2).
\end{equation}
This scenario would lead to a  nonzero contribution to the $B_s$-$\bar{B}_s$ mixing coefficient as
\begin{equation}
\mc{C}^{U_1}_{box} = \frac{|V_{cb}|^2|V_{cs}|^2(\lambda^L_{23})^4 }{8\pi^2M^2_{U_1}}.
\end{equation}
The dominant decay modes of $U_1$ in this scenario are $U_1\to c\bar \nu$ and $U_1\to s\tau^+$, and both of them share almost $50$\% BR.

\vspace{0.2cm}
\hypertarget{sce:rd1b}{\bulletsubsec{~Scenario RD1B}} In this scenario, only $\lm_{33}^L$ is assumed to be nonzero, thus generating the $\bar b\ta U_1$ and $\bar t\n U_1$ couplings. Assuming the $U_1$ interaction to be aligned with the physical basis of the down-type quarks, we generate $\bar c\n U_1$ coupling $V_{cb}\lm_{33}^L$ through the mixing in the up-type quarks. The interaction Lagrangian is given by
\begin{align}
\mathcal{L} &\supset \lambda_{33}^{L}[\bar{t}_L \gamma_{\mu}\nu_{L} + \bar{b}_L\gamma_{\mu}\tau_{L}] U_{1}^{\mu}\nn\\
&= \lambda_{33}^{L}[(V_{ub}\bar{u}_L + V_{cb}\bar{c}_L + V_{tb}\bar{t}_L) \gamma_{\mu}\nu_{L}) + \bar{b}_L\gamma_{\mu}\tau_{L}] U_{1}^{\mu},
\end{align}
and the contributions to the Wilson coefficients are given by
\begin{equation}
\mc{C}_{V_L}^{RD1B} = \frac{1}{2\sqrt{2}G_F}\frac{\left(\lm_{33}^{L}\right)^2}{M_{U_1}^2}\,,\quad\mc{C}^{RD1B}_{S_L} = 0.
\end{equation}
Like in \hyperlink{sce:rd1a}{Scenario RD1A}, the observables $R_{D^{(*)}}$, $F_L(D^*)$, $P_\tau(D^*)$, and $\mathcal{B}(B_{(c)}\to \tau\nu)$ would receive contribution from $U_1$ in this case too.
Here, the dominant decay modes of $U_1$ are
$U_1\to t \bar \nu$ and $U_1\to b\tau^+$ with $50$\% BR each.

\vspace{0.2cm}
\hypertarget{sce:rd2a}{\bulletsubsec{~Scenario RD2A}} In this scenario, we assume $\lm_{23}^L$ and $\lm_{33}^L$ to be nonzero, and the interaction of $U_1$ is aligned with the physical basis of the down-type quarks. The interaction Lagrangian can be written as
\begin{align}
\mathcal{L} &\supset [\lambda_{23}^{L}(\bar{c}_L\gamma_{\mu}\nu_{L} +
\bar{s}_L \gamma_{\mu}\tau_{L}) +
\lambda_{33}^{L}(\bar{t}_L\gamma_{\mu}\nu_{L} + \bar{b}_L\gamma_{\mu}\tau_{L})] U_{1}^{\mu} \nn\\
& = [\lambda_{23}^{L}(V_{us}\bar{u}_L\gamma_{\mu}\nu_{L} + V_{cs}\bar{c}_L\gamma_{\mu}\nu_{L} + V_{ts}\bar{t}_L\gamma_{\mu}\nu_{L} +
\bar{s}_L \gamma_{\mu}\tau_{L}) \nn\\
&\quad+
\lambda_{33}^{L}(V_{ub}\bar{u}_L\gamma_{\mu}\nu_{L} + V_{cb}\bar{c}_L\gamma_{\mu}\nu_{L} + V_{tb}\bar{t}_L\gamma_{\mu}\nu_{L} + \bar{b}_L\gamma_{\mu}\tau_{L})] U_{1}^{\mu},
\end{align}
where, in the second step, we have assumed mixing among the up-type quarks.
In the absence of $\lm_{23}^R$, in this case, $\mc{C}_{V_L}^{RD2A}$ is the only nonzero Wilson coefficient,  i.e., 
\begin{equation}
\mc{C}_{V_L}^{RD2A} = \frac{1}{2\sqrt{2}G_FV_{cb}}\frac{(V_{cs}\lm_{23}^{L} + V_{cb}\lm_{33}^{L})\lm_{33}^L}{M_{U_1}^2}\, ,\quad \quad\mc{C}^{RD2A}_{S_L} = 0\, .
\end{equation}
In addition to the contribution to the $R_{D^{(*)}}$, $F_L(D^*)$, $P_\tau(D^*)$, and $\mathcal{B}(B_{(c)}\to \tau\nu)$ processes, we consider the lepton flavour-universal contribution
\begin{equation}
\mc{C}_9^{\rm univ} = -\frac{1}{V_{tb} V_{ts}^\ast} \frac{\lambda^L_{23} \lambda^{L}_{33}}{3\sqrt{2} G_F M_{U_1}^2}  \log(m_b^2/M_{U_1}^2).
\end{equation}
In this scenario, the $B_s$-$\bar{B}_s$ mixing coefficient would receive a contribution from $U_1$
\begin{equation}
\mc{C}^{U_1}_{box} = \frac{\lt(\lambda^L_{23}\rt)^2\lt(\lambda^L_{33}\rt)^2 }{8
\pi^2M^2_{U_1}}.
\end{equation}
Here, $U_1$ can decay to $c\bar\nu$, $s\tau^+$, $t\bar\nu$ and $b\tau^+$ final states with comparable BRs.

\vspace{0.2cm}
\hypertarget{sce:rd2b}{\bulletsubsec{~Scenario RD2B}} Here, both $\lm_{23}^L$ and $\lm_{33}^R$ are nonzero. Ignoring possible CKM-suppressed couplings, the interaction Lagrangian is given by
\begin{align}
\mathcal{L} & \supset [\lambda_{23}^{L}(\bar{c}_L\gamma_{\mu}\nu_{L} +
\bar{s}_L \gamma_{\mu}\tau_{L}) +
\lambda_{33}^{R} \bar{b}_R\gamma_{\mu}\tau_{R}] U_{1}^{\mu}\nn\\
& = [\lambda_{23}^{L}(V_{us}\bar{u}_L\gamma_{\mu}\nu_{L} + V_{cs}\bar{c}_L\gamma_{\mu}\nu_{L} + V_{ts}\bar{t}_L\gamma_{\mu}\nu_{L} +
\bar{s}_L \gamma_{\mu}\tau_{L}) +
\lambda_{33}^{R} \bar{b}_R\gamma_{\mu}\tau_{R}] U_{1}^{\mu}
\end{align}
where, once again in the second step we have assumed mixing among the up-type quarks.
This gives the following contribution to $\mc{C}_{S_L}$:
\begin{align}
\mc{C}_{V_L}^{RD2B} = 0\, ,\quad
\mc{C}_{S_L}^{RD2B} =  \displaystyle-\frac{V_{cs}}{\sqrt{2}G_FV_{cb}}\frac{\lm_{23}^{L}\lm_{33}^R}{M_{U_1}^2}.
\end{align}
Here, $R_{D^{(*)}}$, $F_L(D^*)$, $P_\tau(D^*)$, and $\mathcal{B}(B_{(c)}\to \tau\nu)$  would receive contributions from $U_1$.
The dominant decay modes of $U_1$ are $U_1\to c \bar \nu$, $U_1\to s\tau^+$, and $U_1\to b\tau^+$.
Note that even though $\lm^L_{33}=0$ in this scenario, a small $\mc{C}_{V_L}$ can be generated from effective $\lm^L_{33}$ coupling if, instead of up-type quark mixing, one assumes mixing in the down sector (like in \hyperlink{sce:rd1a}{Scenario RD1A}).

\vspace{0.2cm}
\hypertarget{sce:rd3}{\bulletsubsec{~Scenario RD3}} All the three free couplings $\lambda_{23}^{L}$, $\lambda_{33}^{L}$, and $\lambda_{33}^{R}$ are free to vary. Assuming mixing in the up-type quark sector, the interaction Lagrangian is given by
\begin{align}
\mathcal{L} &\supset [\lambda_{23}^{L}(\bar{c}_L\gamma_{\mu}\nu_{L} +
\bar{s}_L \gamma_{\mu}\tau_{L}) + \lambda_{33}^{L}(\bar{t}_L\gamma_{\mu}\nu_{L} + \bar{b}_L\gamma_{\mu}\tau_{L}) 
+\lambda_{33}^{R} \bar{b}_R\gamma_{\mu}\tau_{R}] U_{1}^{\mu}\nn\\
&= [\lambda_{23}^{L}(V_{us}\bar{u}_L\gamma_{\mu}\nu_{L} + V_{cs}\bar{c}_L\gamma_{\mu}\nu_{L} + V_{ts}\bar{t}_L\gamma_{\mu}\nu_{L} +
\bar{s}_L \gamma_{\mu}\tau_{L}) \nn\\
&\quad+
\lambda_{33}^{L}(V_{ub}\bar{u}_L\gamma_{\mu}\nu_{L} + V_{cb}\bar{c}_L\gamma_{\mu}\nu_{L} + V_{tb}\bar{t}_L\gamma_{\mu}\nu_{L} + \bar{b}_L\gamma_{\mu}\tau_{L})+
\lambda_{33}^{R} \bar{b}_R\gamma_{\mu}\tau_{R}] U_{1}^{\mu}.
\end{align}
This Lagrangian contributes to $\mc{C}_{V_L}$ and $\mc{C}_{S_L}$ as
\begin{align}
\mc{C}_{V_L}^{RD3} &=  \frac{1}{2\sqrt{2}G_FV_{cb}}\frac{\left( V_{cb}\lambda^L_{33} + V_{cs} \lambda^L_{23} \right)\lm_{33}^L}{M_{U_1}^2},\quad
\mc{C}_{S_L}^{RD3} =  -\frac{1}{\sqrt{2}G_FV_{cb}}\frac{\left( V_{cb}\lambda^L_{33} + V_{cs} \lambda^L_{23} \right)\lm_{33}^R}{M_{U_1}^2}.
\end{align}
The lepton flavour-universal contribution through the off-shell photon penguin diagram is
\begin{equation}
\mc{C}_9^{\rm univ} = -\frac{1}{V_{tb} V_{ts}^\ast} \frac{\lambda^L_{23} \lambda^{L}_{33}}{3\sqrt{2} G_F M_{U_1}^2}  \log(m_b^2/M_{U_1}^2).
\end{equation}
The contribution of $U_1$ to the $B_s$-$\bar{B}_s$ mixing coefficient is given  as
\begin{equation}
\mc{C}^{U_1}_{box} = \frac{\lt(\lambda^L_{23}\rt)^2\lt(\lambda^L_{33}\rt)^2 }{8\pi^2M^2_{U_1}}.
\end{equation}
In this scenario, $U_1$  dominantly decays to $c\bar\nu$, $s\tau^+$, $t\bar\n$, and $b\tau^+$ final states.
\bigskip

\subsection*{$R_{K^{(*)}}$ scenarios}
\noindent
A general Lagrangian for $b\to s\m^+\m^-$ transition can be written as~\cite{Bobeth:1999mk,Bobeth:2001jm}
\begin{align}
\mc L &\supset \frac{4G_F}{\sqrt{2}} V_{tb}V_{ts}^*\sum_{i=9,10,S,P}\left(\mc C_i\mc O_i+\mc C^\prime_i\mc O^\prime_i\right)
\end{align}
where the Wilson coefficients are evaluated at $\mu_{ren}=m_b$. The operators are given by
\begin{eqnarray*}\begin{array}{llll}
\mc O_9~ =& \displaystyle\frac\al{4\pi}\left(\bar s_L\gm_\al b_L\right)\left(\bar \mu\gm^\al \m\right),& \mc O^\prime_9~ 
=& \displaystyle\frac\al{4\pi}\left(\bar s_R\gm_\al b_R\right)\left(\bar \mu\gm^\al \m\right),\\~\\
\mc O_{10} =& \displaystyle\frac\al{4\pi}\left(\bar s_L\gm_\al b_L\right)\left(\bar \mu\gm^\al \gm_5\m\right),& \mc O^\prime_{10} 
=& \displaystyle\frac\al{4\pi}\left(\bar s_R\gm_\al b_R\right)\left(\bar \mu\gm^\al\gm_5 \m\right),\\~\\
\mc O_S~ =& \displaystyle\frac\al{4\pi}\left(\bar s_L b_R\right)\left(\bar \mu \m\right),&\mc O^\prime_S~ 
=& \displaystyle\frac\al{4\pi}\left(\bar s_R b_L\right)\left(\bar \mu \m\right),\\~\\
\mc O_P~ =& \displaystyle\frac\al{4\pi}\left(\bar s_L b_R\right)\left(\bar \m\gm_5 \m\right),& \mc O^\prime_P~ 
=& \displaystyle\frac\al{4\pi}\left(\bar s_R b_L\right)\left(\bar \mu\gm_5 \m\right)\\
\end{array}\end{eqnarray*}
where $\al$ is the fine-structure constant. Keeping the $R_{K^{(*)}}$ observables in mind, we make the following simple Ansatz:
\begin{equation}
x^{LL}_1 = 
\begin{pmatrix}
0 & 0 & 0 \\
0 & \lm^L_{22} & 0 \\
0 & \lm^L_{32} & 0
\end{pmatrix};~~
x^{RR}_1 = 
\begin{pmatrix}
0 & 0 & 0 \\
0 & \lm^R_{22} & 0 \\
0 & \lm^R_{32} & 0
\end{pmatrix}. \label{eq:rkcouplings}
\end{equation}
The $U_1$ contribution to the Wilson coefficients can be written in terms of the $\bar b\m U_1$ and $\bar s\m U_1$ couplings in general as
\begin{eqnarray}
\left.\begin{array}{lcccl}
\mc{C}_9^{U_1} &=& -\mc{C}_{10}^{U_1} &=& \displaystyle \frac{\pi }{\sqrt{2}G_FV_{tb}V_{ts}^{*}\al}\frac{\lambda^L_{s\mu}(\lambda^L_{b\mu})^*}{M_{U_1}^2} \\\\
\mc{C}_S^{U_1} &=& -\mc{C}_{P}^{U_1} &=& \displaystyle \frac{\sqrt{2}\pi }{G_FV_{tb}V_{ts}^{*}\al}\frac{\lm_{s\mu}^L(\lambda^R_{b\mu})^*}{M_{U_1}^2}  \\
\\
\mc{C}_9^{\prime~U_1} &=& \mc{C}_{10}^{\prime~U_1} &=& \displaystyle \frac{\pi }{\sqrt{2}G_FV_{tb}V_{ts}^{*}\al}\frac{\lm_{s\mu}^R(\lm_{b\mu}^{R*})}{M_{U_1}^2}  \\
\\
\mc{C}_S^{\prime~U_1} &=& \mc{C}_{P}^{\prime~U_1} &=& \displaystyle \frac{\sqrt{2}\pi }{G_FV_{tb}V_{ts}^{*}\al}\frac{\lm_{s\mu}^R(\lm_{b\mu}^{L*})}{M_{U_1}^2} 
\end{array}\right\}.\label{eq:rkstops}
\end{eqnarray}
Like in the $R_{D^{(*)}}$ scenarios, the relationship between $\{\lm_{s\m}^{L/R}, \lm_{b\mu}^{L/R}\}$ with $\{\lm^{L/R}_{22},\lm^{L/R}_{32}\}$ would depend on the particulars of the scenario we consider.
The relevant global fits of the Wilson coefficients to the $b\to s\mu^+\mu^-$ data are taken from Refs.~\cite{Altmannshofer:2017fio,Aebischer:2019mlg, Alguero:2021anc} and are listed in Table \ref{tab:globalfit}.

\begin{table}[t!]
\caption{Global fits of relevant combinations of Wilson coefficients in $b\to s\m\m$ observables~\cite{Altmannshofer:2017fio,Aebischer:2019mlg, Alguero:2021anc}.}
\begin{center}
{\linespread{1.3}\footnotesize
\begin{tabular*}{\textwidth}{l@{\extracolsep{\fill}}cccc}
\hline
$\vphantom{\Big|}$
Combinations & Best fit& $1\sg$ & $2\sg$& Corresponding scenarios  \\
\hline\hline
$\mc{C}_9^{U_1} = -\mc{C}_{10}^{U_1}$ & $-0.44$& $[-0.52, -0.37]$ &$[-0.60, -0.29]$ 
&\hyperlink{sce:rk1a}{RK1A}, \hyperlink{sce:rk1b}{RK1B}, \hyperlink{sce:rk2a}{RK2A}\\
$\mc{C}_S^{U_1} = -\mc{C}_{P}^{U_1}$ & $-0.0252$ & $[-0.0378, -0.126]$ &$[-0.0588, -0.0042]$ 
& \hyperlink{sce:rk2b}{RK2B}\\
$\mc{C}_9^{\prime~U_1}= \mc{C}_{10}^{\prime~U_1}$ & $+0.06$ & $[-0.18, +0.30]$ &$[-0.42, +0.55]$ 
& \hyperlink{sce:rk1c}{RK1C}, \hyperlink{sce:rk1d}{RK1D}, \hyperlink{sce:rk2d}{RK2D}\\
$\mc{C}_S^{\prime~U_1} = \mc{C}_{P}^{\prime~U_1}$ & $-0.0252$ & $[-0.0378, -0.126]$ &$[-0.0588, -0.0042]$ 
& \hyperlink{sce:rk2c}{RK2C}\\
\hline
\end{tabular*}}
\label{tab:globalfit}
\end{center}
\end{table}

\vspace{0.2cm}
\hypertarget{sce:rk1a}{\bulletsubsec{~Scenario RK1A}} In this scenario, only $\lm_{22}^L$ is nonzero. This generates the $\bar s\m U_1$ coupling. The $\bar b\m U_1$ coupling is generated via CKM mixing in the down-quark sector (as in \hyperlink{sce:rd1a}{Scenario RD1A} and \hyperlink{sce:rd1b}{Scenario RD1B}). The interaction Lagrangian can be written as
\begin{align}
\mathcal{L} \supset \lambda_{22}^{L}[\bar{c}_L\gamma_{\mu}\nu_{L} + (V_{cd}^*\bar{d}_L + V_{cs}^*\bar{s}_L + V_{cb}^*\bar{b}_L) \gamma_{\mu}\mu_{L})] U_{1}^{\mu}.
\end{align}
This Lagrangian contributes to the following coefficients:
\begin{align}
\mc{C}_9^{RK1A} = -\mc{C}_{10}^{RK1A} = \frac{\pi V_{cb}V_{cs}^*}{\sqrt{2}G_FV_{tb}V_{ts}^{*}\al}\frac{(\lm_{22}^L)^2}{M_{U_1}^2}.
\end{align}
The contribution to the $B_s$-$\bar{B}_s$ mixing coefficient is
\begin{equation}
\mc{C}^{U_1}_{box} = \frac{|V_{cb}|^2|V_{cs}|^2(\lambda^L_{22})^4 }{8\pi^2M^2_{U_1}}.
\end{equation}
The dominant decay modes of $U_1$ in this case are $U_1\to c\bar \nu$ and $U_1\to s\mu^+$ with almost $50\%$ BR each.

\vspace{0.2cm}
\hypertarget{sce:rk1b}{\bulletsubsec{~Scenario RK1B}} Only $\lm_{32}^L$ is nonzero. The interaction Lagrangian is given by
\begin{align}
\mathcal{L} \supset \lambda_{32}^{L}[\bar{t}_L\gamma_{\mu}\nu_{L} + (V_{td}^*\bar{d}_L + V_{ts}^*\bar{s}_L + V_{tb}^*\bar{b}_L) \gamma_{\mu}\mu_{L})] U_{1}^{\mu}.
\end{align} 
The relevant Wilson coefficients are given by
\begin{align}
\mc{C}_9^{RK1B} = -\mc{C}_{10}^{RK1B} = \frac{\pi }{\sqrt{2}G_F\al}\frac{(\lm_{32}^L)^2}{M_{U_1}^2},\label{eq:wcoeffrk1b}
\end{align}
and the contribution to the $B_s$-$\bar{B}_s$ mixing coefficient is given as
\begin{equation}
\mc{C}^{U_1}_{box} = \frac{|V_{tb}|^2|V_{ts}|^2(\lambda^L_{32})^4 }{8\pi^2M^2_{U_1}}.
\end{equation}
Here, the $\bar s\mu U_1$ coupling  is $V_{ts}^*$-suppressed. The coupling $\lm_{32}^L$ alone, however, cannot explain the $R_K^{(*)}$ anomalies. From Table~\ref{tab:globalfit} we see that the anomalies need a negative $\mc C_9$, whereas the r.h.s. of Eq.~\eqref{eq:wcoeffrk1b} is always positive (even if we consider a complex $\lm^L_{32}$). 
The dominant decay modes of $U_1$ in this case are $U_1\to t\bar \nu$ and $U_1\to b\mu^+$, and they share almost $50$\% BR each.

\vspace{0.2cm}
\hypertarget{sce:rk1c}{\bulletsubsec{~Scenario RK1C}} In this scenario, we assume only $\lm_{22}^R$ to be nonzero. The interaction Lagrangian is given by,
\begin{align}
\mathcal{L} \supset \lambda_{22}^{R}[(V_{cd}\bar{d}_R + V_{cs}\bar{s}_R + V_{cb}\bar{b}_R) \gamma_{\mu}\mu_{R}] U_{1}^{\mu}.
\end{align}
The nonzero Wilson coefficients from Eq.~\eqref{eq:rkstops} are
\begin{align}
\mc{C'}_9^{RK1C} = \mc{C'}_{10}^{RK1C} = \frac{\pi V_{cb}^{*} V_{cs}}{\sqrt{2}G_FV_{tb}V_{ts}^{*}\al}\frac{(\lm_{22}^R)^2}{M_{U_1}^2}\ ,
\end{align}
and the contribution to the $B_s$-$\bar{B}_s$ mixing coefficient is
\begin{equation}
\mc{C}^{U_1}_{box} = \frac{|V_{cb}|^2|V_{cs}|^2(\lambda^R_{22})^4 }{8\pi^2 M^2_{U_1}}.
\end{equation}
Here, the  $\bar b\mu U_1$ coupling is $V_{cb}^*$ suppressed.  In this scenario, the $U_1\to s\mu^+$ decay mode has almost $100$\% BR. 

\vspace{0.2cm}
\hypertarget{sce:rk1d}{\bulletsubsec{~Scenario RK1D}} We assume $\lm_{32}^R$ to be nonzero and the rest of the couplings to be SM-like. The interaction Lagrangian is given by
\begin{align}
\mathcal{L} \supset \lambda_{32}^{R}[(V_{td}\bar{d}_R + V_{ts}\bar{s}_R + V_{tb}\bar{b}_R) \gamma_{\mu}\mu_{R}] U_{1}^{\mu}
\end{align}
where the $\bar s\mu U_1$ coupling is $V_{ts}$ suppressed. The nonzero Wilson coefficients are
\begin{align}
\mc{C'}_9^{RK1D} = \mc{C'}_{10}^{RK1D} = \frac{\pi V_{tb}^{*} V_{ts}}{\sqrt{2}G_FV_{tb}V_{ts}^{*}\al}\frac{(\lm_{32}^R)^2}{M_{U_1}^2}.
\end{align}
In this scenario, the $U_1\to b\mu^+$ decay mode is dominant with almost $100$\% BR. The contribution to the $B_s$-$\bar{B}_s$ mixing coefficient is given as
\begin{equation}
\mc{C}^{U_1}_{box} = \frac{|V_{tb}|^2|V_{ts}|^2(\lambda^R_{32})^4 }{8\pi^2 M^2_{U_1}}.
\end{equation}

\vspace{0.2cm}
\hypertarget{sce:rk2a}{\bulletsubsec{~Scenario RK2A}} In this scenario, two couplings, namely, $\lm_{22}^L$ and $\lm_{32}^L$ are nonzero. The interaction Lagrangian is given by
\begin{align}
\mathcal{L} \supset [\lambda_{22}^{L}(\bar{c}_L\gamma_{\mu}\nu_{L} +
\bar{s}_L\gamma_{\mu}\mu_{L}) + \lambda_{32}^{L}(\bar{t}_L\gamma_{\mu}\nu_{L} +
\bar{b}_L\gamma_{\mu}\mu_{L})] U_{1}^{\mu}.
\end{align}
Here, we have not shown the CKM-suppressed couplings. The Wilson coefficients getting the dominant contributions are
\begin{align}
\mc{C}_9^{RK2A} = -\mc{C}_{10}^{RK2A} \approx \frac{\pi}{\sqrt{2}G_FV_{tb}V_{ts}^{*}\al}\frac{\lm_{22}^L \lm_{32}^L}{M_{U_1}^2}.
\end{align}
The contribution to the $B_s$-$\bar{B}_s$ mixing coefficient is
\begin{equation}
\mc{C}^{U_1}_{box} = \frac{(\lambda^L_{22})^2(\lambda^L_{32})^2 }{8\pi^2 M^2_{U_1}}.
\end{equation}
In this scenario, the dominant decay modes for $U_1$ are $b\mu^+$, $s\mu^+$, $c\bar\n$, and $t\bar\n$.

\vspace{0.2cm}
\hypertarget{sce:rk2b}{\bulletsubsec{~Scenario RK2B}} In this scenario, only $\lm_{22}^L$ and $\lm_{32}^R$ are nonzero. The interaction Lagrangian is given by
\begin{align}
\mathcal{L} \supset [\lambda_{22}^{L}(\bar{c}_L\gamma_{\mu}\nu_{L} +
\bar{s}_L\gamma_{\mu}\mu_{L}) + \lambda_{32}^{R}
\bar{b}_R\gamma_{\mu}\mu_{R}] U_{1}^{\mu}
\end{align}
Here, once again, the CKM-suppressed couplings are ignored. The Wilson coefficients getting the dominant contributions are
\begin{align}
-\mc{C}_P^{RK2B} = \mc{C}_{S}^{RK2B} \approx \frac{\sqrt{2}\pi}{G_FV_{tb}V_{ts}^{*}\al}\frac{\lm_{22}^L \lm_{32}^R}{M_{U_1}^2}.
\end{align}
For this scenario, $B_s$-$\bar{B}_s$ mixing is not relevant. The dominant decay modes of $U_1$ are $b\mu^+$, $s\mu^+$, and $c\bar\n$. 

\vspace{0.2cm}
\hypertarget{sce:rk2c}{\bulletsubsec{~Scenario RK2C}} Only $\lm_{22}^R$ and $\lm_{32}^L$ are nonzero. Ignoring the CKM-suppressed couplings, we get the following interaction Lagrangian:
\begin{align}
\mathcal{L} \supset [\lambda_{22}^{R}\bar{s}_R\gamma_{\mu}\mu_{R} + \lambda_{32}^{L}(\bar{t}_L\gamma_{\mu}\nu_{L} +
\bar{b}_L\gamma_{\mu}\mu_{L})] U_{1}^{\mu}
\end{align}
The Wilson coefficients getting the dominant contributions are
\begin{align}
\mc{C'}_P^{RK2C} = \mc{C'}_{S}^{RK2C} \approx \frac{\sqrt{2}\pi}{G_FV_{tb}V_{ts}^{*}\al}\frac{\lm_{22}^R \lm_{32}^L}{M_{U_1}^2}.
\end{align}
In this case also $B_s$-$\bar{B}_s$ mixing is not relevant. The dominant decay modes of $U_1$ are $b\mu^+$, $s\mu^+$, and $t\bar\n$. 

\vspace{0.2cm}
\hypertarget{sce:rk2d}{\bulletsubsec{~Scenario RK2D}} Only $\lm_{22}^R$ and $\lm_{32}^R$ are nonzero. Ignoring the CKM-suppressed couplings, we get
\begin{align}
\mathcal{L} \supset (\lambda_{22}^{R}\bar{s}_R\gamma_{\mu}\mu_{R} + \lambda_{32}^{R}\bar{b}_R\gamma_{\mu}\mu_{R}) U_{1}^{\mu}.
\end{align}
The Wilson coefficients getting the dominant contributions are
\begin{align}
\mc{C'}_9^{RK2D} = \mc{C'}_{10}^{RK2D} \approx \frac{\pi}{\sqrt{2}G_FV_{tb}V_{ts}^{*}\al}\frac{\lm_{22}^R \lm_{32}^R}{M_{U_1}^2}.
\end{align}
The contribution to the $B_s$-$\bar{B}_s$ mixing coefficient is given as
\begin{equation}
\mc{C}^{U_1}_{box} = \frac{(\lambda^R_{22})^2 (\lambda^R_{32})^2 }{8\pi^2 M^2_{U_1}}.
\end{equation}
The dominant decay modes of $U_1$ are $b\mu^+$ and $s\mu^+$.

\vspace{0.2cm}
\hypertarget{sce:rk4}{\bulletsubsec{~Scenario RK4}} All couplings are nonzero.
In this scenario, the interaction Lagrangian is given by,
\begin{align}
\mathcal{L} \supset& \big[\lambda_{22}^{L}(\bar{c}_L\gamma_{\mu}\nu_{L} +
\bar{s}_L\gamma_{\mu}\mu_{L}) + \lambda_{32}^{L}(\bar{t}_L\gamma_{\mu}\nu_{L} +
\bar{b}_L\gamma_{\mu}\mu_{L})+ \lambda_{22}^{R}\bar{s}_R\gamma_{\mu}\mu_{R} + \lambda_{32}^{R}\bar{b}_R\gamma_{\mu}\mu_{R}\big] U_{1}^{\mu},
\end{align}
and the dominant contributions to the Wilson coefficients can be read from Eq.~\eqref{eq:rkstops}. The main decay modes of $U_1$ are $c\bar\n$, $t\bar\n$, $s\m^+$, and $b\m^+$.

Our selection of scenarios motivated by the $R_K^{(*)}$ anomalies is not exhaustive. For example, we do not consider any three-coupling scenarios. (One can define RK3X scenarios by taking combinations of three couplings at a time for completeness. We, however, skip the three-coupling-$R_{K^{(*)}}$ scenarios since they would not add anything significant to our study.) The  single-coupling scenarios can be thought of as templates that can help us read bounds on scenarios where more than one couplings are nonzero~\cite{Mandal:2018kau,Aydemir:2019ynb}. In Table \ref{tab:globalfit}, we show the relevant global fits for the one- and two-coupling scenarios.  We have summarised the  couplings that contribute to the $R_{D^{(*)}}$ and $R_{K^{(*)}}$ observables in different scenarios in Table~\ref{tab:lmqlmi}.
\bigskip

\noindent
As mentioned earlier, one of the reasons for considering the $R_{D^{(*)}}$ and $R_{K^{(*)}}$ scenarios is that they can have different signatures at the LHC. We are now in a position to illustrate that point further. Let us consider the first two $R_{D^{(*)}}$-motivated one-coupling scenarios -- \hyperlink{sce:rd1a}{Scenario RD1A} and \hyperlink{sce:rd1b}{Scenario RD1B}. In both cases,  $\mc{C}_{V_L}$ receives a nonzero contribution proportional to the square of an unknown new coupling 
(either $\lm^L_{23}$ or $\lm^L_{33}$). Hence, from an effective theory perspective, these two look almost the same. However, the dominant decay modes of $U_1$ in these two scenarios are different -- in the first one, they are $U_1\to c\nu$ and $U_1\to s\ta$, whereas in the second one, they are $U_1\to t\nu$ and $U_1\to b\ta$.\footnote{From here on, unless necessary, we do not distinguish between particles and their antiparticles as it is not important for the LHC analysis.} As a result, a $U_1$ can produce $t+\slashed E_T$ 
or $\ta+b$ signatures in the second scenario, as opposed to the  
$\textrm{jet}+\slashed E_T$ or $\ta+\textrm{jet}$ signatures in the first one. Not only that, in the first scenario, a $U_1$ can be produced via $c$- or $s$-quark-initiated processes, as compared to the $b$-quark-initiated processes in the second one. Hence, in these two scenarios, $U_1$ would have different single production processes. Moreover, since the $b$-quark parton distribution function (PDF) is smaller than the second-generation ones, $U_1$ production cross sections would be higher in \hyperlink{sce:rd1a}{Scenario RD1A} than those in 
\hyperlink{sce:rd1b}{Scenario RD1B}.
Hence, one needs to analyse the LHC bounds  for the scenarios differently. 

\begin{table}[t!]
\caption{Summary of the coupling combinations that contribute to the $R_{D^{(*)}}$ and $R_{K^{(*)}}$ observables in different one-, two- and multi-coupling scenarios.}
\begin{center}
{\linespread{1.3}\footnotesize
\begin{tabular*}{\textwidth}{l@{\extracolsep{\fill}}ccclcccc}
\hline
$R_{D^{(*)}}$ scenarios & $\lambda_{c\nu}^L$ & $\lambda_{b\tau}^L$ & $\lambda_{b\tau}^R$ &$R_{K^{(*)}}$ scenarios&$\lambda_{s\mu}^L$ & $\lambda_{b\mu}^L$ & $\lambda_{s\mu}^R$ & $\lambda_{b\mu}^R$ \\  

\hline\hline
\hyperlink{sce:rd1a}{RD1A}  & $\lambda_{23}^L$ & $V_{cb}^*\lambda_{23}^L$ & $-$ & \hyperlink{sce:rk1a}{RK1A}  & $V_{cs}^*\lambda_{22}^L$ & $V_{cb}^*\lambda_{22}^L$ & $-$ & $-$ \\
\hyperlink{sce:rd1a}{RD1B}  & $V_{cb}\lambda_{33}^L$  & $\lambda_{33}^L$ & $-$ & \hyperlink{sce:rd1a}{RK1B}  & $V_{ts}^*\lambda_{32}^L$ & $V_{tb}^*\lambda_{32}^L$ & $-$ & $-$ \\
  &   &    &   & \hyperlink{sce:rk1c}{RK1C}  & $-$ & $-$ & $V_{cs}\lambda_{22}^R$ & $V_{cb}\lambda_{22}^R$ \\
&   &    &   & \hyperlink{sce:rk1d}{RK1D}  & $-$ & $-$ & $V_{ts}\lambda_{32}^R$ & $V_{tb}\lambda_{32}^R$   \\
\hline
\hyperlink{sce:rd2a}{RD2A}  & $V_{cs}\lambda_{23}^{L} + V_{cb}\lambda_{33}^{L}$ & $\lambda_{33}^{L}$ & $-$ &  \hyperlink{sce:rk2a}{RK2A}  & $\lambda_{22}^L$ & $\lambda_{32}^L$ & $-$ & $-$  \\
\hyperlink{sce:rd2b}{RD2B}  & $V_{cs}\lambda_{23}^{L}$  &  $-$  &  $\lambda_{33}^{R}$ & \hyperlink{sce:rk2b}{RK2B}  & $\lambda_{22}^L$ & $-$ & $-$ & $\lambda_{32}^R$  \\
&   &    &   & \hyperlink{sce:rk2c}{RK2C} & $-$ & $\lambda_{32}^L$ & $\lambda_{22}^R$ & $-$  \\
&   &   &   & \hyperlink{sce:rk2d}{RK2D}  & $-$ & $-$ & $\lambda_{22}^{R}$ & $\lambda_{32}^{R}$ \\\hline
\hyperlink{sce:rd3}{RD3}  & $V_{cb}\lambda^L_{33} + V_{cs} \lambda^L_{23}$  &  $\lambda^L_{33}$  &  $\lambda_{33}^{R}$ & \hyperlink{sce:rk4}{RK4}  & $\lambda_{22}^{L} $ & $\lambda_{32}^{L}$ & $\lambda_{22}^{R}$ & $\lambda_{32}^{R}$ \\
\hline
\end{tabular*}}
\label{tab:lmqlmi}
\end{center}
\end{table}


\section{Production modes and decays}
\label{sec:pheno}

\begin{figure*}[!t]
\captionsetup[subfigure]{labelformat=empty}
\subfloat[(a)]{\includegraphics[width=0.22\textwidth]{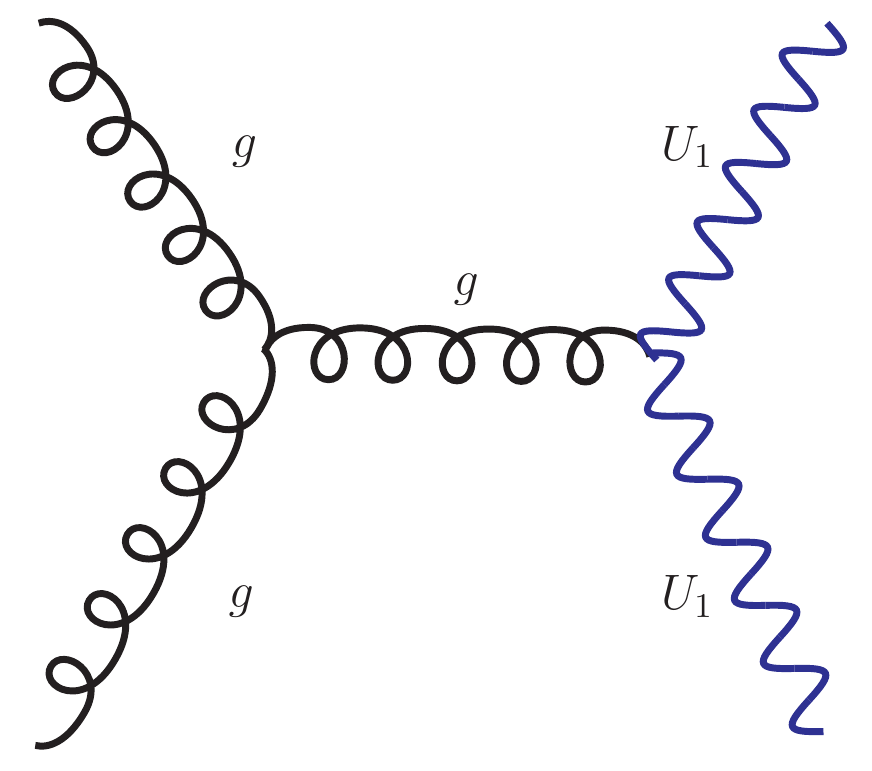}\label{fig:feynpg}}\hfill
\subfloat[(b)]{\includegraphics[width=0.22\textwidth]{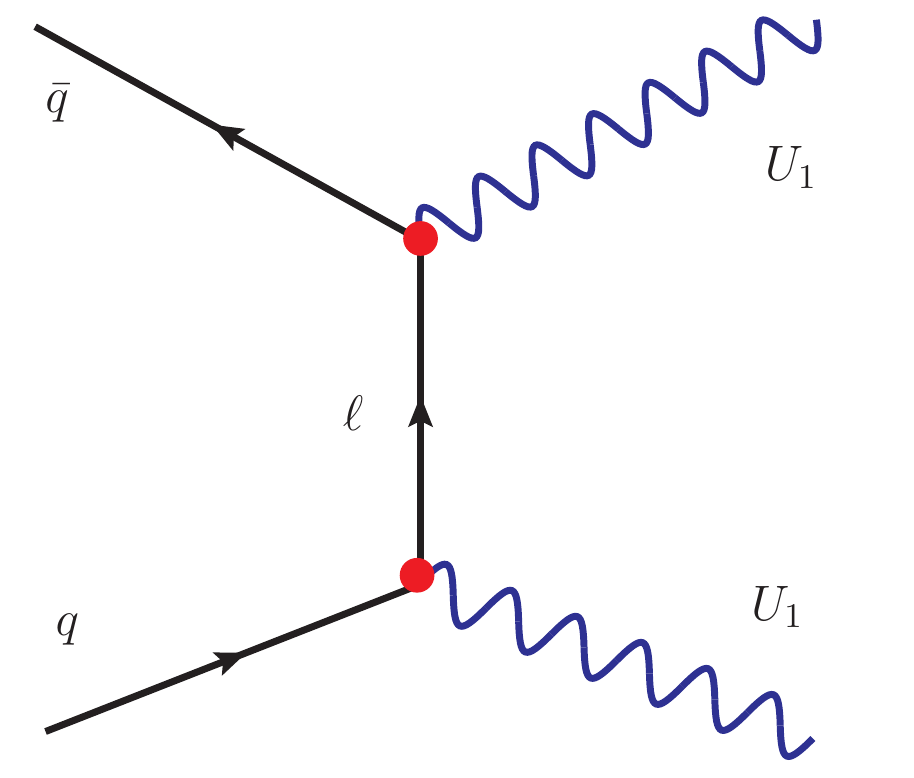}\label{fig:feynpq}}\hfill
\subfloat[(c)]{\includegraphics[width=0.22\textwidth]{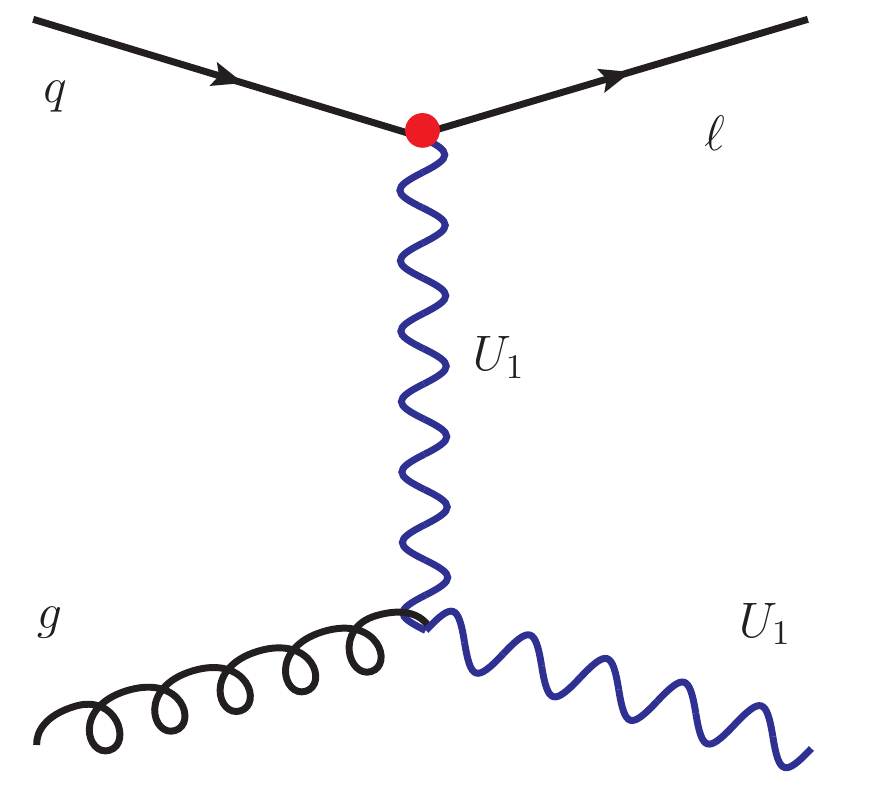}\label{fig:feyns}}\hfill
\subfloat[(d)]{\includegraphics[width=0.22\textwidth]{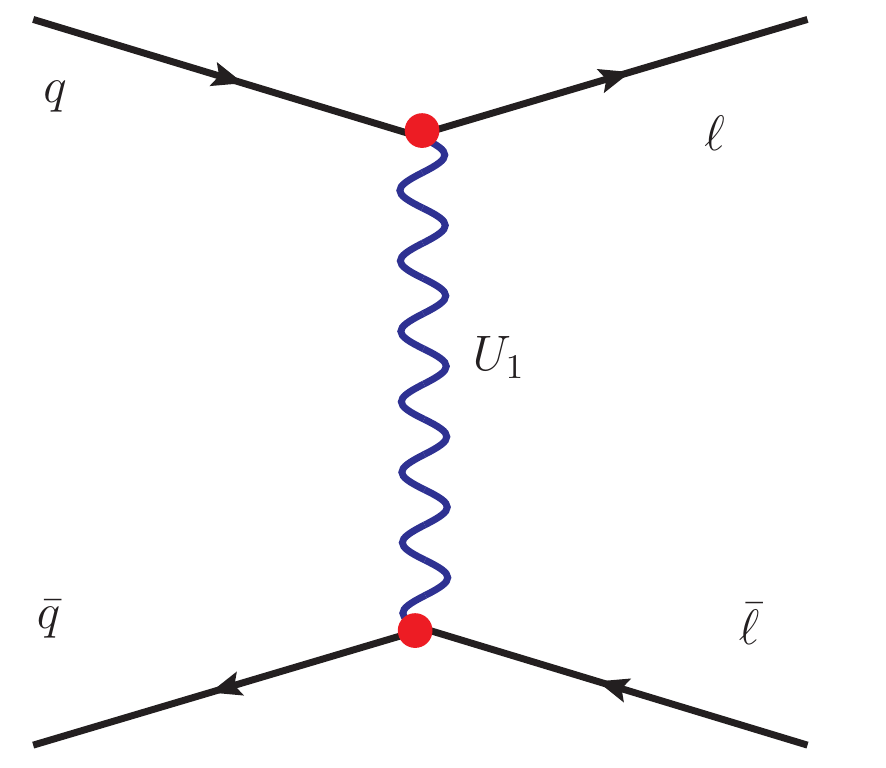}\label{fig:feynt}}
\caption{Representative Feynman diagrams for various $U_1$ production processes: (a) gluon-initiated pair production, (b) quark-initiated pair production, (c) single production, and (d) $t$-channel (nonresonant) production. The $q\ell U_1$ vertices ($\lm$) are marked with red colour.}
\label{fig:Feynman}
\end{figure*}

\noindent
We now explore the possible LHC signatures of the minimal scenarios with only one free coupling and the next-to-minimal scenarios with more than one nonzero couplings we constructed in the previous section. There are different ways to produce $U_1$ at the LHC (see Fig.~\ref{fig:Feynman}) -- resonantly (through pair and single productions) and nonresonantly (through $t$-channel $U_1$ exchange). Below, we briefly discuss various production channels and the subsequent decay modes of $U_1$ that can arise in the flavour-motivated scenarios. We also discuss how different production modes with similar final states can contribute to
the exclusion limits.
\bigskip

\subsection*{Pair production}\label{sce:PhenoPair}
\noindent
We have classified the $R_{D^{(*)}}$ scenarios with the three free couplings, $\lm_{23}^L$, $\lm_{33}^L$, and $\lm_{33}^R$. In 
\hyperlink{sce:rd1a}{Scenario RD1A} (where only $\lm_{23}^L$ is nonzero), $U_1\to s\tau$ and $U_1\to c\nu$ are the main decay modes of $U_1$ with roughly equal (about $50$\%) BRs. In this case the pair production of $U_1$ leads to the following final states (we ignore the CKM-suppressed effective couplings in the discussions on the LHC phenomenology of $U_1$ as they do not play any important role): 
\begin{equation}
\label{eq:pairRD1A}
pp\to\left\{\begin{array}{lclcl}
  U_1 U_1 &\rightarrow& s\tau \, s\tau &\equiv& \tau\tau + 2j \\
  U_1 U_1 &\rightarrow& s\tau \, c\nu &\equiv& \tau + \slashed{E}_T + 2j \\
  U_1 U_1 &\rightarrow& c\nu \, c\nu &\equiv& \slashed{E}_T + 2j
\end{array}\right\}
\end{equation}
where $j$ denotes a light jet or a $b$-jet.
Among the three channels,  the second one (i.e., $\tau + \slashed{E}_T + 2j$) has almost two times the cross section of the first or the third (a factor of $2$ comes from combinatorics), but due to the presence of missing energy, it is not fully reconstructable (or, is  difficult to reconstruct). As a result, both the first and second channels have comparable sensitivities. However, the sensitivity of the third channel, $\slashed{E}_T + 2j$, is very poor because of the two neutrinos in the final state. So far, these channels with cross-generation couplings have not been used in any LQ search at the LHC. 

In \hyperlink{sce:rd1B}{Scenario RD1B} (where only $\lm_{33}^L$ is nonzero), the pair production of $U_1$ mostly leads to the following final states: 
\begin{equation}
\label{eq:pairRD1B}
pp\to\left\{\begin{array}{lclcl}
  U_1 U_1 &\rightarrow& b\tau \, b\tau &\equiv& \tau\tau + 2j \\
  U_1 U_1 &\rightarrow& b\tau \, t\nu &\equiv& \tau + \slashed{E}_T + j_t + j  \\
  U_1 U_1 &\rightarrow& t\nu \, t\nu &\equiv& \slashed{E}_T + 2j_t
\end{array}\right\}.
\end{equation}
Here, $j_t$ represents a fat-jet originating from a top quark decaying hadronically (one can also consider the top quark's leptonic decay modes with lower cross section). It is possible to tag the (boosted) top-jets with sophisticated jet-substructure techniques and thus improve the second and third channels' prospects. The symmetric 
$\slashed{E}_T + 2j_t$ channel has been considered in Refs.~\cite{Biswas:2018snp,Vignaroli:2018lpq}. The asymmetric channel, the one with single $\tau$, one top-jet, and missing energy ($\tau + \slashed{E}_T + j_t + b$), has started receiving attention only very recently~\cite{Sirunyan:2020zbk}. Due to the factor of $2$ coming from combinatorics, this channel has a bigger cross section. Hence, its unique final state might act as a smoking-gun signature for this type of scenarios (i.e., ones with non-negligible $\lm_{33}^L$).

If only $\lm_{33}^R$ is nonzero, $U_1$ cannot resolve the $R_{D^{(*)}}$ anomalies anymore as it is not possible to generate the necessary couplings in that case. Here, $U_1$ entirely decays through the $U_1\to b\tau$ mode and
contributes to the $b\tau \, b\tau \equiv \tau\tau + 2j$ final state~\cite{Sirunyan:2018kzh}.

When two or more couplings are nonzero simultaneously (\hyperlink{sce:rd2a}{Scenario RD2A}, \hyperlink{sce:rd2b}{Scenario RD2B} and \hyperlink{sce:rd3}{Scenario RD3}) with comparable strengths, numerous possibilities arise (Reference~\cite{Aydemir:2019ynb} discusses this in the context of scalar LQ searches). It is then possible to have all the final states shown in Eqs.~\eqref{eq:pairRD1A} and~\eqref{eq:pairRD1B}. One can have more asymmetric channels like $pp\to U_1U_1\to s\ta b\ta$ etc. The strength of any particular channel would depend on the couplings involved in production (if we do not ignore the small $t$-channel lepton exchange) as well as the BRs involved (the dependence of the pair production signal on multiple couplings is made explicit in Appendix~\ref{sec:appendixA}).

The $R_{K^{(*)}}$ scenarios have similar signatures with muons in the final states. 
When only $\lm_{22}$ is nonzero (\hyperlink{sce:rk1a}{Scenario RK1A}), we can easily obtain the possible final states 
by replacing $\tau\to \mu$ in Eq.~\eqref{eq:pairRD1A}. In \hyperlink{sce:rk1B}{Scenario RK1B},
the possible final states are obtained by replacing $\tau\to \mu$ in 
Eq.~\eqref{eq:pairRD1B}. In \hyperlink{sce:rk1c}{Scenario RK1C}, the BR of the $U_1\to s\mu$ decay is $100$\% leading to the process, $U_1 U_1 \rightarrow s\mu \, s\mu \equiv \mu\mu + 2j$. Similarly, in \hyperlink{sce:rk1d}{Scenario RK1D}, the BR of the $U_1\to b\mu$ decay is $100$\% leading to the same two-muon$+$two-jet final states through the $U_1 U_1 \rightarrow b\mu \, b\mu \equiv \mu\mu + 2j$ process. Like the $R_{D^{(*)}}$ scenarios with more than one nonzero couplings, these scenarios also lead to numerous interesting possibilities~\cite{Aydemir:2019ynb}.
The LHC is yet to perform searches for LQs in most of the asymmetric channels and some of the symmetric channels.

\begin{table}[!t]
\caption{Effect of branching ratios on different final states generated from the $pp\to U_1U_1$ process in various one and two-coupling scenarios. Here, we show the possible final states and the fraction of $U_1$ pairs producing them. One multiplies the pair production cross section with the fractions shown in the table to estimate its contribution to various channels in the narrow width approximation. Here, $0\leq\xi\leq\frac12$ is a free parameter. We have ignored the mass differences among the daughter particles.}
\centering{\linespread{1.5}\footnotesize
\begin{tabular*}{\textwidth}{l@{\extracolsep{\fill}}cccccc}
\hline
Nonzero couplings & \multicolumn{6}{c}{Signatures}\\\hline\hline
&$\tau\tau + 2j$  & $\tau + \slashed{E}_T + 2j$ & $\slashed{E}_T + 2j$ & $\tau + \slashed{E}_T + j_t + j$ & $\slashed{E}_T + 2j_t$ & $\slashed{E}_T + j_t +j$\\ \cline{2-7}
$\lm_{23}^L$ (\hyperlink{sce:rd1a}{Scenario RD1A}) & $0.25$ & $0.50$ & $0.25$ & $-$ & $-$& $-$  \\
$\lm_{33}^L$ (\hyperlink{sce:rd1b}{Scenario RD1B}) & $0.25$ & $-$ & $-$ & $0.50$ & $0.25$& $-$  \\ 
$\lm_{33}^R$ & $1.00$ & $-$ & $-$ & $-$ & $-$ & $-$ \\ 
$\lm_{23}^L,\lm_{33}^L$ (\hyperlink{sce:rd2a}{Scenario RD2A}) & $0.25$ & $\xi$ & $\xi^2$ & $\frac12-\xi$ & $\lt(\frac12-\xi\rt)^2$& $2\xi\lt(\frac12-\xi\rt)$  \\
$\lm_{23}^L,\lm_{33}^R$ (\hyperlink{sce:rd2b}{Scenario RD2B}) & $\lt(\frac12+\xi\rt)^2$ & $2\lt(\frac14-\xi^2\rt)$ & $\lt(\frac12-\xi\rt)^2$ & $-$ & $-$& $-$  \\
\cline{2-7}
 & $\mu\mu + 2j$  & $\mu + \slashed{E}_T + 2j$ & $\slashed{E}_T + 2j$ & $\mu + \slashed{E}_T + j_t + j$ & $\slashed{E}_T + 2j_t$  & $\slashed{E}_T + j_t +j$\\ \cline{2-7}
$\lm_{22}^L$ (\hyperlink{sce:rk1a}{Scenario RK1A})& $0.25$ & $0.50$ & $0.25$ & $-$ & $-$ & $-$  \\ 
$\lm_{32}^L$ (\hyperlink{sce:rk1b}{Scenario RK1B})& $0.25$ & $-$ & $-$ & $0.50$ & $0.25$ & $-$  \\ 
$\lm_{22}^R$ (\hyperlink{sce:rk1c}{Scenario RK1C})& $1.00$ & $-$ & $-$ & $-$ & $-$ & $-$  \\
$\lm_{32}^R$ (\hyperlink{sce:rk1d}{Scenario RK1D})& $1.00$ & $-$ & $-$ & $-$ & $-$ & $-$  \\ 
$\lm_{22}^L,\lm_{32}^L$ (\hyperlink{sce:rk2a}{Scenario RK2A}) & $0.25$ &  $\xi$ & $\xi^2$ & $\frac12-\xi$ & $\lt(\frac12-\xi\rt)^2$& $2\xi\lt(\frac12-\xi\rt)$  \\
$\lm_{22}^L,\lm_{32}^R$ (\hyperlink{sce:rk2b}{Scenario RK2B}) & $\lt(\frac12+\xi\rt)^2$ & $2\lt(\frac14-\xi^2\rt)$ & $\lt(\frac12-\xi\rt)^2$ & $-$ & $-$& $-$ \\
$\lm_{22}^R,\lm_{32}^L$ (\hyperlink{sce:rk2c}{Scenario RK2C}) & $\lt(\frac12+\xi\rt)^2$ & $-$ & $-$ & $2\lt(\frac14-\xi^2\rt)$ & $\lt(\frac12-\xi\rt)^2$& $-$  \\
$\lm_{22}^R,\lm_{32}^R$ (\hyperlink{sce:rk2d}{Scenario RK2D}) & $1.00$ & $-$ & $-$ & $-$ & $-$ & $-$ \\
\hline
\end{tabular*}}
\label{tab:process}
\end{table}

In Table~\ref{tab:process}, we have summarized the possible final states from $U_1$ pair production and the fraction of $U_1$ pairs producing the final states in the one- and two-coupling scenarios. The  
fractions depend on combinatorics and the relevant $U_1$ BRs. 
(Here, we have ignored the possible minor correction due the the mass differences between different final states, i.e., assumed all final state particles are much lighter than  $U_1$.) 
For example, in  \hyperlink{sce:rd1a}{Scenario RD1A}, since $\bt(U_1\to s\ta)\approx\bt(U_1\to c\n)\approx50\%$,  
only $25\%$ of the produced $U_1$ pairs would decay to either  $\ta\ta+2j$ or $\slashed E_T+2j$, whereas, as explained above, $50\%$ of them would decay to the $\ta+\slashed E_{T}+2j$ final state. Interestingly, we see that even in some two-coupling scenarios the fractions corresponding to the $\ta\ta/\m\m+2j$ final states are constant irrespective of the relative magnitudes of the couplings -- for example, it is $25\%$ in \hyperlink{sce:rd2a}{Scenario RD2A} or $100\%$ in \hyperlink{sce:rk2d}{Scenario RK2D}. This is interesting, because  in the presence of two nonzero couplings, one normally expects the fraction corresponding to a particular final state to depend on their relative strengths. This, of course, happens because we sum over the possible flavours of the jets. Moreover, we show that it is possible to parametrise all final states with just one free parameter ($\xi)$. Such simple parametrisations could guide us in future $U_1$ searches at the LHC.

Note that the model dependence of the pair production of $U_1$ appears in two places. One occurs  through the free parameter $\kp$ present in the kinetic terms ($ig_s\kp U_{1\mu}^\dagger T^a U_{1\nu}G^{a~\mu\nu}$)~\cite{Dorsner:2016wpm,Bhaskar:2020gkk}. The  pair production cross section depends on $\kp$. The other occurs in the contribution of the $t$-channel
lepton/neutrino exchange. The amplitudes of these diagrams grow as $\lm^2$, and the cross section
grows as $\lm^4$. Although the $\lm$ dependence of the pair production is negligible for small $\lm$ values, it can become significant for larger couplings. As we see later, the pair production channels produce a relatively minor contribution to the final exclusion limits. Therefore, we take a benchmark value for $\kp$ by setting $\kp=0$ in our analysis. However, we keep the $\lm$-dependent terms in the pair production contributions (see Appendix~\ref{sec:appendixA}).
\bigskip

\subsection*{Single production}
\noindent
In the single-production channels, a $U_1$ is produced in association with other SM particles. There are two types of single productions of our interest: (a) where a $U_1$ is produced in association with a lepton, i.e., $U_1 \m$, $U_1 \tau$ or $U_1 \nu$ and (b) where a $U_1$ is produced with a lepton and a jet, i.e., $U_1 \m j$, $U_1\tau j$ or $U_1\nu j$. One has to be careful while computing the second type of process as the set of Feynman diagrams for them might overlap with the pair production ones when the lepton-jet pair originates in a LQ decay. We keep the two types of single production contributions in our analysis by carefully avoiding any double-counting with the pair production contribution~\cite{Mandal:2015vfa,Mandal:2012rx,Mandal:2016csb}. Single productions of $U_1$ are fully model-dependent processes; they depend on the coupling $\lm$ as well as $\kp$~\cite{Bhaskar:2020gkk}. Like the pair production, the single production processes can also be categorised into symmetric and asymmetric channels~\cite{Aydemir:2019ynb}. In
\hyperlink{sce:rd1a}{Scenario RD1A}, we have the following single production channels: 
\begin{equation}
\label{eq:singRD1A}
pp\to\left\{\begin{array}{lclcl}
U_1\tau + U_1\tau j & \rightarrow &(s\tau)\tau + (s\tau)\tau j &\equiv& \tau\tau + n\,j \\
U_1\nu + U_1\nu j & \rightarrow &(c\nu)\nu + (c\nu)\nu j &\equiv& \slashed{E}_T + n\,j \\
U_1\tau + U_1\tau j & \rightarrow &(c\nu)\tau + (c\nu)\tau j &\equiv& \tau + \slashed{E}_T + n\,j \\
U_1\nu + U_1\nu j & \rightarrow &(s\tau)\nu + (s\tau)\nu j &\equiv& \tau + \slashed{E}_T + n\,j
\end{array}\right\}.
\end{equation}
Notice that the single production processes produce similar final states as the pair production. In the above equation, the first and the second channels are symmetric, whereas the third and the forth  are asymmetric. In the $\tau + \slashed{E}_T + n\,j$ final state, both $pp\to U_1\tau + U_1\tau j$ and $pp\to U_1\nu + U_1\nu j$ contribute. This channel also has not been considered for LQ searches so far.
\hyperlink{sce:rd1b}{Scenario RD1B} is very similar to \hyperlink{sce:rd1a}{Scenario RD1A} and gives the same final states if we treat the $b$-jet as a light jet. If only $\lm_{33}^R$ is nonzero, $U_1$ decays only to $b\tau$. Thus this scenario only leads to the $ (b\tau)\tau + (b\tau)\tau j \equiv \tau\tau + n\,j$ final state.

The possible final states in case of \hyperlink{sce:rk1a}{Scenario RK1A} can be obtained 
by replacing $\tau\to \mu$ in Eq.~\eqref{eq:singRD1A}. In \hyperlink{sce:rk1b}{Scenario RK1B}, we have some interesting signatures from  boosted top quarks in the final states,
\begin{equation}
\label{eq:singRD1B}
pp\to\left\{\begin{array}{lclcl}
U_1\mu + U_1\mu j &\rightarrow& (b\mu)\mu + (b\mu)\mu j &\equiv& \mu\mu + n\,j \\
U_1\mu + U_1\mu j &\rightarrow& (t\nu)\mu + (t\nu)\mu j &\equiv& \mu + \slashed{E}_T + j_t +n\, j  \\
U_1\n + U_1\nu j &\rightarrow& (b\mu)\nu + (b\mu)\nu j &\equiv& \mu + \slashed{E}_T + n\,j \\
U_1\n + U_1\nu j &\rightarrow& (t\nu)\nu + (t\nu)\nu j &\equiv& \slashed{E}_T + j_t + n\,j \\
\end{array}\right\}.
\end{equation}
These final states can also come from pair production in \hyperlink{sce:rk1b}{Scenario RK1B}.
In \hyperlink{sce:rk1c}{Scenario RK1C}, the $U_1\to s\mu$ decay has $100$\% BR, and it leads to the process $U_1\mu+U_1\mu j \rightarrow (s\mu)\mu+ (s\mu)\mu j \equiv \mu\mu + n\,j$. In  \hyperlink{sce:rk1d}{Scenario RK1D}, the $U_1\to b\mu$ decay mode has $100$\% BR. It leads to the process $U_1\mu+U_1\mu j \rightarrow (b\mu)\mu+(b\mu)\mu j \equiv \mu\mu + n\,j$.
\bigskip

\subsection*{Nonresonant production and interference}
\noindent
A $U_1$ can be exchanged in the $t$-channel and give rise to both dilepton and lepton$+$missing-energy final states [see e.g., Fig.~\ref{fig:feynt}]. As the cross sections of the nonresonant production grows as $\lm^4$, this
channel becomes important for large values of the new couplings. Especially when the mass of the $U_1$ is large, the nonresonant production contributes more than the resonant pair and single productions. There is a possibility of large interference of the nonresonant processes with the SM backgrounds like $pp\to \gm/Z (W)\to \ell\ell~(\ell+\slashed{E}_T)$. The interference contribution grows as $\lm^2$ but the contribution can be significant due to the large SM background. For $U_1$, the interference is destructive in nature. 
However, depending on the parameter/kinematic region we consider, the cross section of the exclusive $pp\to\ell\ell~(\ell+\slashed{E}_T)$ process can be bigger or smaller than the SM-only contribution~\cite{Bansal:2018eha} because the total nonresonant contribution, including the term proportional to $\lm^4$ and the destructive $\lm^2$ term, can be both positive or negative.
\bigskip

\begin{figure*}[!t]
\captionsetup[subfigure]{labelformat=empty}
\subfloat[\quad\quad\quad(a)]{\includegraphics[width=0.48\textwidth]{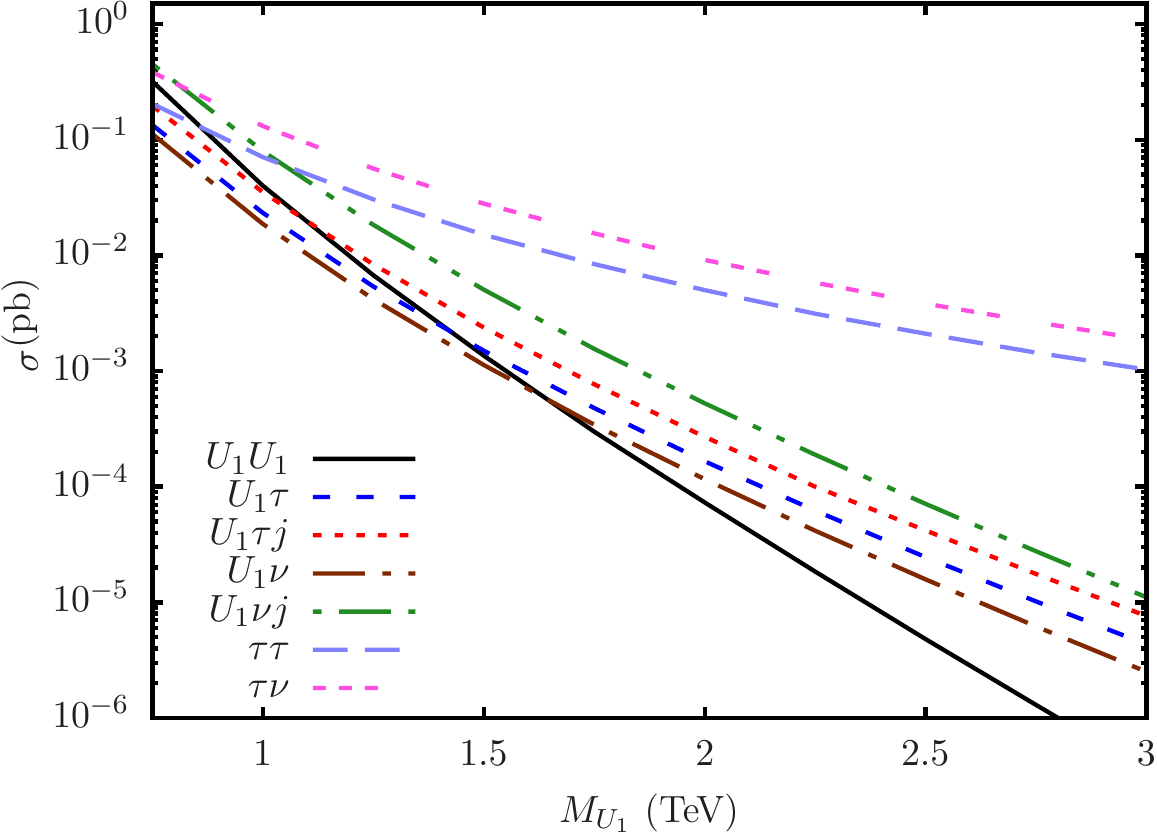}\label{fig:sigma_mass_tata}}\hfill
\subfloat[\quad\quad\quad(b)]{\includegraphics[width=0.48\textwidth]{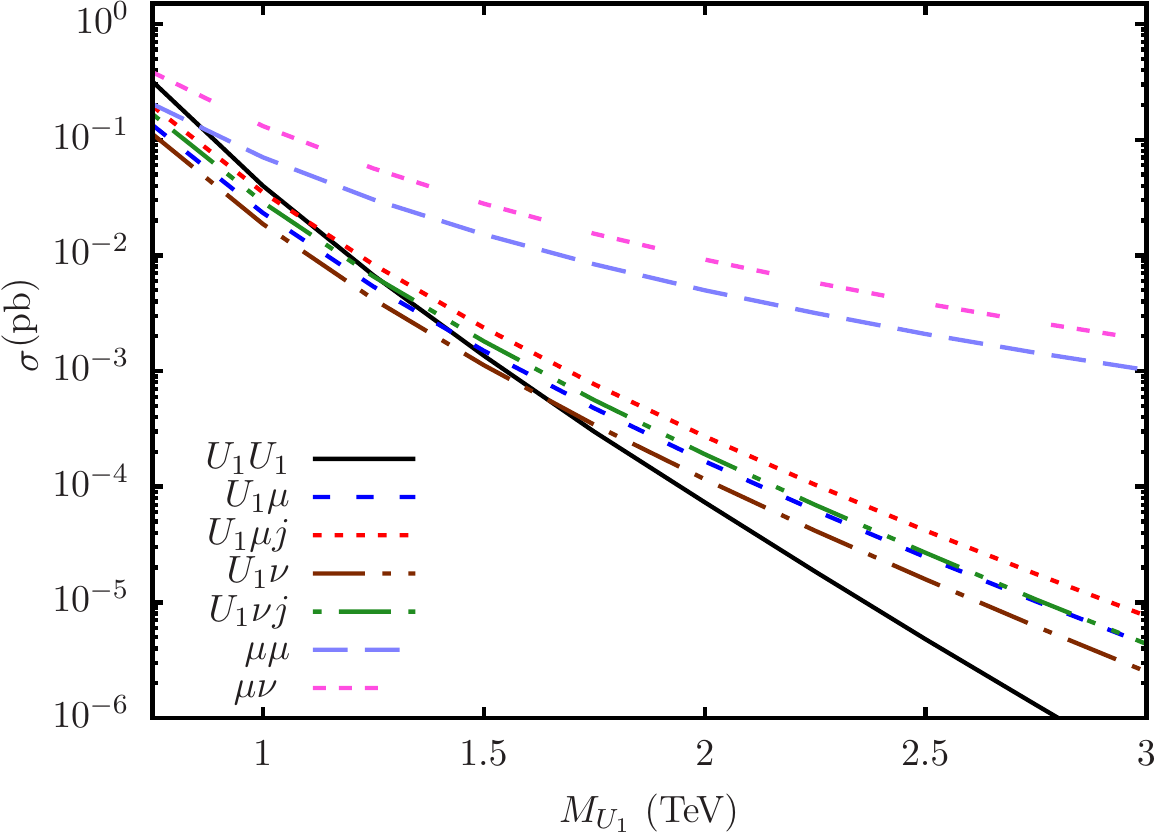}\label{fig:sigma_mass_mumu}}
\caption{Parton-level cross sections of various production modes of $U_1$ LQ as functions of $M_{U_1}$. These cross sections are computed at the 13 TeV LHC for benchmark couplings, $\lm_{23}^L=1$ (left) and $\lm_{22}^L=1$ (right) with 
$\kappa=0$. Here, $j$ stands for all light jets including the $b$-jet. A basic generation-level cut, $p_{\rm T}> 20$ GeV is applied on the jets and leptons.}
\label{fig:sigma_mass}
\end{figure*}

\noindent
In Fig.~\ref{fig:sigma_mass}, we show the parton-level cross sections of various production modes of $U_1$ as a function of $M_{U_1}$. In Figs.~\ref{fig:sigma_mass_tata} and~\ref{fig:sigma_mass_mumu} the cross sections have been obtained by setting $\kappa$=0 and the new couplings, $\lambda_{23}^L=1$ and $\lambda_{22}^L=1$ respectively. The pair production cross section is the same in both figures as it is insensitive to the $\lambda$ couplings. As expected, the single production cross sections are more significant at higher mass values. Processes like $U_1\tau$j, $U_1\mu$j, $U_1\nu$j are generated after ensuring that no more than one onshell LQ contributes to the cross section to avoid contamination from the pair production process. The nonresonant LQ production cross section does not depend very strongly on the LQ mass. With nonzero $\lambda_{23}^L$ and $\lambda_{22}^L$, we now have the possibility of producing $U_1$ (that couples to the third-generation fermions) through charm- and/or strange-initiated processes at the LHC.

There are some phenomenological consequences of having more than one coupling. The presence of multiple couplings affects the BRs. For example, we see from Table~\ref{tab:process} that BRs for one-coupling scenarios are different from those in two coupling ones. Then, different single and nonresonant production (including its interference with the SM background) processes may or may not become significant depending on the strength of various couplings. All these can significantly affect the exclusion limits. 

\section{Recast of dilepton data}\label{sec:recast} 
\noindent
From the different production mechanisms of $U_1$ discussed in the previous section, it is evident that pair, single and nonresonant productions can give rise to dilepton 
($\ell\ell+jets$) and/or monolepton plus missing-energy, $\ell+\slashed{E}_T+jets$ signatures. 
However, as pointed out in Ref.~\cite{Mandal:2018kau} for $S_1$ LQ, the bounds on the LQ model parameter space from the dilepton resonance search data are more stringent. Therefore, apart from the direct search bounds, we rely only on the resonant dilepton searches ($pp\to Z^\prime\to\ell\ell$)~\cite{Aad:2020zxo,Sirunyan:2021khd} and recast the bounds in terms of $U_1$ parameters for various scenarios. Note that the number of jets are not restricted in those searches, and hence all production modes of $U_1$ with $\ell\ell+jets$ final states would contribute in the exclusion limits. As shown in~\cite{Mandal:2018kau}, the interference of the $t$-channel $U_1$ exchange process with the SM background play the leading role in determining the exclusion limits. However, pair, single, and nonresonant productions also contribute non-negligibly, especially in the lower mass region. Since, the kinematics of different $U_1$ contributions to the $\ell\ell+jets$ channel are different from those of the resonant dilepton production ($pp\to Z^\prime\to\ell\ell$), recasting is nontrivial, especially when multiple new couplings are present. Possible interference among different signal processes complicate the recasting further. We systematically take care of all these factors in our analysis. We explain our method in Appendix~\ref{sec:appendixA}.
\bigskip

\subsection*{ATLAS $\tau\tau$ search}\label{sce:atlastautau}
\noindent
The ATLAS Collaboration searched for a heavy particle decaying to two taus at the $13$ TeV LHC with $139$ fb$^{-1}$ integrated luminosity~\cite{Aad:2020zxo}. The analysis comprised events categorised on the basis of two modes of $\tau$ decays. In the first, one has both $\tau$s decaying hadronically ($\tau_{had}\tau_{had}$). In the second, one tau decays hadronically and the other leptonically ($\tau_{had}\tau_{lep}$). We provide an outline of the basic event selection criteria for the $\tau\tau$ channel.
\begin{itemize}
    \item The $\tau_{had}\tau_{had}$ channel has
    \begin{enumerate}
    \item[--] at least two hadronically decaying $\tau$'s with no additional electrons or muons,
    \item[--] two $\tau_{had}$'s with $p_{\rm T}$ > 65 GeV. They should be oppositely charged and separated in the azimuthal plane by |$\Delta\phi(p_{\rm T}^{\tau_1},p_{\rm T}^{\tau_2})$| > 2.7 rad.
    \item[--] $p_{\rm T}$ of leading $\tau$ must be > 85 GeV.
    
    \end{enumerate}
    \item The $\tau_{lep}\tau_{had}$ channel has one $\tau_{had}$ and only one $\ell = e$ or $\mu$ such that
    \begin{enumerate}
        \item[--] the hadronic $\tau$ has $p_{\rm T}$ > 25 GeV and |$\eta(\tau_{had})$| < 2.5(excluding 1.37 < |$\eta$| < 1.52),
        \item[--] if $\ell = e$, then |$\eta$| < 2.47 (excluding 1.37 < |$\eta$| < 1.52) and if $\ell = \mu$ then |$\eta$| < 2.5,
        \item[--] the lepton has $p_{\rm T}(\ell)$ > 30 GeV with azimuthal separation from the $\tau_{had}$, |$\Delta\phi(p_{\rm T}^{\ell},p_{\rm T}^{\tau_{had}})$| > 2.4.
        \item[--] the transverse mass on the selected lepton and missing transverse momentum, $m_{\rm T}(p_{\rm T}^{\ell},\slashed{E}_{\rm T})$ > 40 GeV.
        \item [--] If $\ell = e$, to reduce the background from $Z\to e e$ events with an invariant mass for $\tau-\ell$ pair between 80 GeV and 110 GeV are rejected.
    \end{enumerate}
\end{itemize} 
The transverse mass is defined as
      \begin{equation}
          m_{\rm T}(p_{\rm T}^{A},p_{\rm T}^{B}) = \sqrt{2p_{\rm T}^Ap_{\rm T}^B \Big\{1 - \cos\Delta\phi(p_{\rm T}^{A},p_{\rm T}^{B})\Big\}}.
      \end{equation}  
The analysis also make use of the total transverse mass defined as
\begin{equation}
    m_{\rm T}^{tot}(\tau_1,\tau_2,\slashed{E}_{\rm T}) = \sqrt{m_T^2(p_{\rm T}^{\tau_1},p_{\rm T}^{\tau_2}) + m_T^2(p_{\rm T}^{\tau_1},\slashed{E}_{\rm T}) + m_T^2(p_{\rm T}^{\tau_2},\slashed{E}_{\rm T})},
\end{equation} 
Here, $\tau_2$ in the $\tau_{lep}\tau_{had}$ channel represents the lepton. We use the distribution of the observed and the SM events with respect to $m_{\rm T}^{tot}$ presented in the analysis.

\subsection*{CMS $\mu\mu$ search}\label{sce:cmsmumu}
\noindent
 A search for nonresonant excesses in the dilepton channel was performed by the CMS experiment at a centre-of-mass energy of $13$ TeV corresponding to a integrated luminosity of $140$ fb$^{-1}$~\cite{Sirunyan:2021khd}. The event selection criteria that we use in our analysis can be summarised as
\begin{itemize}
    \item In the dimuon channel, the requirement is that both of the muons must have $|\eta| < 2.4$ and $p_{\rm T} > 53$ GeV. The invariant mass of the muon pair is $m_{\mu\mu}$ > 150 GeV.
\end{itemize} 
We use the distribution of the observed and the SM events with respect to the invariant mass of the muon pair, $m_{\mu\mu}$ to extract bounds.
\bigskip

\begin{table}[]
\caption{The table displays the cross section ($\sigma$) in fb, efficiency ($\epsilon$) in \% and number of events $(\mc N)$ surviving the cuts applied in the dilepton searches from various production processes. The superscripts are explained in Appendix~\ref{sec:appendixA}. The negative signs in the interference contributions signify destructive interference. }
\label{tab:cross}
\begin{center}
{\linespread{1.3}\footnotesize
\begin{tabular*}{\textwidth}{l@{\extracolsep{\fill}}rrr @{\extracolsep{\fill}}rrr @{\extracolsep{\fill}}rrr @{\extracolsep{\fill}}rrr}\hline
& \multicolumn{3}{c}{Pair production} & \multicolumn{3}{c}{Single production}  & \multicolumn{3}{c}{$t$-channel LQ}   & \multicolumn{3}{c}{Interference}  \\\cline{2-4}\cline{5-7}\cline{8-10}\cline{11-13}
\multirow{-2}{*}{\begin{tabular}[c]{@{}l@{}}Mass\\ (Tev)\end{tabular}} & $\sigma^p$& $\epsilon^p$& $\mc N^p$  & $\sigma^s$& $\epsilon^s$& $\mc N^s$ & $\sigma^{nr4}$& $\epsilon^{nr4}$& $\mc N^{nr4}$ & $\sigma^{nr2}$& \multicolumn{1}{r}{$\epsilon^{nr2}$} & \multicolumn{1}{r}{$\mc N^{nr2}$} \\\hline
\multicolumn{13}{c}{Contribution to $\ta\ta$ signal~\cite{Aad:2020zxo}}\\\hline
\multicolumn{13}{l}{$\displaystyle\lm_{23}^L=1$ (\hyperlink{sce:rd1a}{Scenario RD1A})}
\\\hline
1.0  & 40.87  & 2.33  & 8.59 & 58.80  & 3.30  & 35.07   & 70.57  & 7.22  & 183.33  & -232.63   & 3.17  & -266.21   \\
1.5  & 1.39  & 1.50  & 0.19  & 3.91  & 2.74  & 1.93  & 14.94  & 7.00   & 37.77  & -104.31   & 3.34  & -125.62  \\
2.0  & 0.08  & 1.01   & 0.01  & 0.44  & 2.50  & 0.20   & 5.04   & 7.25  & 13.19   & -58.79  & 3.28  & -69.57  
\\\hline
\multicolumn{13}{l}{$\displaystyle\lm_{33}^L=1$ (\hyperlink{sce:rd1b}{Scenario RD1B})}
 \\\hline
1.0  & 35.67  & 1.69  & 5.43 & 29.00 & 2.57 & 13.46 & 20.20  & 6.21  & 45.26  &   -75.02 & 3.08  & -83.41   \\
1.5  & 1.17  & 1.09  & 0.11  & 1.72  & 2.16  & 0.67   & 4.31  & 6.22  & 9.68  & -33.62    &  2.88 & -33.01  \\
2.0  & 0.06  & 0.81  & 0.00  & 0.17  & 1.98  & 0.06   & 1.39   & 6.27  & 3.15   & -18.97  & 2.88  & -19.71  
\\\hline
\multicolumn{13}{l}{$\displaystyle\lm_{33}^R=1$}
 \\\hline
1.0  & 35.67  & 1.74  & 22.45 & 29.18 & 2.43 & 25.62 & 20.17 & 6.45 & 46.97 & -27.4 & 3.32 & -32.83   \\
1.5  & 1.17  & 1.10  & 0.46  & 1.69  & 1.88 & 1.15  & 4.31  & 6.47 & 10.06  & -12.31 & 3.27 & -14.54  \\
2.0  & 0.06  & 0.84    & 0.02   & 0.17  & 1.57 & 0.10  & 1.39  & 6.33 & 3.18  & -6.94 & 3.26 & -8.17  
\\\hline
\multicolumn{13}{c}{Contribution to $\m\m$ signal~\cite{Sirunyan:2021khd}}\\\hline
\multicolumn{13}{l}{$\displaystyle\lm_{22}^L=1$ (\hyperlink{sce:rk1a}{Scenario RK1A})}
\\\hline
1.0  & 40.89  & 71.88  & 265.27 & 58.68  & 72.66  & 769.52 & 70.40  & 62.77  & 1595.21  & -233.00   & 42.73  & -3594.15   \\
1.5  & 1.39  & 64.44 & 8.10  & 3.91  & 71.35  & 50.30 & 15.20  & 64.33   & 352.97  & -105.00   & 42.59  & -1614.37 \\
2.0  & 0.08  & 52.62 & 0.36  & 0.44  & 70.15  & 5.60  & 5.00   & 64.22  & 115.92   & -58.80  & 43.08  & -914.54  
\\\hline
\multicolumn{13}{l}{$\displaystyle\lm_{22}^R=1$ (\hyperlink{sce:rk1b}{Scenario RK1B})}
 \\\hline
1.0  & 38.91  & 71.74  & 1007.69 & 58.29  & 72.36 & 1522.36 & 70.43  & 62.69  & 1593.99  & -82.52   & 49.17  & -1464.79   \\
1.5  & 1.32  & 64.18  & 30.64  & 3.81  & 68.62  & 94.40  & 15.21  & 64.20   & 352.57  & -37.33  &  49.09  & -661.52  \\
2.0  & 0.07  & 52.50  & 1.36  & 0.42  & 63.79  & 9.78  & 5.00   & 64.53  & 116.48  & -21.0  & 48.62  & -368.53  
\\\hline
\multicolumn{13}{l}{$\displaystyle\lm_{32}^L=1$ (\hyperlink{sce:rk1c}{Scenario RK1C})}
 \\\hline
1.0  & 35.67  & 71.59  & 230.45 & 28.93 & 72.74 & 379.76 & 20.00 & 63.49 & 458.17 & -75.30 & 39.10 & -1062.87   \\
1.5  & 1.17  & 64.46 & 6.78  & 1.72  & 72.33 & 22.44 & 4.29  & 64.58 & 100.49 & -33.70 & 39.82 & -484.39  \\
2.0  & 0.06  & 52.47  & 0.29  & 0.17  & 71.77 & 2.22 & 1.41  & 64.90 & 33.04 & -19.00 & 40.12 & -275.17  
\\\hline
\multicolumn{13}{l}{$\displaystyle\lm_{32}^R=1$ (\hyperlink{sce:rk1d}{Scenario RK1D})}
 \\\hline
1.0  & 35.67  & 71.75  & 923.90 & 29.04 & 72.37 & 758.73   & 20.05  & 63.73  & 461.36  & -26.29   & 45.77  & -434.43   \\
1.5  & 1.17  & 64.60  & 27.19 & 1.69  & 69.28  & 42.27 & 4.29  & 64.43   & 99.74  & -11.84   & 46.32  & -197.94 \\
2.0  & 0.06  & 52.00  & 1.14 & 0.17  & 65.35 & 3.95 & 1.41   & 65.37  & 33.25  & -6.69  & 46.64  & -112.60  
\\\hline
\end{tabular*}
}
\end{center}
\end{table}
\noindent
We implement the above cuts in our analysis codes after validating them with the efficiencies given in the experimental papers. As explained in Ref.~\cite{Mandal:2018kau}, we generated $pp\to Z^\prime\to \ell\ell$ events for validation and compared our cut efficiencies ($\varepsilon$) with the experimental $\textrm{efficiencies}\times \textrm{detector acceptance}$ to ensure they agree with each other.
In Table~\ref{tab:cross}, we show the production cross sections, cut efficiencies, and number of events surviving the cuts for different signal contributions for the $R_{D^{(*)}}$-motivated and $R_{K^{(*)}}$-motivated one-coupling scenarios, respectively. We obtain these numbers by setting the concerned coupling to unity. There are a few points to note here. Pair production is, in general, insensitive to new physics couplings. However, a mild sensitivity arises due to the model-dependent $t$-channel lepton exchange diagram that contributes to the pair production [see Fig.~\ref{fig:feynpq}]. In \hyperlink{sce:rd1a}{Scenario RD1A} where only $\lm_{23}^L$ is nonzero, the pair production cross section 
is $40.87$ fb for $M_{U_1}=1$~TeV, whereas in \hyperlink{sce:rd1b}{Scenario RD1B}, it is $35.67$ fb. This is because the $t$-channel lepton exchange contribution is larger in 
\hyperlink{sce:rd1a}{Scenario RD1A}. In this scenario,  the second-generation quarks contribute in the initial states with PDFs bigger than the $b$-PDF contributing in \hyperlink{sce:rd1b}{Scenario RD1B}. A similar minor difference can be seen between \hyperlink{sce:rk1a}{Scenario RK1A} and \hyperlink{sce:rk1b}{Scenario RK1B}. In \hyperlink{sce:rk1a}{Scenario RK1A}, the process $cc\to U_1U_1$ through a neutrino exchange is present, but it is absent in \hyperlink{sce:rk1b}{Scenario RK1B} causing the minor difference. The single production 
cross sections are relatively larger in scenarios where the second-generation quarks appear in the initial states than those where only $b$-quarks can appear.

The cut efficiencies for different production modes for $R_{K^{(*)}}$ scenarios are generally much higher compared to $R_{D^{(*)}}$ scenarios. This is mainly because the selection efficiency of the $\tau$ in the final state is much lower compared to the muons. For instance, in the $R_{K^{(*)}}$ scenarios, the efficiency for pair production processes $\varepsilon^p$ can be as high as $71\%$ for $M_{U_1}=1$~TeV, whereas for $R_{D^{(*)}}$ scenarios, it is only $\sim 2\%$. The hadronic BR of $\tau$ is $\sim 64\%$, and the $\tau$-tagging efficiency is about $60\%$. Combining just these two factors we get a factor of $0.64^2\times 0.6^2\sim 1/7$ reduction in the efficiency for the two $\tau_{had}$'s in the pair production final state.   
Note that all of pair and single productions, and $t$-channel $U_1$ exchange contributes positively towards the dilepton signal, whereas the signal-background interference contribute negatively as it is destructive in nature. The minus signs in $\sg^{nr2}$ and $\mc{N}^{nr2}$ indicate the destructive nature of the interference. 
\bigskip

\noindent
Before presenting our results, we list the 
publicly available HEP packages used at various stages of our analysis.
\begin{itemize}
\item 
\emph{Lagrangian and model files:} The Lagrangian terms defined in the previous section are implemented in 
the {\tt FeynRules}
package~\cite{Alloul:2013bka}  to obtain the UFO model files~\cite{Degrande:2011ua}. 

\item
\emph{Event generation:} Using the UFO model files, we generate signal events using the {\tt MadGraph5} Monte-Carlo event generator~\cite{Alwall:2014hca} at the leading order (LO). 
The NNPDF2.3LO PDFs~\cite{Ball:2012cx} are used with default dynamical scales.\footnote{The NNPDF2.3LO PDF for the heavy quarks might
have considerable uncertainties. However, our results, i.e., the limits on the $U_1$ parameters, are largely insensitive to these.}
Higher-order QCD corrections for the vLQ are not considered in this analysis as they are not available in the literature.

\item 
\emph{Showering and hadronization:} This is performed by passing the parton-level
events to {\tt Pythia6}~\cite{Sjostrand:2006za}. We use the {\tt MLM} matching scheme~\cite{Mangano:2006rw,Hoche:2006ph} (up to two additional jets) with 
virtuality-ordered Pythia showers to avoid double counting 
of the matrix-element partons with parton showers. 

\item
\emph{Detector simulation:} We use {\tt Delphes} $3.4.2$~\cite{deFavereau:2013fsa} (with ATLAS and CMS cards) to perform the detector simulations. The jet-clusterings are done using the FastJet package~\cite{Cacciari:2011ma}. We use the
anti-$k_T$ algorithm~\cite{Cacciari:2008gp} with the radius parameter $R = 0.4$.
\end{itemize}

\section{Exclusion limits}\label{sec:results}

\begin{figure*}[!t]
\captionsetup[subfigure]{labelformat=empty}
\subfloat[(a)]{\includegraphics[width=0.5\textwidth]{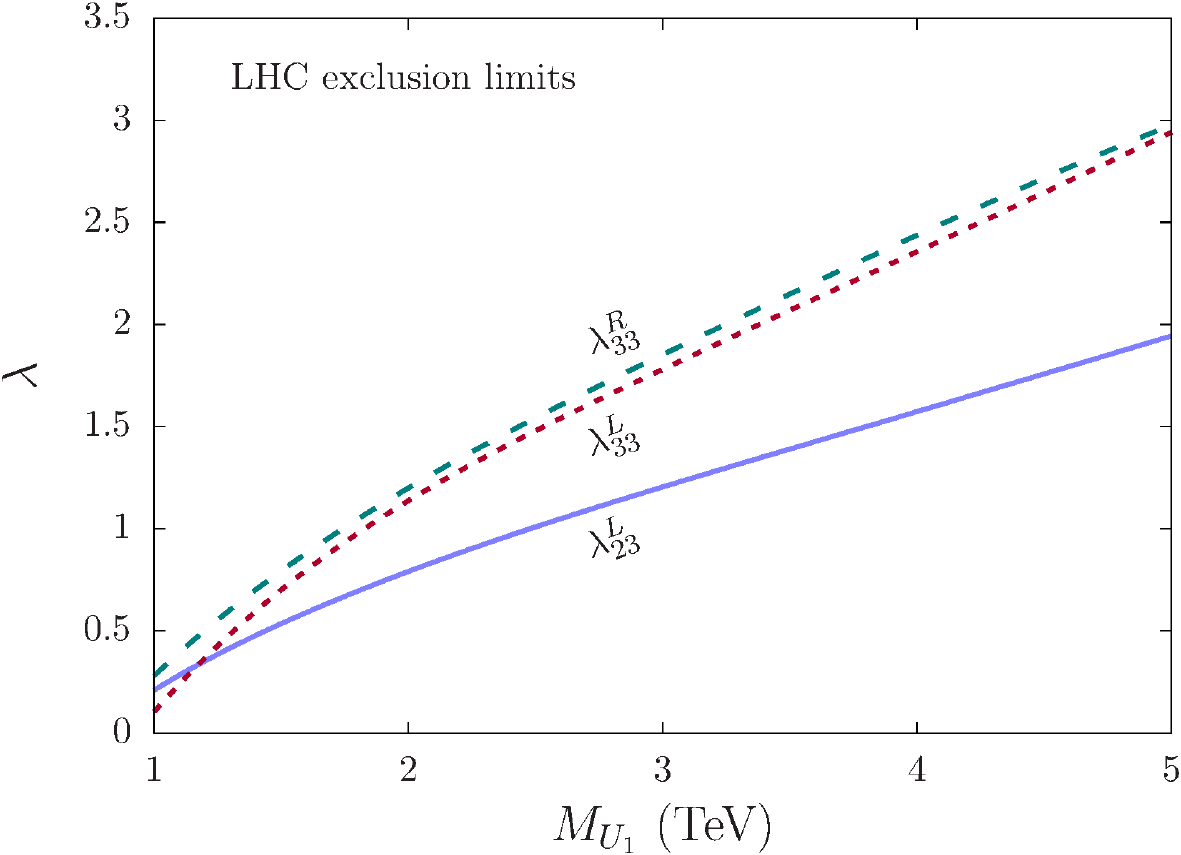}\label{fig:RD1lm}}
\subfloat[(b)]{\includegraphics[width=0.5\textwidth]{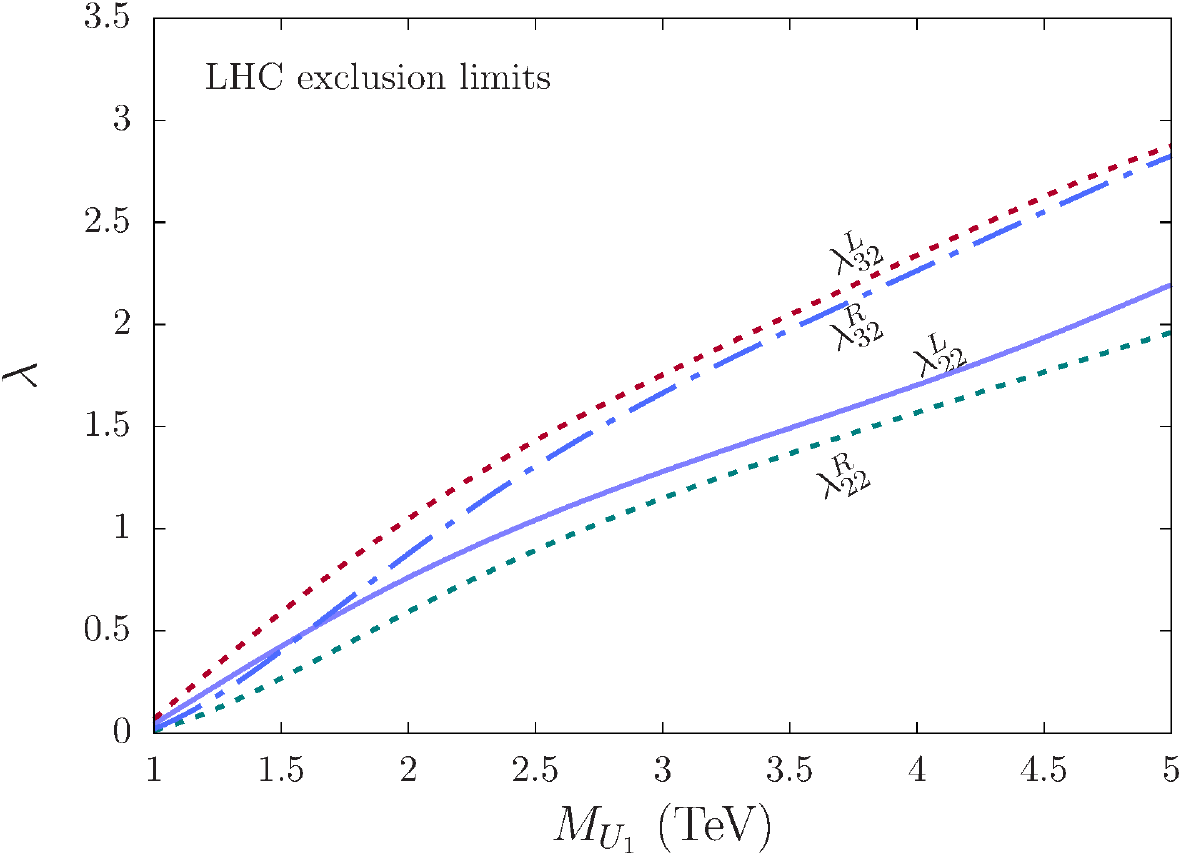}\label{fig:RK1lm}}
\caption{The $2\sg$ LHC exclusion limits on the couplings participating in the (a) $R_{D^{(*)}}$ and (b) $R_{K^{(*)}}$ scenarios. The regions above these lines are excluded. These exclusion limits are obtained by recasting the dilepton search data ~\cite{Aad:2020zxo,Sirunyan:2021khd} with a combination of all possible $U_1$ production processes that can contribute to the dilepton final states.}
\label{fig:RDRK_LHC}
\end{figure*}

\begin{figure*}[!t]
\captionsetup[subfigure]{labelformat=empty}
\subfloat[(a)]{\includegraphics[width=0.5\textwidth]{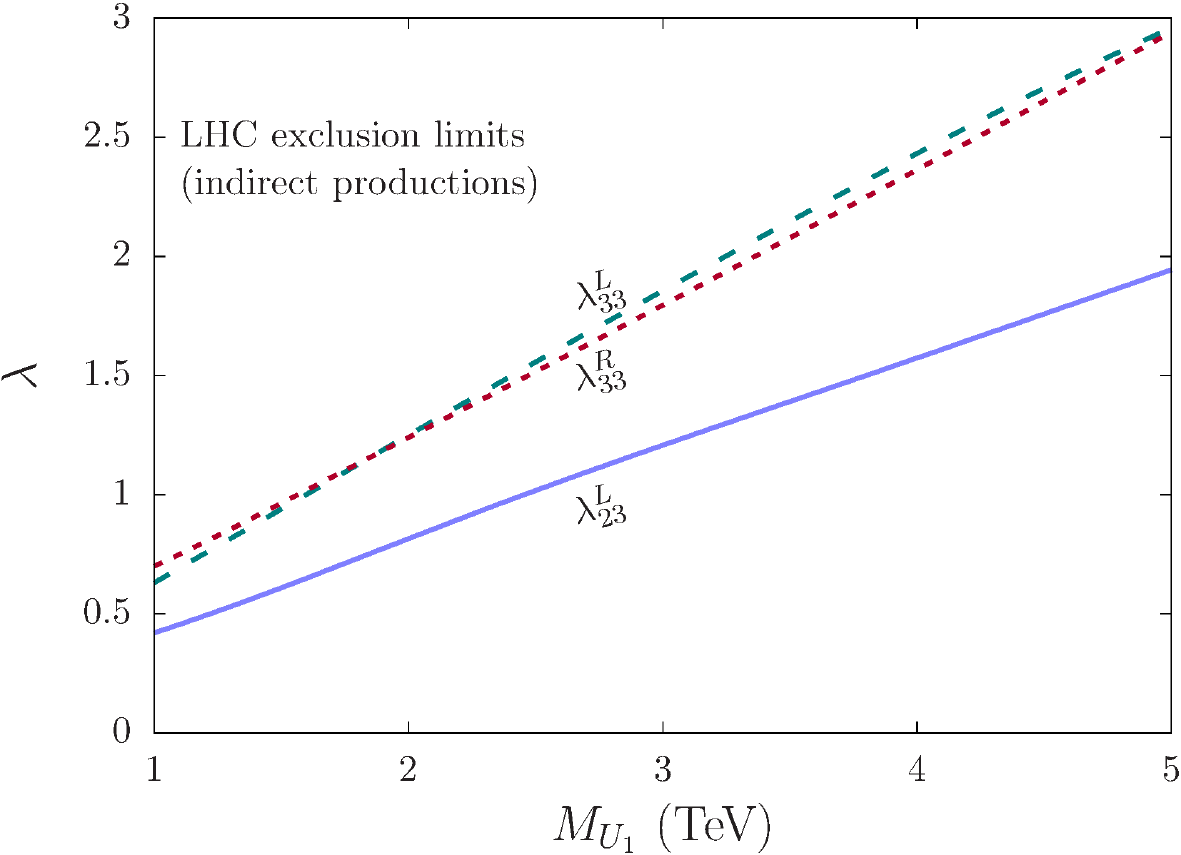}\label{fig:RDI1lm}}
\subfloat[(b)]{\includegraphics[width=0.5\textwidth]{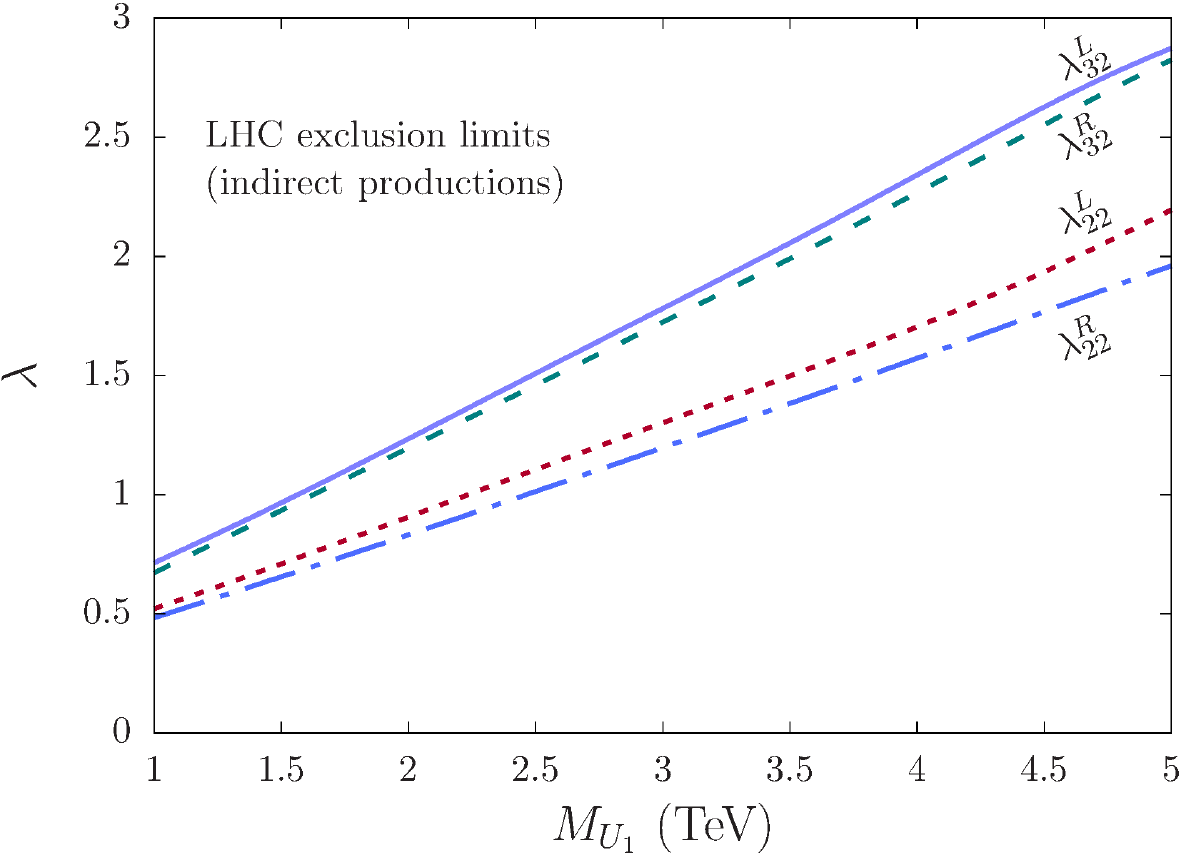}\label{fig:RKIlm}}
\caption{These exclusion plots are similar to the ones in Fig.~\ref{fig:RDRK_LHC} but they are obtained only for the nonresonant production and its interference with the SM background. Comparing with Fig.~\ref{fig:RDRK_LHC}, they show that in the lower mass region, the contribution of the resonant $U_1$ productions is significant. However, in the higher mass region, the nonresonant production and its interference with the SM background determine the limits.}
\label{fig:RKcomparison}
\end{figure*}
\begin{figure*}[!t]
\captionsetup[subfigure]{labelformat=empty}
\subfloat[\quad\quad\quad(a)]{\includegraphics[width=0.48\textwidth]{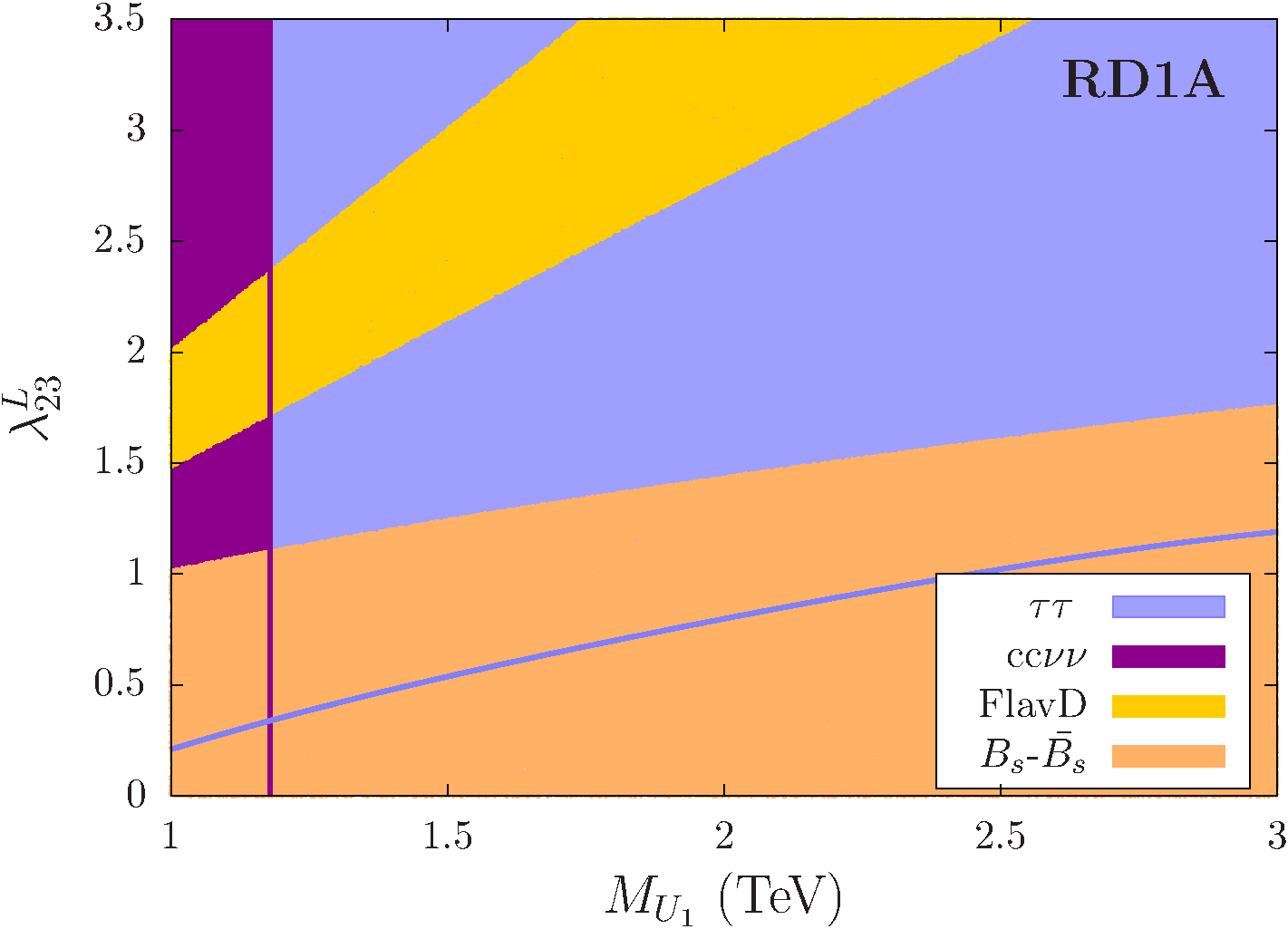}\label{fig:RD1Aa}}\quad
\subfloat[\quad\quad\quad(b)]{\includegraphics[width=0.48\textwidth]{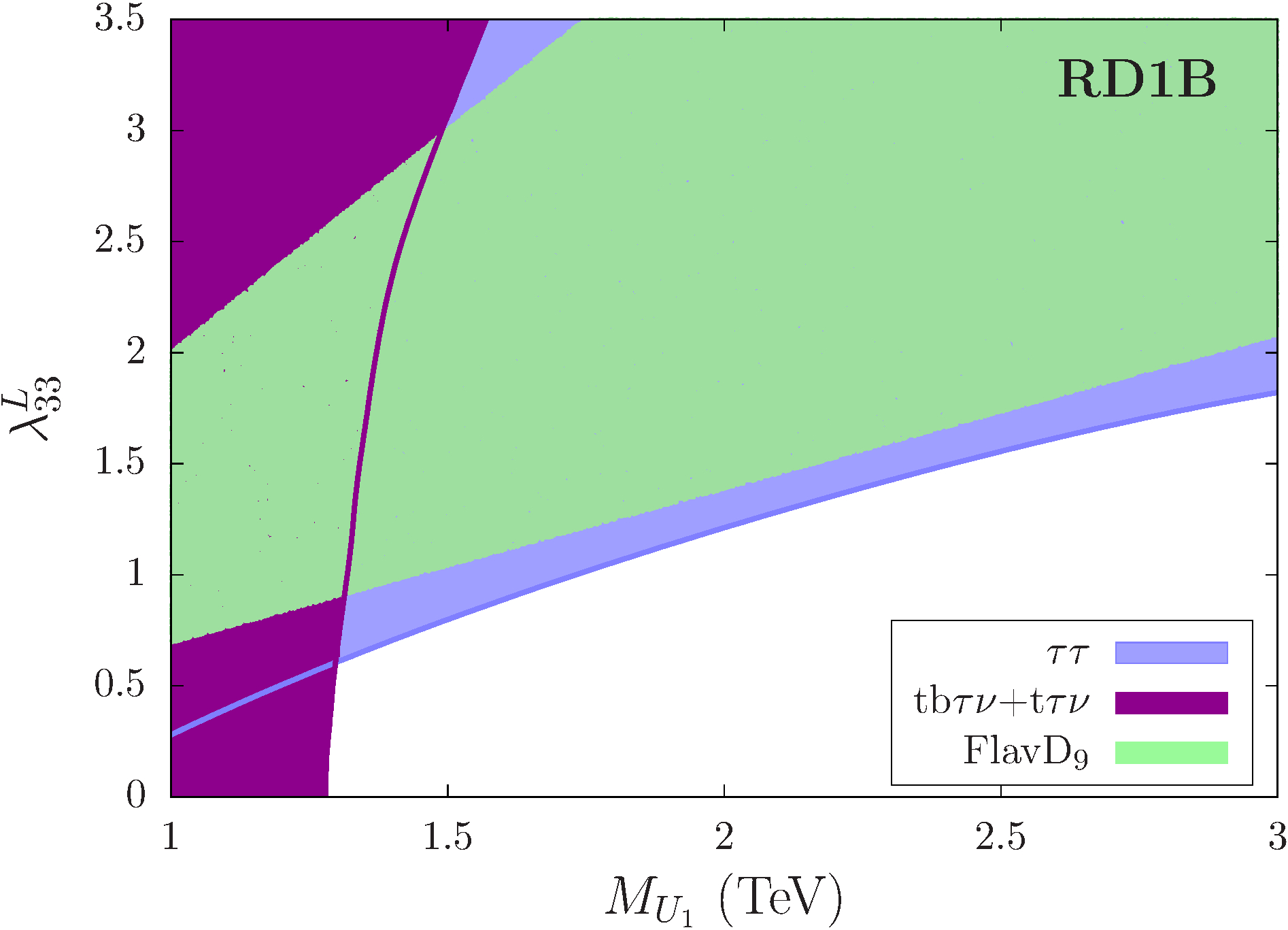}\label{fig:RD1Bb}}\\
\subfloat[\quad\quad\quad(c)]{\includegraphics[width=0.48\textwidth]{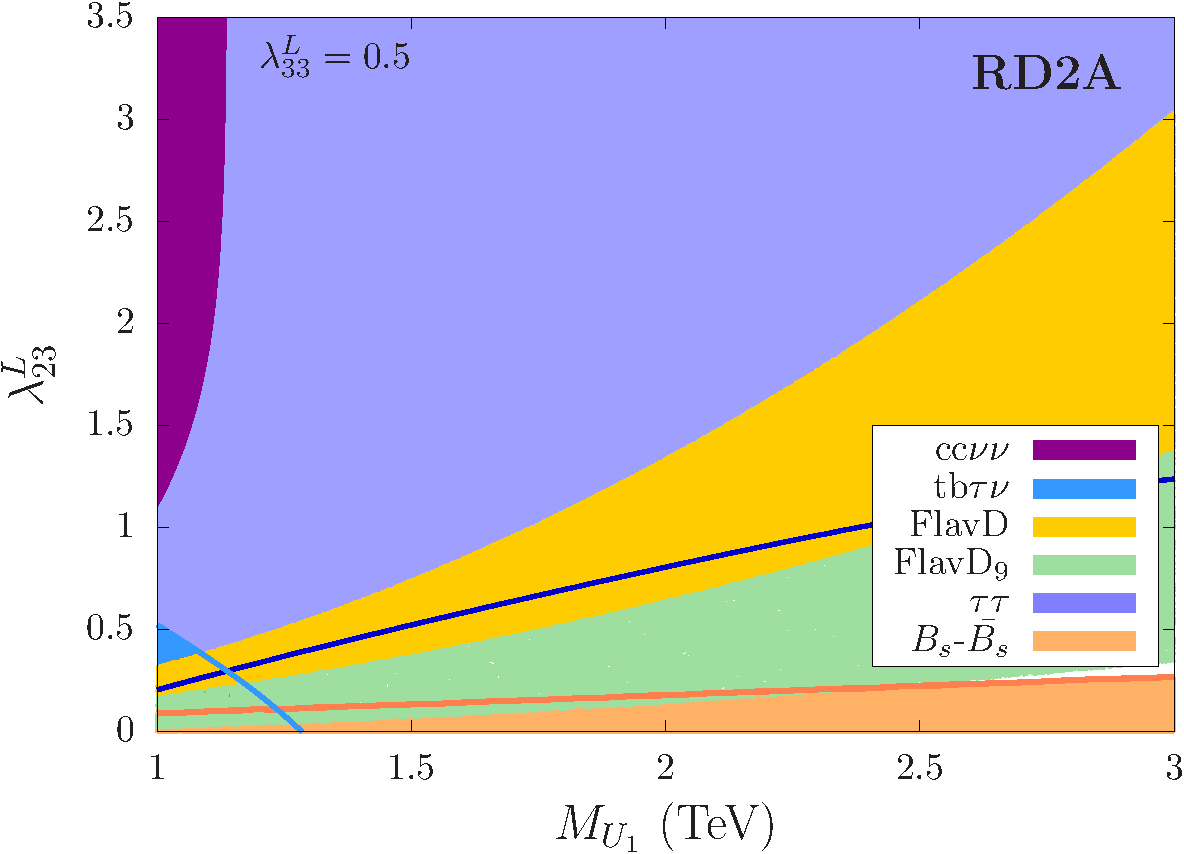}\label{fig:RD2Ac}}\quad
\subfloat[\quad\quad\quad(d)]{\includegraphics[width=0.48\textwidth]{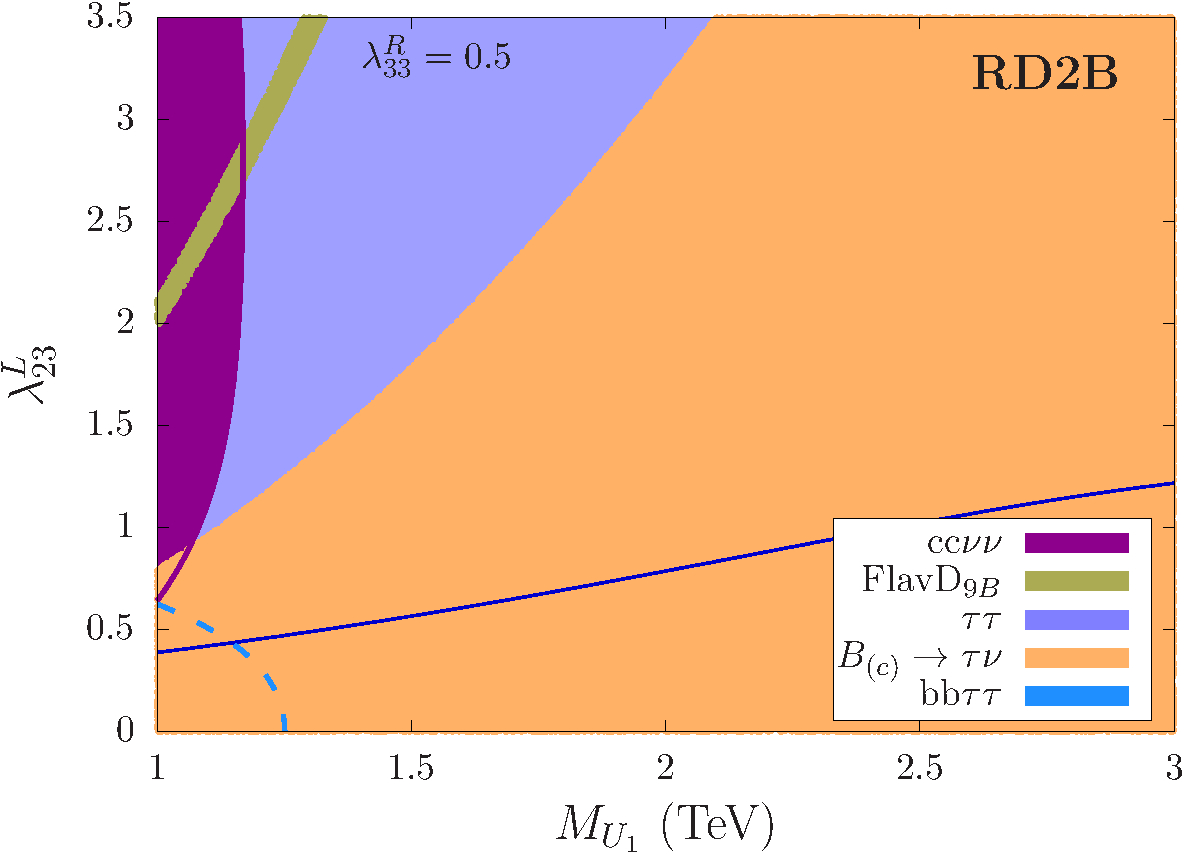}\label{fig:RD2Bd}}\\
\caption{The $2\sg$ exclusion limits from the LHC and the preferred regions by the flavour anomalies. The purple regions are excluded at 2$\sigma$ by the ATLAS $\tau\tau$ data~\cite{Aad:2020zxo}. The magenta and blue regions depict the excluded mass ranges from direct searches. 
(a) {\protect\hyperlink{sce:rd1a}{Scenario RD1A}}:  Only $\lambda_{23}^L$ is nonzero.  The FlavD region (yellow), defined in Eq.~\eqref{eq:flavD}, is favoured by the $R_{D^{(*)}}$ anomalies. The light orange region is favoured by ${B_s}$-$\bar{B}_s$ mixing. The direct detection mass limit (magenta) is from the CMS search in the $cc\nu\nu$ channel~\cite{Sirunyan:2018kzh}.
 (b) {\protect\hyperlink{sce:rd1b}{Scenario RD1B}}:  Only $\lambda_{33}^L$ is nonzero. The FlavD$_9$ region (green) agrees with all the constraints in FlavD except $\mc{C}_9^{\rm{univ}}$ [Eq.~\eqref{eq:deltac9a}] since the corresponding operator, $\mc{O}_9^{\rm{univ}}$, cannot be generated in this scenario. The magenta region here depicts the limits from the very recent CMS combined search in the $tb\ta \nu+ t\ta \n$ channel~\cite{Sirunyan:2020zbk}. We have recast the observed events data to extrapolate the limit for $\lm^L_{33}>2.5$. There is no ${B_s}$-$\bar{B}_s$ mixing in this scenario. 
 (c) {\protect\hyperlink{sce:rd2a}{Scenario RD2A}}:  $\lambda_{33}^L=0.5$ (benchmark choice), and $\lambda_{23}^L$ is free.
The magenta and blue regions are the limits from the direct searches at CMS~\cite{Sirunyan:2018kzh,Sirunyan:2020zbk}.
 (d) {\protect\hyperlink{sce:rd2b}{Scenario RD2B}}:
$\lambda_{33}^R=0.5$ (benchmark choice) and $\lambda_{23}^L$ is free. 
The blue dashed line shows the limits from the ATLAS $bb\tau\tau$ direct search data~\cite{Aaboud:2019bye}. The dashed line implies that the recast has been done from a scalar LQ direct search. The FlavD$_{9B}$ region (olive green), defined in Eq.~\eqref{eq:flavd9b}, agrees with all the constraints in the FlavD$_9$ region without the constraints from the $B_{(c)}\rightarrow\tau\nu$ decays. The light orange colour shows the region preferred by  $B_{(c)}\rightarrow\tau\nu$.}
\label{fig:RD1all}
\end{figure*}

\noindent
There are three free couplings $\lm_{23}^L,\lm_{33}^L$, and $\lm_{33}^R$ that take part in the  $R_{D^{(*)}}$ scenarios. 
We show the current exclusion limits on these couplings taken one at a time in Fig.~\ref{fig:RD1lm} from the latest LHC $\tau\tau$ resonance search data~\cite{Aad:2020zxo}. 
Similarly, the $R_{K^{(*)}}$ scenarios have four free couplings in total: $\lm_{22}^L,\lm_{32}^L$, $\lm_{22}^R$, and $\lm_{32}^R$. We use the latest CMS $\mu\mu$ resonance search data~\cite{Sirunyan:2021khd} to obtain exclusion limits on these couplings by considering one of them at a time as shown in Fig.~\ref{fig:RK1lm}.
These are the $95\%$ ($2\sg$) confidence level (CL) exclusion limits. To obtain the limits, we set all other couplings except the one under consideration to zero. The method we follow is the same as 
the one used in Ref.~\cite{Mandal:2018kau} and is elaborated on in Appendix~\ref{sec:appendixB}. From the left plot, we see that the limit
on $\lm_{23}^L$ is more severe than on $\lm_{33}^{L/R}$. This is because, for nonzero 
$\lm_{23}^L$, there is a $s$-quark-initiated contribution to the $t$-channel $U_1$ exchange that interferes with the SM $ss\to \gm^{*}/Z^{*}\to\tau\tau$ process. In the case of nonzero $\lm_{33}^{L/R}$, the process is $b$-quark-initiated 
and, therefore, is suppressed by the small $b$-PDF. Among the offshell photon and $Z$-boson contributions to the signal-background (SB) interference, the second one dominates. Similarly, one can also understand the relatively weaker limits on $\lm^{L/R}_{32}$ in Fig.~\ref{fig:RK1lm}.

Among the four sources of dilepton events [the pair and single production, $t$-channel $U_1$ exchange, 
and the SB interference], different processes play the dominating roles in deciding the limits in different mass ranges. We observe an interesting role switch in 
these plots. In the high mass region where the limits on $\lm$ reach high values ($\gtrsim 1$), the resonant productions are relatively less 
important (also see Fig.~\ref{fig:sigma_mass})], and the nonresonant productions
play the determining roles. However,for $M_{U_1}\lesssim 1.5$ TeV, mainly the resonant productions determine the limits, and all the $\lm$-dependent contributions are small. This switch of roles can also be inferred from Fig.~\ref{fig:RKcomparison} where we show the same  limits as in Fig.~\ref{fig:RDRK_LHC} but ignore the resonant contributions in the dilepton signal. Comparing Figs.~\ref{fig:RDRK_LHC} and~\ref{fig:RKcomparison} we see that for $M_{U_1}\lesssim 2.5$ TeV, the limits can vary significantly depending on whether one considers the resonant productions or not.  Since the nonresonant processes do not depend on the $U_1$ branching ratios, for a low mass $U_1$ the limits thus obtained are  
strictly valid when the branching ratio of the decay $U_1\to\tau j/\tau b$ mediated by the coupling $\lm$ is small. However, the limits in Fig.~\ref{fig:RDRK_LHC} are obtained assuming only one coupling is nonzero, i.e., maximum branching ratio for the $U_1\to\tau j/\tau b$ decay mediated by the coupling $\lm$. For a very heavy $U_1$ ($\gtrsim3$ TeV) the resonant productions become negligible, and hence, the limits of Figs.~\ref{fig:RDRK_LHC} and~\ref{fig:RKcomparison} converge.
We also observe that in the low mass region ($M_{U_1}\sim 1$ TeV) where the limit is determined almost solely by the pair production, the data are badly fitted by the $U_1$ events (i.e., $\chi^2_{min}/d.o.f.\gg 1$). This suggests that the dimuon data disfavour pair production of a $\sim 1$ TeV $U_1$ -- this is similar to but independent of the direct LQ search limits.

\begin{figure*}[]
\captionsetup[subfigure]{labelformat=empty}
\subfloat[\quad\quad\quad(a)]{\includegraphics[width=0.48\textwidth]{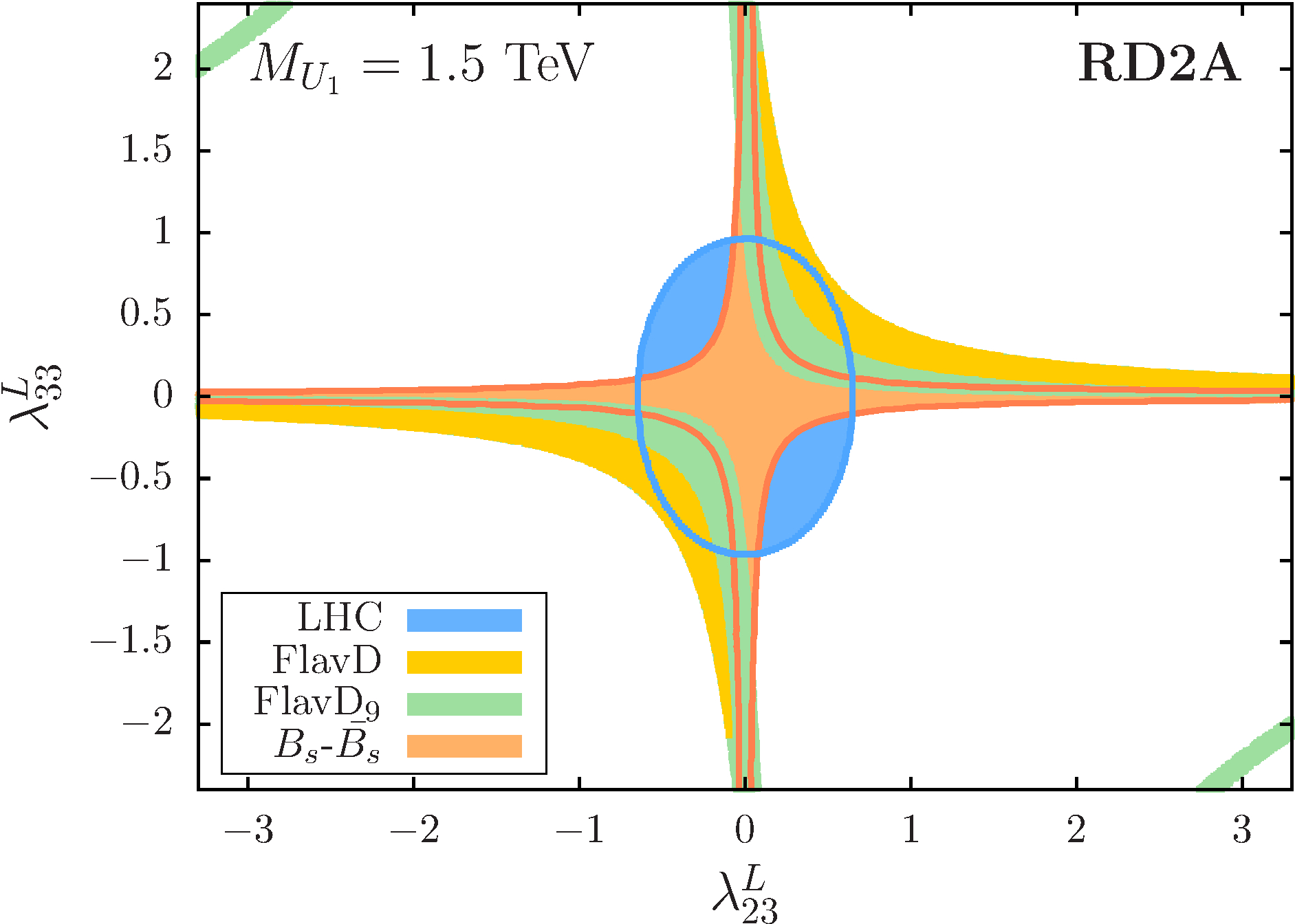}\label{fig:RD2A15}}\quad
\subfloat[\quad\quad\quad(b)]{\includegraphics[width=0.48\textwidth]{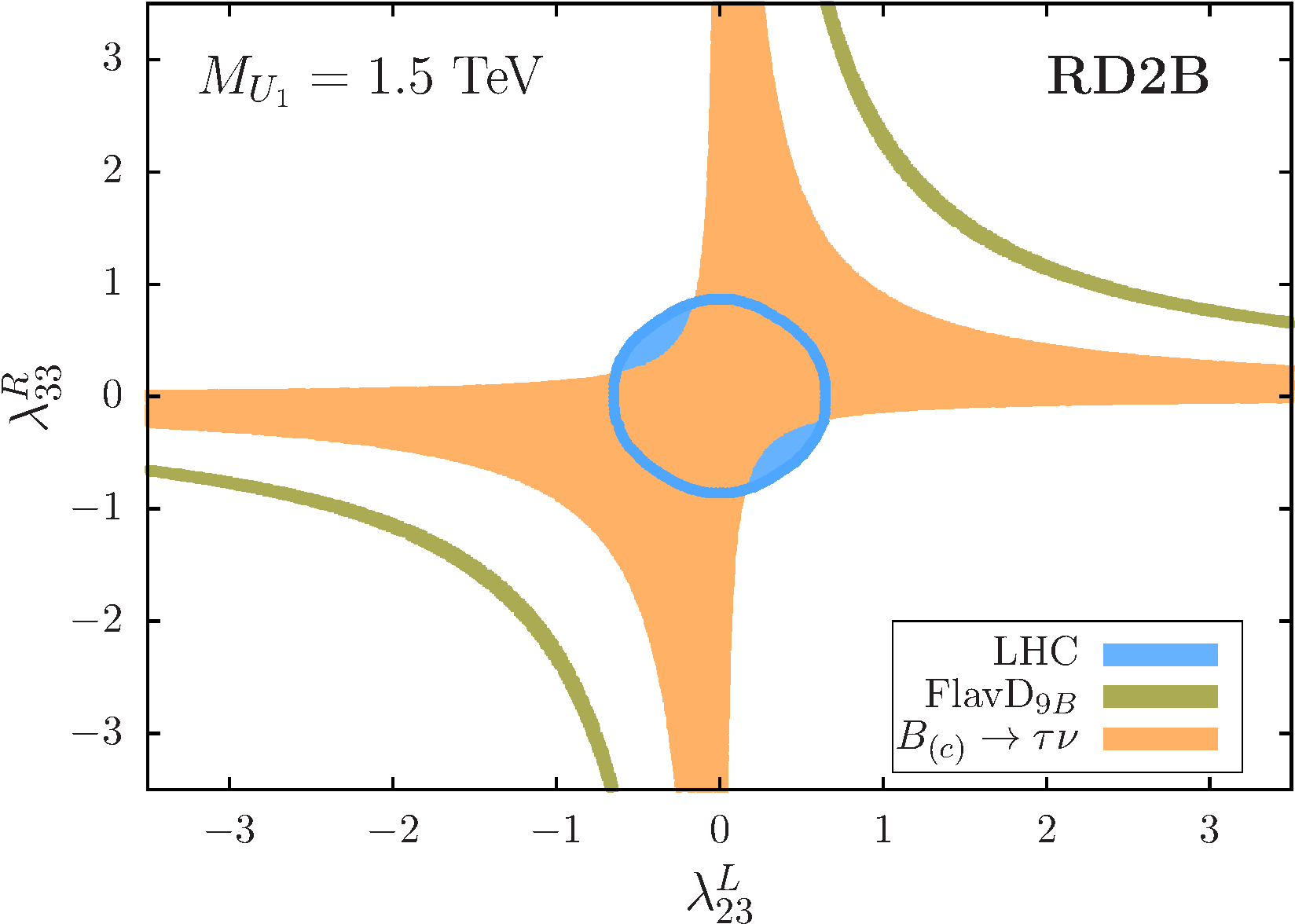}\label{fig:RD2B15}}\\
\subfloat[\quad\quad\quad(c)]{\includegraphics[width=0.48\textwidth]{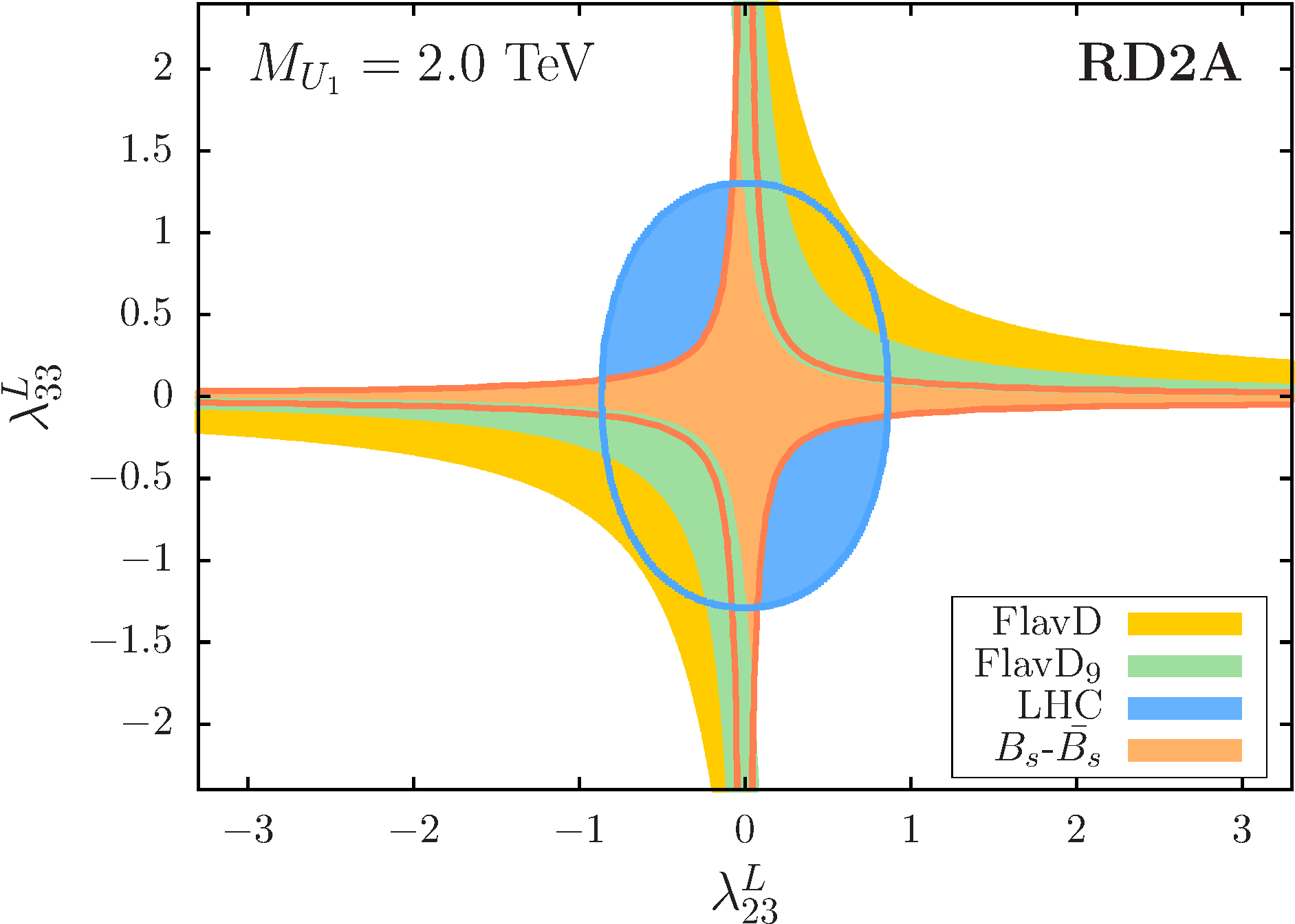}\label{fig:RD2A20}}\quad
\subfloat[\quad\quad\quad(d)]{\includegraphics[width=0.48\textwidth]{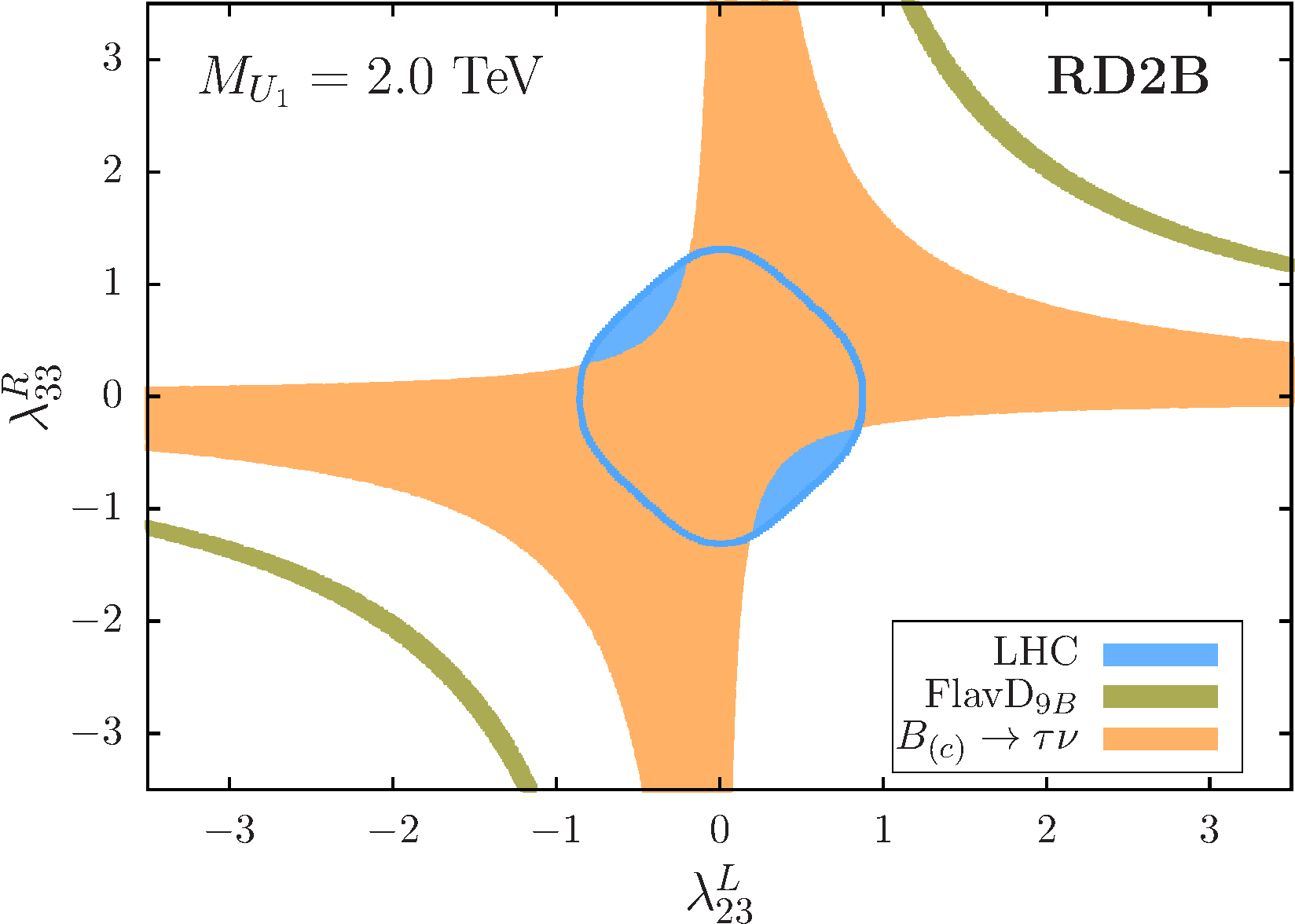}\label{fig:RD2B20}}
\caption{The regions favoured by the flavour observables (yellow, green, and orange) and allowed by the LHC data (blue). We consider two benchmark mass values, $M_{U_1}=1.5$ TeV and $2$ TeV. (a) and (c) {\protect\hyperlink{sce:rd2a}{Scenario RD2A}} ($\lambda_{33}^R=0$): FlavD (yellow) and FlavD$_9$ (green) are the regions preferred by the flavour anomalies with and without the $\mc{O}_9^{\rm{univ}}$ operator, respectively. The light orange region is favoured by the ${B_s}$-$\bar{B}_s$ mixing. (b) and (d) {\protect\hyperlink{sce:rd2b}{Scenario RD2B}} ($\lambda_{33}^L=0$): The light orange color marks the regions preferred by the $B_{(c)}\to\tau\nu$ decay. FlavD$_{9B}$ (olive green) shows the region preferred by the flavour anomalies except $\mc{O}_9^{\rm{univ}}$ and the $B_{(c)}\to\tau\nu$ decay.}
\label{fig:RD2all}
\end{figure*}

\begin{figure*}[]
\captionsetup[subfigure]{labelformat=empty}
\subfloat[\quad\quad\quad(a)]{\includegraphics[width=0.48\textwidth]{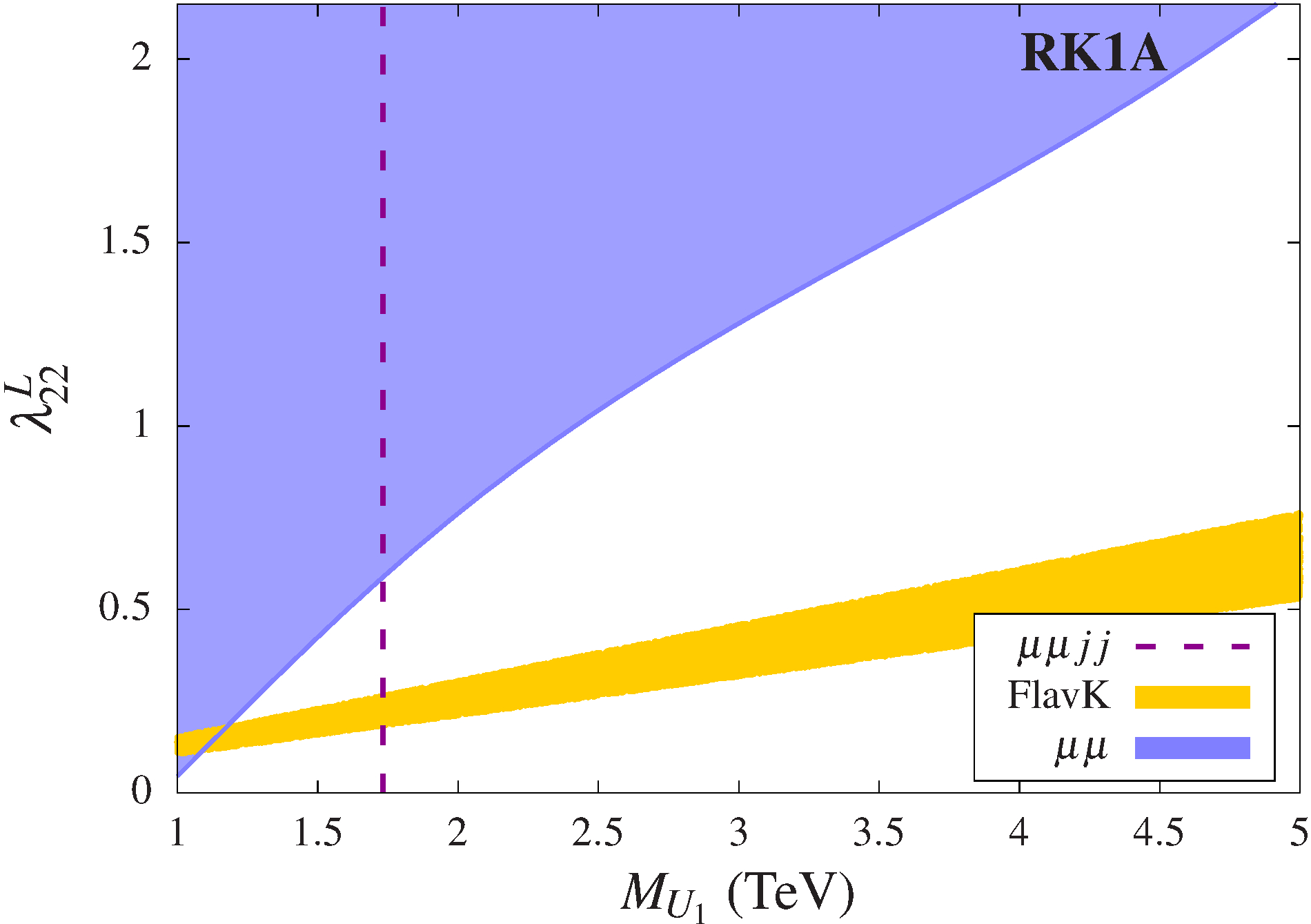}\label{fig:RK1A}}\quad
\subfloat[\quad\quad\quad(b)]{\includegraphics[width=0.48\textwidth]{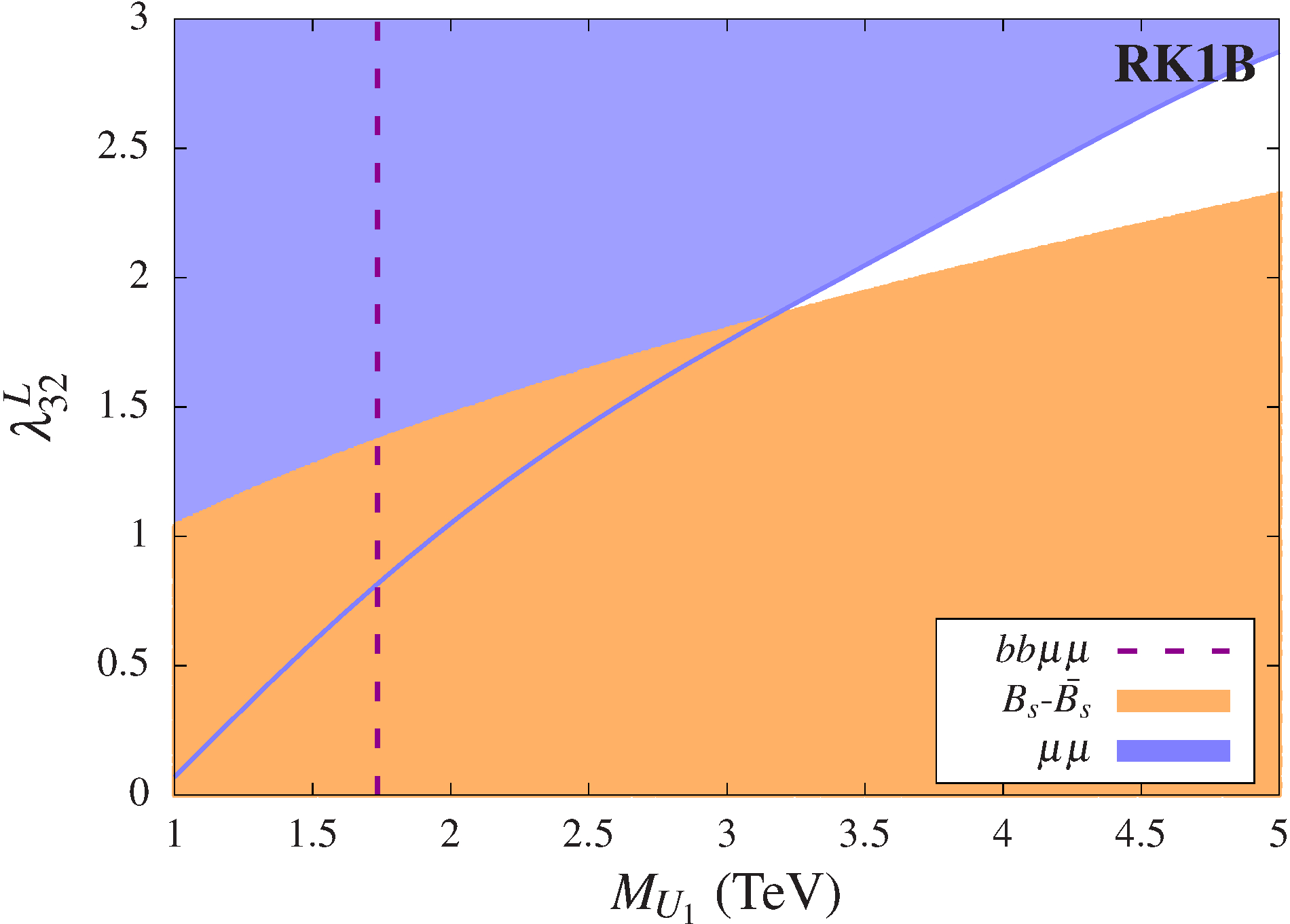}\label{fig:RK1B}}\\
\subfloat[\quad\quad\quad(c)]{\includegraphics[width=0.48\textwidth]{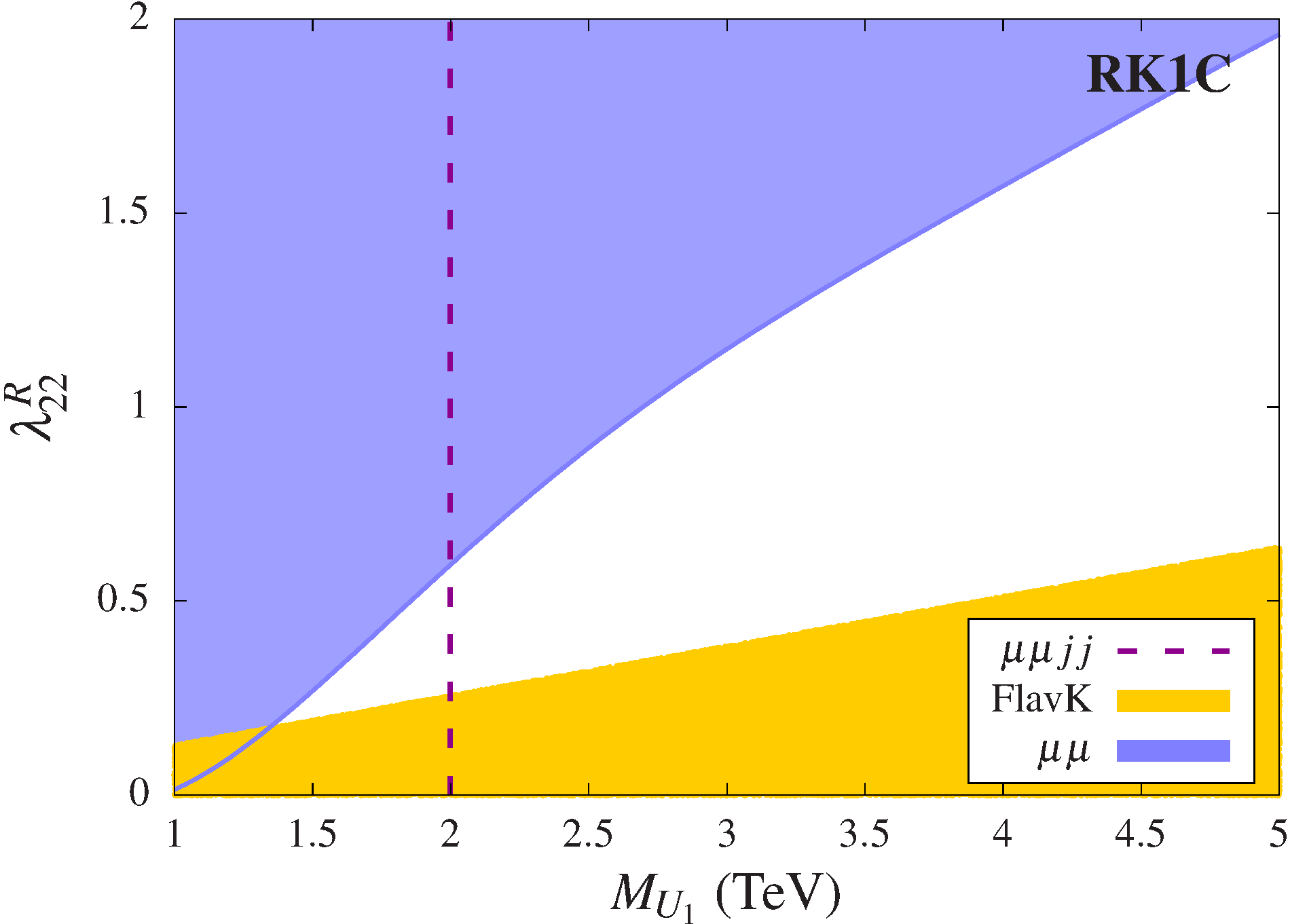}\label{fig:RK1C}}\quad
\subfloat[\quad\quad\quad(d)]{\includegraphics[width=0.48\textwidth]{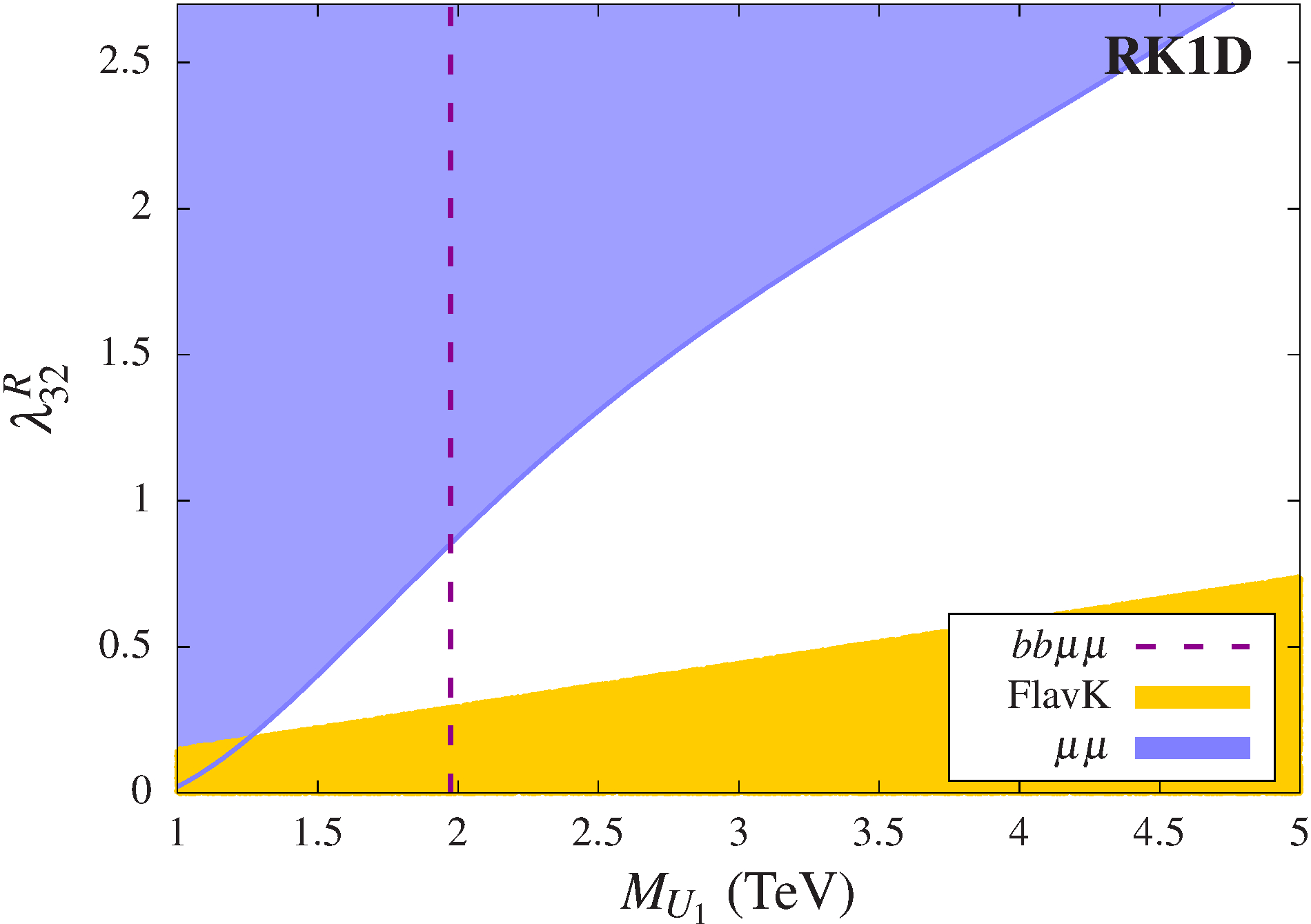}\label{fig:RK1D}}\\

\caption{The regions excluded at 2$\sigma$ by the CMS $\mu\mu$ search data~\cite{Sirunyan:2021khd}  in the minimal $R_{K^{(*)}}$ scenarios (violet). FlavK (yellow), defined in Eq.~\eqref{eq:flavk}, is the region favoured by the global fits to the $\rm{b}\rightarrow \rm{s}\mu\mu$ and $\rm{B_s}$-$\bar{\rm{B_s}}$ mixing.(a) {\protect\hyperlink{sce:rk1a}{Scenario RK1A}}: Only $\lambda_{22}^L$ is nonzero. (b) {\protect\hyperlink{sce:rk1b}{Scenario RK1B}}: Only $\lambda_{32}^L$ is nonzero. Since, $\lm^L_{32}$ alone cannot explain the  $R_{K^{(*)}}$ anomalies,  we only show the region allowed by the $B_s$-$\bar B_s$ mixing data.
(c) {\protect\hyperlink{sce:rk1c}{Scenario RK1C}}: Only $\lambda_{22}^R$ is nonzero. (d) {\protect\hyperlink{sce:rk1d}{Scenario RK1D}}: Only $\lambda_{32}^R$ is nonzero. The magenta dashed lines denote the recast exclusion limits from the ATLAS direct search for pair production of scalar LQs in the $\mu\mu + jj/bb$ channels~\cite{Aad:2020iuy}. }
\label{fig:exrk2}
\end{figure*}

We plot the relevant direct search limits from the $jj+\slashed{E}_{\rm T}$ and $tt+\slashed{E}_{\rm T}$ channels~\cite{Sirunyan:2018kzh} together with 
the limits on the $R_{D^{(*)}}$ scenarios with one coupling in Figs.~\ref{fig:RD1Aa} and \ref{fig:RD1Bb}. 
We also show the parameter regions that are favoured by the $R_{D^{(*)}}$ observables and consistent with the relevant flavour observables (as discussed in Section~\ref{sec:model}) in the same plots. We see that in \hyperlink{sce:rd1a}{Scenario RD1A}, the region marked as FlavD, which is defined as
\begin{equation}
{\rm FlavD}\equiv  \mbox{the region allowed by }\left\{R_{D^{(*)}} + F_L(D^*)+P_\tau(D^*)+\mathcal{B}(B_{(c)}\to \tau\nu)+\mc C_9^{\rm univ}\right\},\label{eq:flavD}
\end{equation}
is in tension with the $B_s$-$\bar B_s$ mixing data and is independently and entirely excluded by the $\ta\ta$ data. The tension between FlavD and the $B_s$-$\bar B_s$ mixing data arises since the $B_s$-$\bar B_s$ mixing data favours a smaller $\mc C_{V_L}^{U_1}$ (via $\mc C^{U_1}_{box}$ which roughly goes as the square of $\mc C_{V_L}^{U_1}$) than the $R_{D^{(*)}}$ observables.
\hyperlink{sce:rd1b}{Scenario RD1B} does not contribute to Eq.~\eqref{eq:deltac9} and hence, cannot accommodate a nonzero $\mc C_9^{\rm univ}$. We mark the parameter region favoured by the $R_{D^{(*)}}$ observables in this scenario as FlavD$_9$ which stands for the region allowed by all the constraints included in FlavD except $\mc C_9^{\rm univ}$, i.e., 
\begin{equation}
{\rm FlavD}_9\equiv \mbox{the region allowed by }\left\{R_{D^{(*)}} + F_L(D^*)+P_\tau(D^*)+\mathcal{B}(B_{(c)}\to \tau\nu)\right\}.\label{eq:flavd9}
\end{equation}
From Fig.~\ref{fig:RD1Bb}, we see that for this strictly one-coupling scenario, the entire FlavD$_9$ is excluded by the latest $\ta\ta$ data.

\begin{figure*}[]
\captionsetup[subfigure]{labelformat=empty}
\subfloat[\quad\quad\quad(a)]{\includegraphics[width=0.45\textwidth]{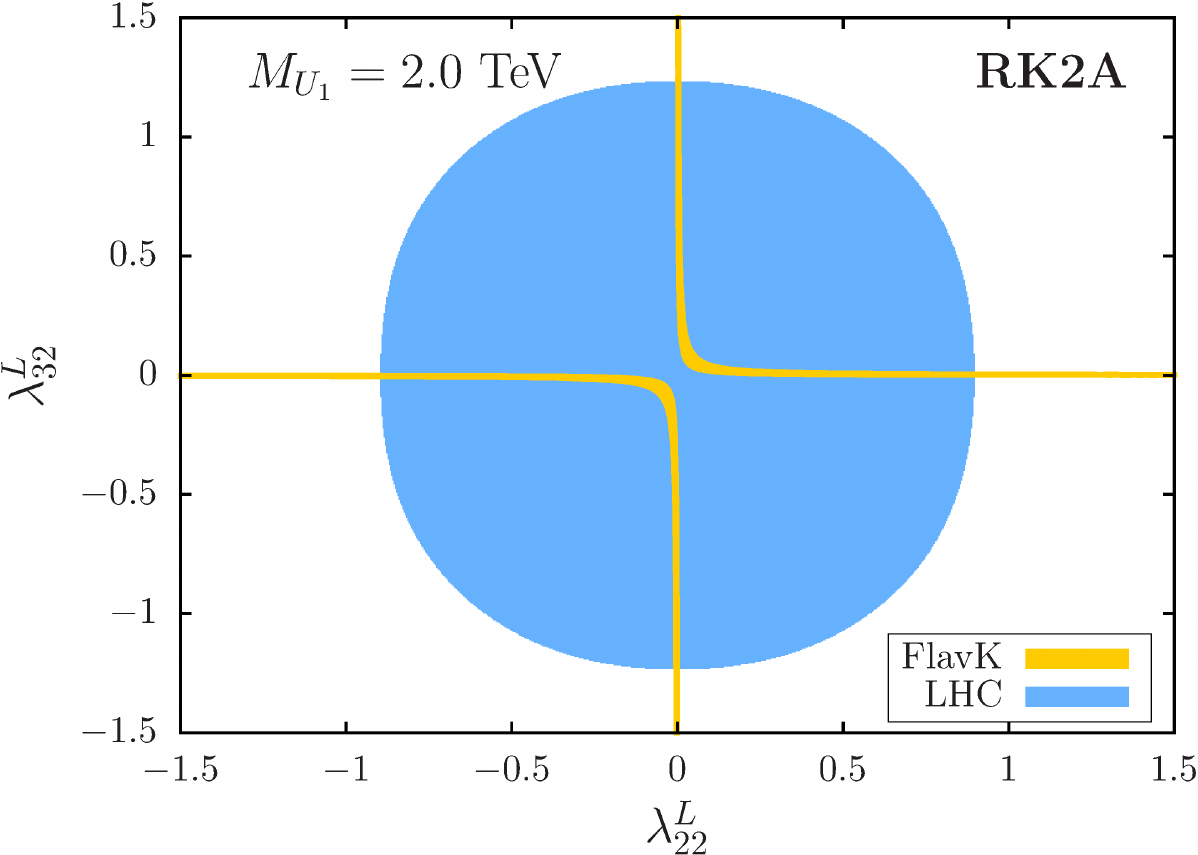}\label{fig:RK2A20}}\hfill
\subfloat[\quad\quad\quad(b)]{\includegraphics[width=0.45\textwidth]{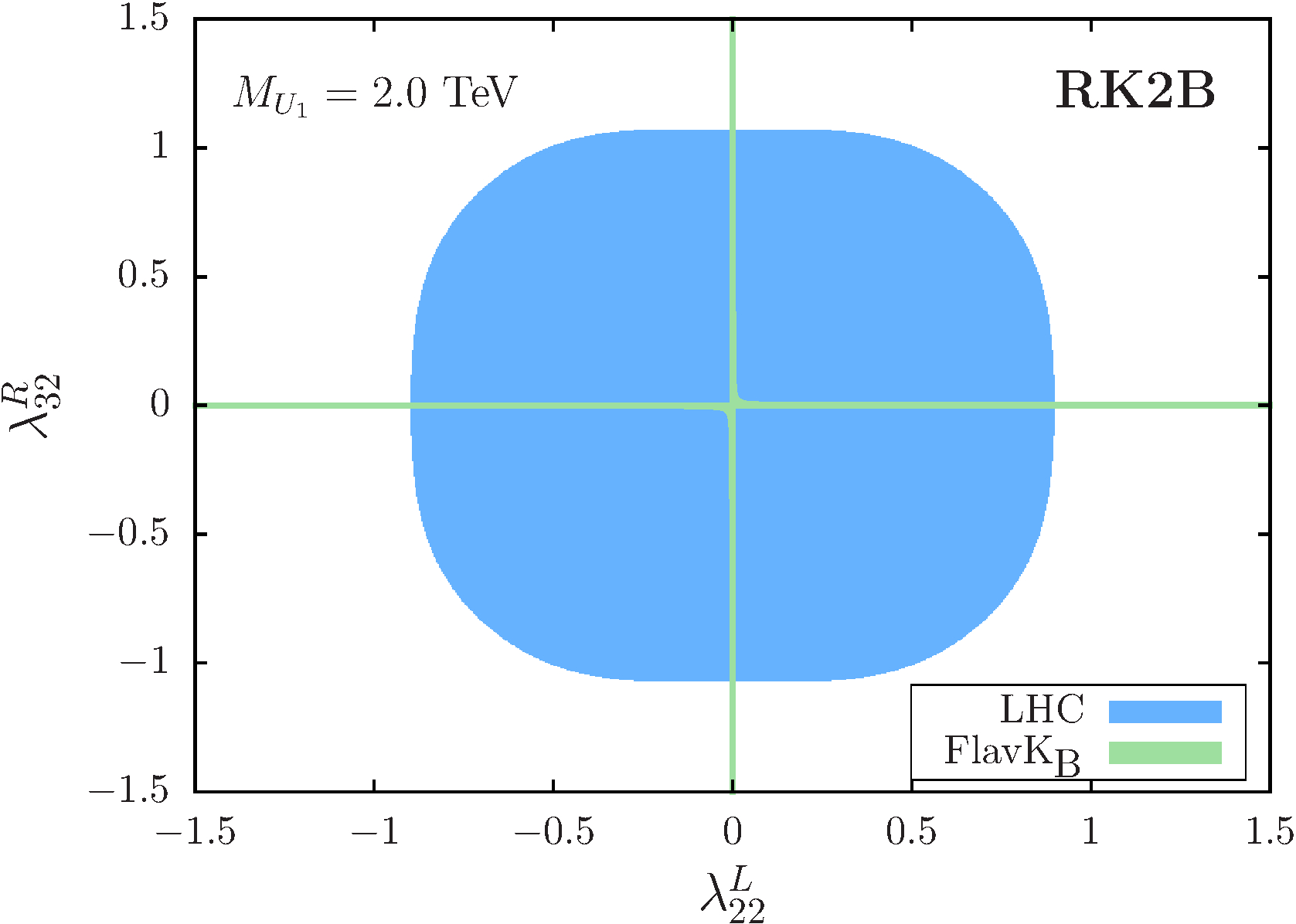}\label{fig:RK2B20}}\\
\subfloat[\quad\quad\quad(c)]{\includegraphics[width=0.45\textwidth]{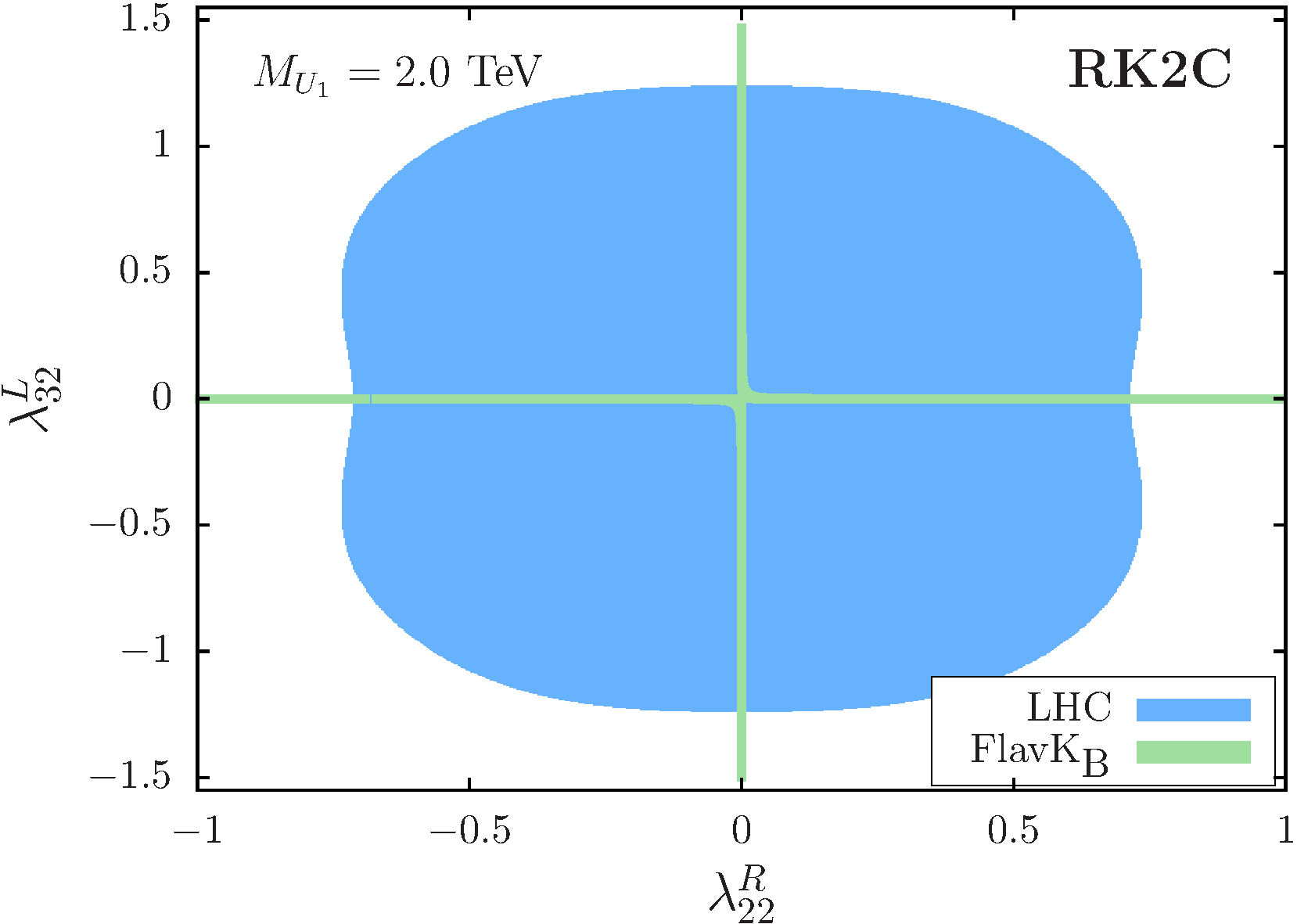}\label{fig:RK2C20}}\hfill
\subfloat[\quad\quad\quad(d)]{\includegraphics[width=0.45\textwidth]{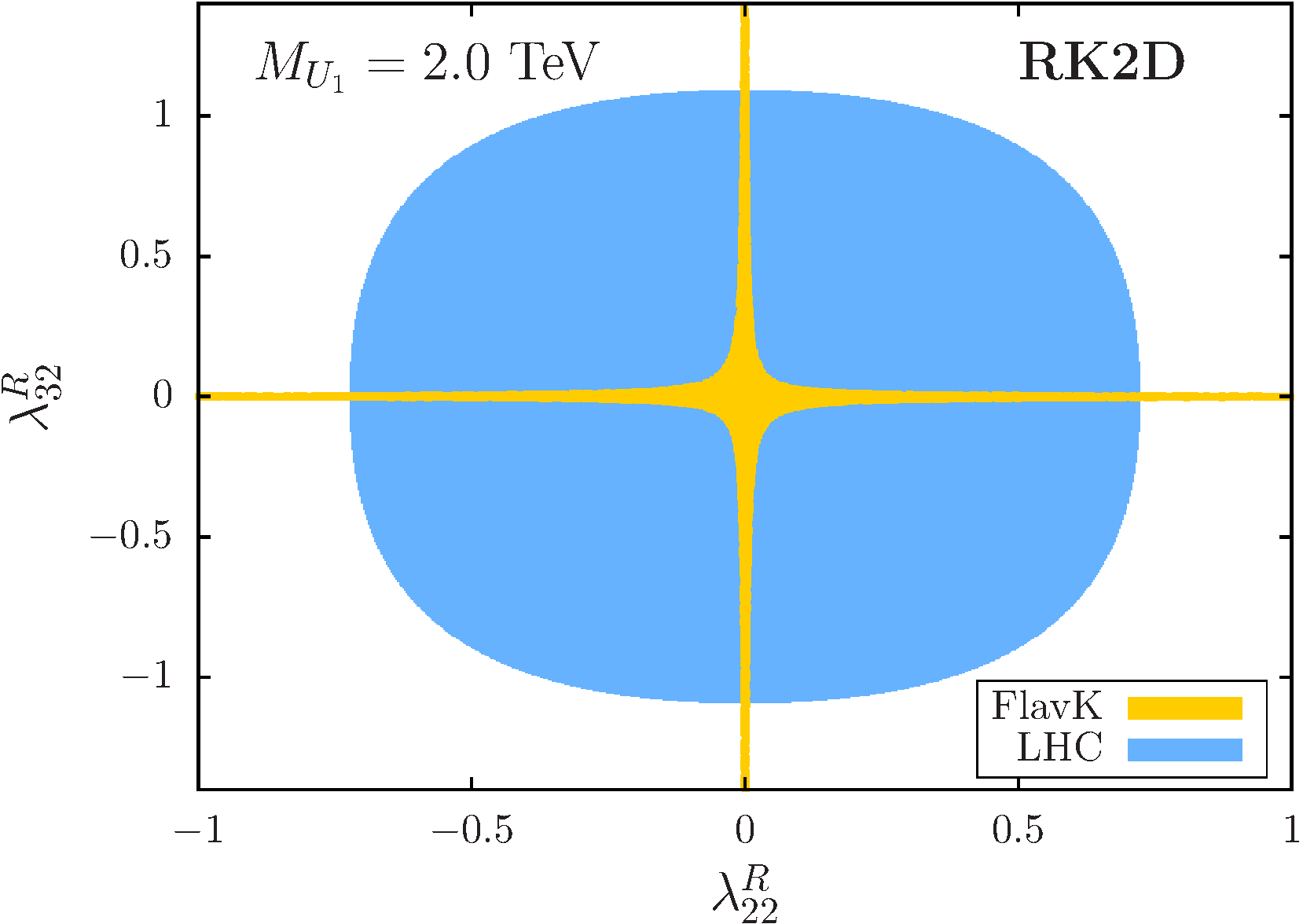}\label{fig:RK2D20}}
\caption{The regions favoured by the $R_{K^{(*)}}$ observables (yellow and green) and allowed by the LHC data (blue). The blue regions are obtained by recasting the CMS $\mu\mu$ search data~\cite{Sirunyan:2021khd}. (a) and (d) In {\protect\hyperlink{sce:rk2a}{Scenario RK2A}} and {\protect\hyperlink{sce:rk2d}{Scenario RK2D}}, the FlavK region (yellow) depicts the allowed regions by the $b\rightarrow s\mu\bar{\mu}$ global fits and $B_s$-$\bar{B}_s$ mixing. (b) and (c) In {\protect\hyperlink{sce:rk2b}{Scenario RK2B}} and {\protect\hyperlink{sce:rk2d}{Scenario RK2D}}, there is no restriction from the $B_s$-$\bar{B}_s$ mixing. $\rm{FlavK}_{{B}}$ (green), defined in Eq.~\eqref{eq:flavkb}, stands for the  region preferred by the $b\to s\mu\bar\mu$ global fits alone.}
\label{fig:exrk2_2000}
\end{figure*}

\begin{figure*}[]
\captionsetup[subfigure]{labelformat=empty}
\subfloat[\quad\quad\quad(a)]{\includegraphics[width=0.45\textwidth]{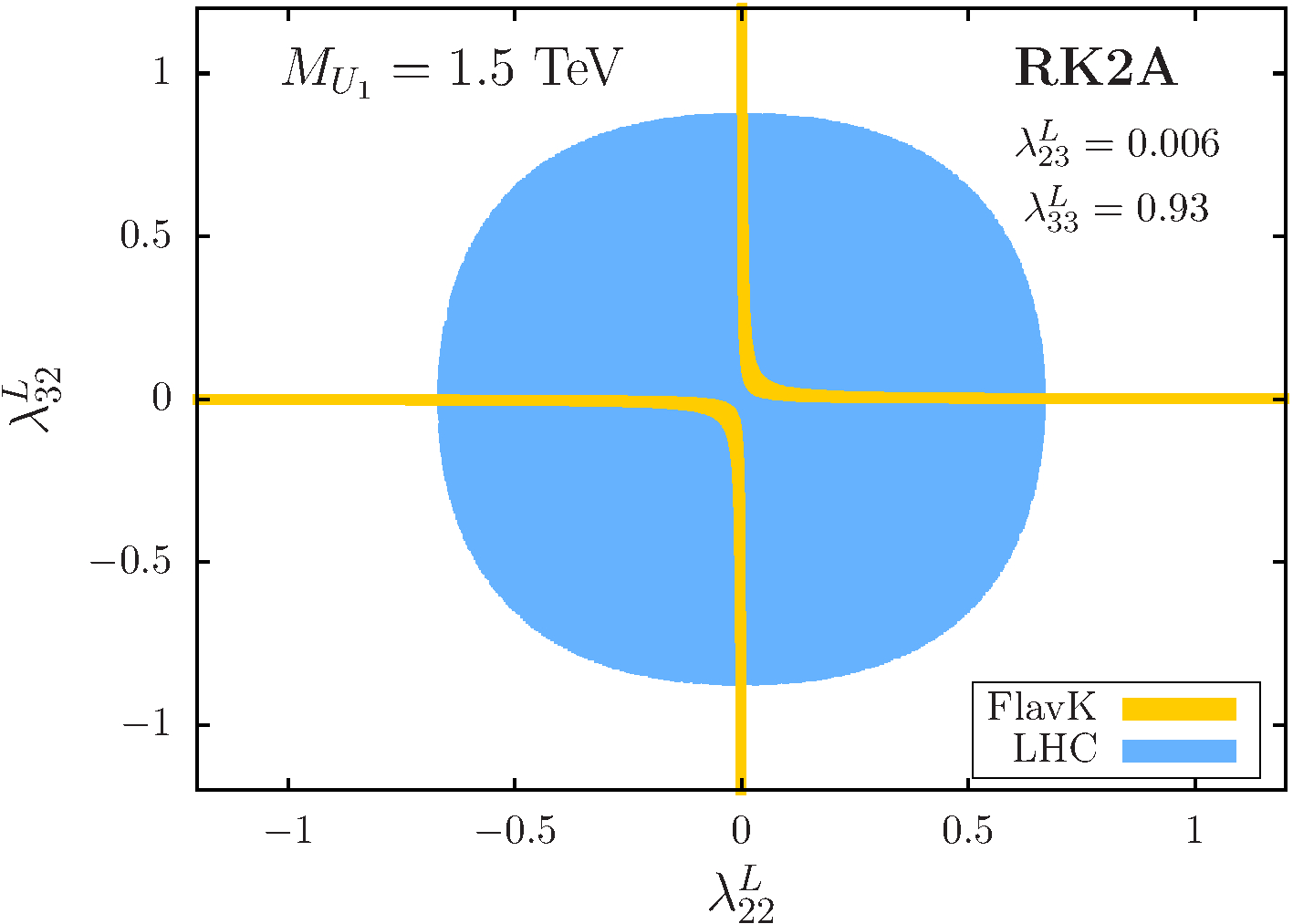}\label{fig:RD1BRK2A20}}\hfill
\subfloat[\quad\quad\quad(b)]{\includegraphics[width=0.45\textwidth]{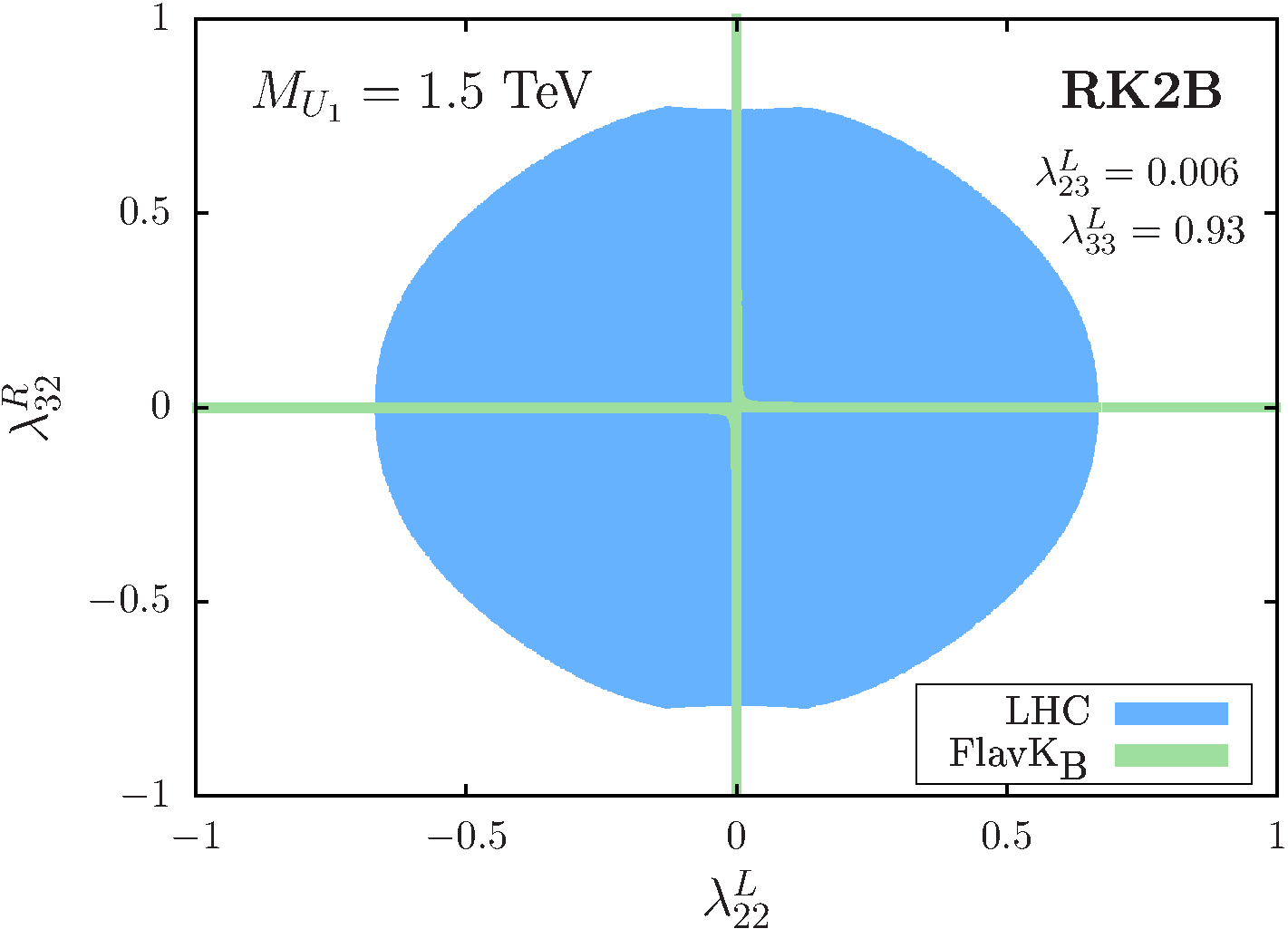}\label{fig:RD1BRK2B20}}\\
\subfloat[\quad\quad\quad(c)]{\includegraphics[width=0.45\textwidth]{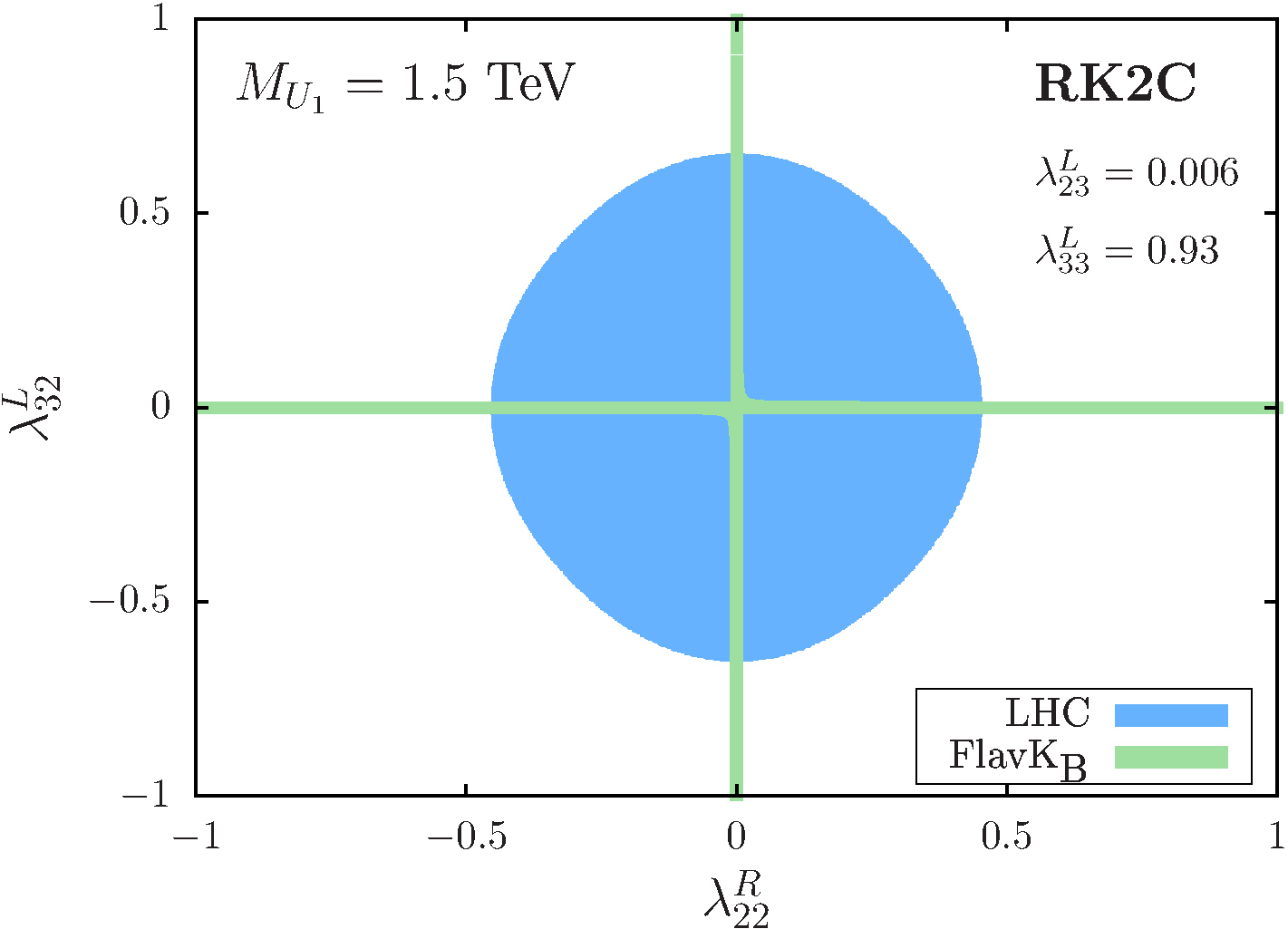}\label{fig:RD1BRK2C20}}\hfill
\subfloat[\quad\quad\quad(d)]{\includegraphics[width=0.45\textwidth]{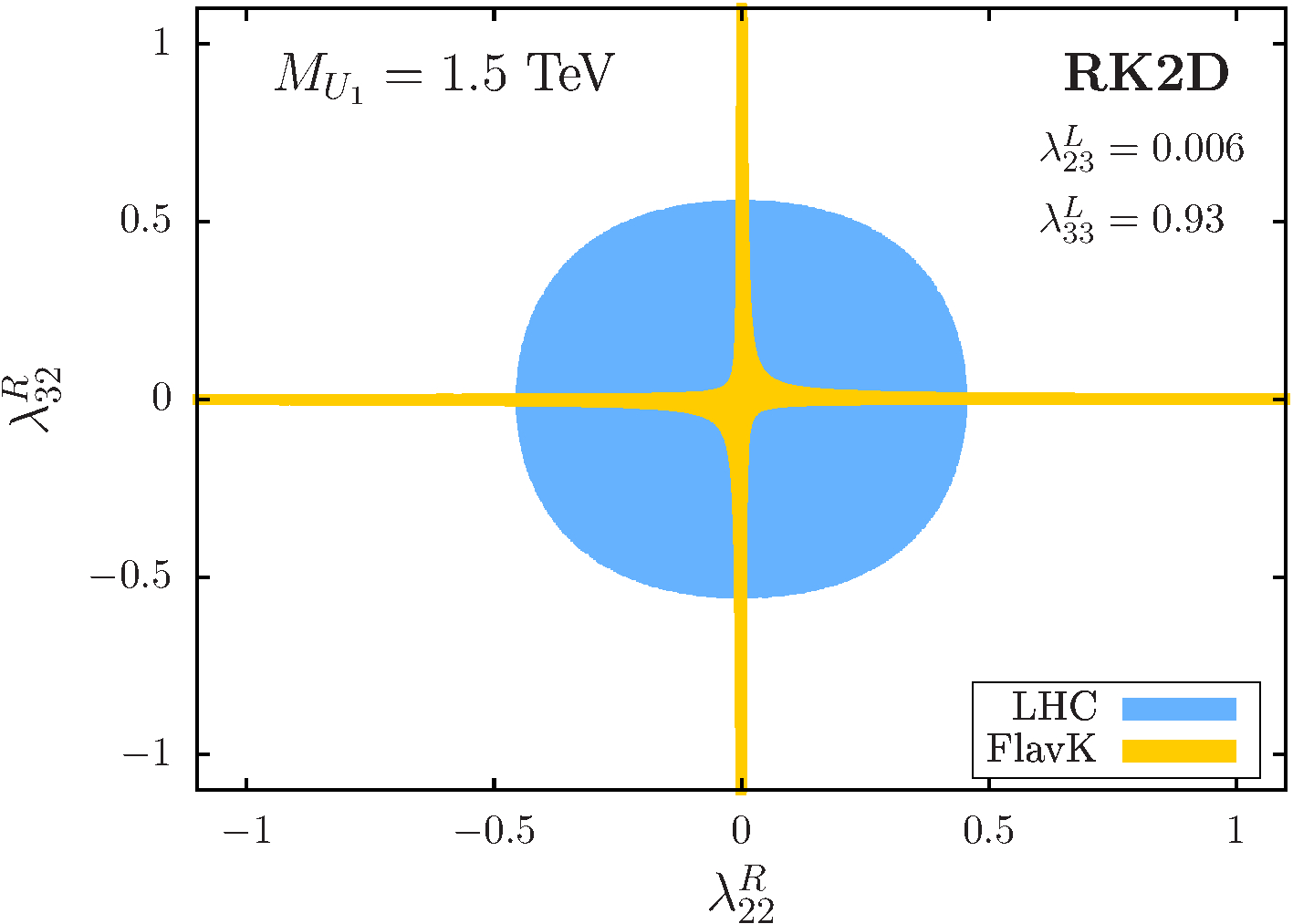}\label{fig:RD1BRK2D20}}
\caption{Examples of regions in the parameter space of a $1.5$ TeV $U_1$ surviving the LHC limits and simultaneously accommodating the 
$R_{D^{(*)}}$ and $R_{K^{(*)}}$ anomalies. The 
FlavK (yellow), FlavK$_{B}$ (green) and the blue regions are identical to the ones in Fig.~\ref{fig:exrk2_2000}.  When we recast the ATLAS search in the $\m\m+bb/jj$ channels~\cite{Aad:2020iuy}, the  presence of the additional couplings, $\lambda_{23}^L=0.006$ and $\lambda_{33}^L=0.93$ [allowed by the LHC and flavour data, see Fig.~\ref{fig:RD2A15}],
relaxes the exclusion limits, thus allowing the otherwise excluded mass value in the R$_{K^{(*)}}$ two-coupling scenarios.}
\label{fig:exrd1brk2_1500}
\end{figure*}

If we look at the two-coupling scenarios, the situation somewhat improves. For example, in Figs.~\ref{fig:RD2Ac} and~\ref{fig:RD2Bd}, we show a projection of the three-dimensional parameter space of 
\hyperlink{sce:rd2a}{Scenario RD2A} and \hyperlink{sce:rd2b}{Scenario RD2B}, respectively. In Fig.~\ref{fig:RD2Ac}, we keep $\lm^L_{33}=0.5$ and let $\lm^L_{23}$ and $M_{U_1}$ vary (\hyperlink{sce:rd2a}{Scenario RD2A}). We see that a good part of the FlavD$_9$ region (note that the FlavD is a subregion within FlavD$_9$)  survives the LHC bounds but a large part of it remains in conflict with the $B_s$-$\bar B_s$ mixing data. However, a small part of FlavD$_9$ does agree with $B_s$-$\bar B_s$ mixing and survives the LHC bounds. We show two more projections of the parameter space of \hyperlink{sce:rd2a}{Scenario RD2A} in Figs.~\ref{fig:RD2A15} and \ref{fig:RD2A20} -- we let $\lm^L_{23}$ and $\lm^L_{33}$ vary and keep the mass of $U_1$ fixed. There we show the region allowed by the LHC data and the relevant flavour 
regions for $M_{U_1} = 1.5~(2.0)$ TeV, respectively.
In absence of $\lm^L_{33}$, \hyperlink{sce:rd2b}{Scenario RD2B} cannot accommodate the allowed $\mc C_9^{\rm univ}$, and, in this case, the region favoured by the $R_{D^{(*)}}$ observable stands in conflict with the  $\mc B(B_{(c)}\to \ta\n)$ constraint -- see the region marked as FlavD$_{9B}$ for the region allowed by all the constraints included in FlavD$_9$ except $\mc B(B_{(c)}\to \ta\n)$, i.e., 
\begin{eqnarray}
{\rm FlavD}_{9B}\equiv \mbox{the region allowed by }\left\{R_{D^{(*)}} + F_L(D^*)+P_\tau(D^*)\right\}.\label{eq:flavd9b}
\end{eqnarray}
From Fig.~\ref{fig:RD2Bd} [where we keep $\lm^R_{33}=0.5$ and let $\lm^L_{23}$ and $M_{U_1}$ vary (\hyperlink{sce:rd2b}{Scenario RD2B})], we see that the entire FlavD$_{9B}$ region is ruled out by the LHC data. This can also be seen from the two coupling plots in Figs.~\ref{fig:RD2B15} and \ref{fig:RD2B20}. 

In Fig.~\ref{fig:exrk2}, we compare the bounds on $\lm_{22}^L,\lm_{32}^L$, $\lm_{22}^R$ and $\lm_{32}^R$ from the CMS $\m\m$ data~\cite{Sirunyan:2021khd}  with the regions favoured by the $R_{K^{(*)}}$ anomalies and allowed by the $B_s$-$\bar B_s$ mixing data marked as
\begin{equation}
{\rm FlavK}\equiv~{\rm the~region~favoured~by~\big\{the~global~fits~to}~b\rightarrow s\mu\mu~{\rm data}
 + \mbox{$B_s$-$\bar B_s$ mixing}\big\}\label{eq:flavk}
\end{equation}
in the one-coupling scenarios [except for \hyperlink{sce:rd2b}{Scenario RD2B} where, as already pointed out, $\lm^L_{32}$ alone cannot explain the  $R_{K^{(*)}}$ anomalies. Hence, in Fig.~\ref{fig:RK1B} we only show the region allowed by the $B_s$-$\bar B_s$ mixing data].
In these plots, we also show the recast limits from the recent pair production search by the ATLAS Collaboration that effectively rule out $U_1$ masses almost up to $2$ TeV. To obtain these limits we have recast the 
recent ATLAS search for scalar LQ in the $\mu\mu+jj/bb$ channel obtained with 
$139$ fb$^{-1}$ of integrated luminosity~\cite{Aad:2020iuy}.\footnote{We show 
the recast ATLAS limits~\cite{Aad:2020iuy} only with dashed lines because, strictly 
speaking, the search was optimised for a scalar LQ. In our recast for $U_1$, we 
have assumed that the selection efficiencies remain unchanged when one switches 
from the pair production of scalar LQs to that of the vector ones.}
We see the LHC $\m\m$ data are much less 
restrictive on the FlavK regions than the $\ta\ta$ data on the FlavD regions. 
This is mainly because the magnitudes of these couplings required to explain the 
$R_{K^{(*)}}$ anomalies are much smaller than those in the $R_{D^{(*)}}$ scenarios. 
We also note that the direct search mass exclusion limits are weaker in the scenarios with left-type 
couplings (i.e., \hyperlink{sce:rk1a}{Scenario RK1A} and \hyperlink{sce:rk1b}{Scenario RK1B}) 
than those with right-type couplings (\hyperlink{sce:rk1c}{Scenario RK1C} and 
\hyperlink{sce:rk1d}{Scenario RK1D}). This is because the decay $U_1\to \m b/\m j$ has $100\%$ BR in the right-type coupling scenarios instead of the $50\%$ in the left-type ones. 
The recast limits imply that a $1.5$ TeV $U_1$ is ruled out in all the two-coupling scenarios. Hence we consider a $2$ TeV
$U_1$ in the two-coupling scenarios in Fig.~\ref{fig:exrk2_2000}. There we show the regions allowed by the LHC data along with the FlavK regions in \hyperlink{sce:rk2a}{Scenario RK2A} and \hyperlink{sce:rk2d}{Scenario RK2D} and FlavK$_B$ regions in 
\hyperlink{sce:rk2a}{Scenario RK2B} and \hyperlink{sce:rk2d}{Scenario RK2C}. In the last two scenarios, the constraints from $B_s$-$\bar B_s$ mixing data are not applicable, and the FlavK$_B$ regions are just the ones favoured by the global fit of $b\to s\m\m$ data 
\begin{eqnarray}
{\rm FlavK}_B&\equiv& {\rm FlavK} +  \mbox{the region exclusively disfavoured by $B_s$-$\bar B_s$ mixing}\nn\\
&\equiv & \mbox{the region favoured by the global fits to}~\rm{b}\rightarrow\rm{s}\mu\mu~\mbox{data}.\label{eq:flavkb}
\end{eqnarray}

The recast ATLAS scalar search limits however does not entirely rule out a $1.5$ TeV $U_1$ solution for the $R_{K^{(*)}}$ anomalies. To see this, one needs to make $\bt(U_1\to \m b/\m j)\lesssim0.25$ by introducing additional nonzero coupling(s). For example, in Fig.~\ref{fig:exrd1brk2_1500} we show that for $\lm^L_{23}=0.006$ and $\lm^L_{33}=0.93$ [a point we chose randomly from the FlavD$_9$ region in Fig.~\ref{fig:RD2A15} that agrees with $B_s$-$\bar B_s$ mixing and is allowed by the LHC data] , all the two-couplings $R_{K^{(*)}}$ scenarios survive the recast bounds for $M_{U_1}=1.5$ TeV. This is interesting, as it explicitly shows four possible parameter choices for which a $1.5$ TeV $U_1$ can account for both the  $R_{D^{(*)}}$ and $R_{K^{(*)}}$ anomalies. These choices are, of course, only illustrative, not unique or special.

\section{Summary and conclusions}\label{sec:conclu}
\noindent
In this paper, we have derived precise limits on a flavour-anomalies-motivated $U_1$
LQ model using the latest LHC and flavour data. We started with a generic coupling texture for $U_1$ (with seven free couplings) that can  contribute to the $R_{D^{(*)}}$ and $R_{K^{(*)}}$ observables. Taking a bottom-up approach, suitable for obtaining bounds from the existing LHC searches, we constructed all possible one- and two-coupling scenarios that can accommodate either the $R_{D^{(*)}}$ or $R_{K^{(*)}}$ anomalies.
In particular, we considered two one-coupling and two two-coupling scenarios that can give rise to the $b\tau U_1$ and $c\nu U_1$ couplings required by the $R_{D^{(*)}}$
observables. Similarly, we considered four one-coupling and four two-coupling scenarios contributing to the $b\to s\mu^+\mu^-$ transition. 
We recast the current LHC dilepton searches ($\ta\ta$ and $\m\m$)~\cite{Aad:2020zxo,Sirunyan:2021khd} to obtain limits on the $U_1$ couplings for a range of $M_{U_1}$ in these scenarios. We also looked at the bounds from the latest direct LQ searches from ATLAS and CMS. Whenever needed, we recast the latest scalar LQ searches in terms of $U_1$ parameters as these were found to give better limits than the existing ones. Put together, our results give the best limits on the $U_1$ parameters currently available from the LHC. These bounds are independent of and complementary to other flavour bounds.

Previously, the high-$p_{\rm T}$ dilepton data were used to put limits on the $U_1$ couplings. Most of these analyses, however, focused only on the nonresonant $t$-channel $U_1$ exchange process. However, we found that this process interferes destructively with the SM background, and, in most cases,  it is this interference that plays the prominent role in setting the exclusion limits, especially for a heavy $U_1$. 
Also, other resonant production processes, namely the pair and the single productions of $U_1$, can also contribute significantly to the high-$p_T$ dilepton tails. We have shown the differences that the inclusion of resonant production processes can make on the exclusion limits in Figs.~\ref{fig:RDRK_LHC} and \ref{fig:RKcomparison}.  The limits we obtained are robust as they depend only on a few assumptions about the underlying model. They are also precise as all the resonant and nonresonant contributions 
including the signal-background interference are systematically incorporated in our statistical recast of the dilepton data. The low mass regions are bounded by the direct pair production search limits that depend only on the BRs. When $U_1$ is heavy, the limits  mostly come from the nonresonant process and its interference with the SM background that depend only on the value of the coupling(s) involved, not on the BRs.

We found that in the minimal (with one free coupling) or some of the next-to-minimal (with two free couplings) scenarios, the parameter spaces required to accommodate the $R_{D^{(*)}}$ anomalies are already ruled out or in tension with the latest LHC data (see e.g. Figs.~\ref{fig:RD1all} and \ref{fig:RD2all}). In some scenarios, the regions favoured by the anomalies are in conflict with other flavour bounds but in the \hyperlink{sce:rd1b}{RD1B} minimal scenario, a part of the parameter space survives the LHC bounds [see Fig.~\ref{fig:RD1Bb}] that can explain the $R_{D^{(*)}}$ anomalies. We found that a good part of the parameter space required to explain the $R_{K^{(*)}}$ anomalies survives the dilepton bounds, except the recent ATLAS search for scalar LQ in the $\mu\mu+jj/bb$ channel~\cite{Aad:2020iuy} put some pressure for $M_{U_1}\lesssim2$ TeV (see Figs.~\ref{fig:exrk2} and~\ref{fig:exrk2_2000}).

Our method for obtaining bounds is generic. It is possible to extend our analysis to scenarios with more nonzero couplings and/or additional degrees of freedom by considering 
our scenarios as templates. 
As an example, we showed the bounds on a combined scenario with three free couplings ($\lm^L_{33}$, $\lm^{L/R}_{32}$, $\lm^{L/R}_{22}$) that can accommodate both the $R_{D^{(*)}}$ and $R_{K^{(*)}}$ anomalies simultaneously with a $1.5$ TeV $U_1$ in Fig.~\ref{fig:exrd1brk2_1500} (even though a simple recast of the recent ATLAS LQ search in the $\m\m+jj/bb$ channel rules out  a $U_1$ of mass $1.5$ TeV that decays to these final states exclusively~\cite{Aad:2020iuy}. One should therefore keep in mind that a $U_1$ LQ with mass $\lesssim 2$ TeV is still allowed and can resolve the $B$-anomalies). To obtain limits on other general scenarios, one can use the parametrization and method elaborated in Appendices~\ref{sec:appendixA} and \ref{sec:appendixB}.

We also identified some possible new search channels of $U_1$ that have not been considered so far. Our simple parametrization of various possible scenarios in terms of a few parameters  can serve as a guide for the future $U_1$ searches at
the LHC. It can also be used for interpreting the results of future bottom-up experimental searches of vLQs.

\acknowledgments 
\noindent 
A.B. and S.M. acknowledge support from the Science and
Engineering Research Board (SERB), DST, India, under Grant No. ECR/2017/000517.
D.D. acknowledges the DST, Government of India for the INSPIRE Faculty Fellowship (Grant No.
IFA16-PH170). T.M. is supported by the intramural grant from IISER-TVM. C.N. is supported by the DST-Inspire Fellowship. 

\appendix
\section{Cross section parametrization for the $\ell\ell$ signal processes}\label{sec:appendixA}
\noindent 
It is not straightforward to obtain precise LHC exclusion limits from the dilepton data when multiple couplings are nonzero simultaneously. This is mainly because different couplings contribute to different topologies with the same final states. In multi-coupling scenarios, the presence of substantial signal-background interference and/or signal-signal interference complicates the picture further. All these possibilities are present in the $U_1$ scenarios
considered here. Therefore, a discussion on a systematic approach to properly take care of these complications might be useful for the reader.
Below, we discuss the method we have used for multi-coupling scenarios in the context of $U_1$. However, this method is not limited to $U_1$ and $\ta\ta$ or $\m\m$ final states but can be adapted easily for any BSM scenarios wherever needed. 

\vspace{0.2cm}
\noindent\textbf{$\blacksquare$~Pair production:} As mentioned in Section~\ref{sec:pheno}, the pair production of $U_1$ is not fully model-independent. It depends on two parameters - $\kp$, parametrizing the new kinetic terms, and $\lm$, the generic coupling for $\ell qU_1$ interactions. In our analysis, we have set $\kp=0$. The dependence on $\lm$ enters in the pair production through the $t$-channel lepton exchange diagrams. If $n$ different new couplings ($\lm_i$ with $i=\{1, 2, 3,\ldots n\}$) are contributing, the total cross section for the process $pp\to U_1U_1$ can be expressed as
\begin{align}
\sg^p\left(M_{U_1},\lm\right) = \sg^{p_0}\left(M_{U_1}\right) +\sum_{i}^n\lm_i^2\sg^{p_2}_i\left(M_{U_1}\right)+ \sum_{i\geq j}^n\lm_i^2\lm_j^2\sg^{p_4}_{ij}\left(M_{U_1}\right)
\end{align}
where the sums go up to $n$. The $\sg^{p_x}$  functions on the r.h.s. depend only on the mass of $U_1$. Here, $\sg^{p_0}\left(M_{U_1}\right)$ is the $\lm$-independent part determined by the strong coupling constant. This part can be computed taking $\lm_i\to 0$ for all the new couplings. The $\sg^{p_2}_i(M_{U_1})$ terms originate from the interference between the QCD-mediated model-independent diagrams and the $t$-channel lepton exchange diagrams. The $\sg^{p_4}_{ij}(M_{U_1})$ terms are from the pure $t$-channel lepton exchange diagrams. 

For a particular $M_{U_1}$, there are $n$ unknown 
$\sg^{p_2}_i$ and $n(n+1)/2$ unknown $\sg^{p_4}_{ij}$ functions that we need to find out. For that, we compute $\sg^p$ for $n(n+3)/2$ different values of $\lm_i$
and solve the resulting linear equations. We repeat the same procedure for different mass points. We can now get 
$\sg^{p}(M_{U_1},\lm)$ for any intermediate value of $M_{U_1}$ either from numerical fits to direct evaluation. 

In the presence of kinematic selection cuts, different $\sg^{p_x}(M_{U_1})$ parts contribute differently to the surviving events. Hence, the overall cut efficiency for the pair production process $\epsilon^p$ depends on both $M_{U_1}$ and $\lm$. The total number of surviving events from the pair production process passing through some selection cuts can, therefore, be expressed as  
\begin{align}
\mc{N}^p &= \sg^p\circ \epsilon^p\ (M_{U_1},\lm)\times {\mc B}^2(M_{U_1},\lm)\times {L}\nn\\
&= \left\{\sg^{p_0}\times\epsilon^{p_0} + \sum_{i}^n\lm_i^2\sg^{p_2}_i\times\epsilon^{p_2}_i+\sum_{i\geq j}^n\lm_i^2\lm_j^2\sg^{p_4}_{ij}\times\epsilon^{p_4}_{ij}\right\}\times {\mc B}^2(M_{U_1},\lm)\times {L}\label{eq:nPair}
\end{align}
where all $\ep^{p_x}$ depend only on $M_{U_1}$. Here ${L}$ is the integrated luminosity, and ${\mc B}(M_{U_1},\lm)$ is the relevant branching ratio (of the decay mode of $U_1$ that contributes to the signal) which can be obtained analytically.
The $\epsilon^{p_x}(M)$ functions can be obtained by computing the fraction of events surviving the selection cuts while computing the $\sg^{p_x}(M_{U_1})$ functions.

\vspace{0.2cm}
\noindent\textbf{$\blacksquare$~Single production:} As discussed earlier, single production of $U_1$ usually contains two types of contributions $U_1x$ and $U_1yz$ (where $x,y,z$ are SM particles). The amplitudes of $U_1x$ type of processes are always proportional to $\lm$. But $U_1yz$ amplitudes can have both linear and cubic terms in $\lm$. Therefore, the most generic form of the single production process $pp\to U_1x+U_1yz$ can be expressed as
\begin{align}
\sg^s(M,\lm_i) = \sum_{i}^n\lm_i^2\sg^{s_2}_i(M_{U_1})+ \sum_{i\geq j\geq k}^n\lm_i^2\lm_j^2\lm_k^2\sg^{s_6}_{ijk}(M_{U_1}).
\end{align}
The $\sg^{s_x}(M)$ functions can be obtained following the same method used for pair production. We can express the total number of single production events as 
\begin{align}
\mc{N}^s &= \sg^s\circ \epsilon^s\ (M_{U_1},\lm)\times \mc B(M_{U_1},\lm)\times {L}\nn\\
&= \left\{\sum_{i}\lm_i^2\sg^{s_2}_i(M_{U_1})\times\epsilon^{s_2}_i(M_{U_1})+\sum_{i\geq j\geq k}\lm_i^2\lm_j^2\lm_k^2\sg^{s_6}_{ijk}(M_{U_1})\times\epsilon^{s_6}_{ijk}(M_{U_1})\right\}
\nn\\
&\quad\times \mc B(M_{U_1},\lm_i)\times{L}.\label{eq:nSing}
\end{align}

\vspace{0.2cm}
\noindent\textbf{$\blacksquare$~Nonresonant production:} The $t$-channel $U_1$ exchange processes fall in this category. There are two different sources of nonresonant contributions one has to consider. One  
is from the signal-SM background interference and  is proportional to $\lm_i^2$.
The other is from the signal-signal interference and hence is quartic in $\lm$. We can express the total nonresonant $pp\to xy$ cross section as
\begin{align}
\sg^{nr}(M_{U_1},\lm) = \sum_{i}^n\lm_i^2\sg^{{nr}_2}_i(M_{U_1})+ \sum_{i\geq j}^n\lm_i^2\lm_j^2\sg^{{nr}_4}_{ij}(M_{U_1}).
\end{align}
Note that $\sg^{nr}(M_{U_1},\lm)$ can be negative when the signal-background interference is destructive. Indeed, this is the case we observe for $U_1$. By introducing the $\epsilon(M_{U_1})$ functions, the total number of surviving events can be written as 
\begin{align}
\mc{N}^{nr} &= \sg^{nr}\circ \epsilon^{nr}\ (M_{U_1},\lm)\times{L}\nn\\
&= \left\{\sum_{i}^n\lm_i^2\sg^{{nr}_2}_i(M_{U_1})\times\epsilon^{{nr}_2}_i(M_{U_1})+\sum_{i\geq j}^n\lm_i^2\lm_j^2\sg^{{nr}_4}_{ij}(M_{U_1})\times\epsilon^{{nr}_4}_{ij}(M_{U_1})\right\}\times{L}.\label{eq:nNR}
\end{align}
Notice, no BR appears in the above equation. A negative $\sg^{nr}(M_{U_1},\lm)$ makes $\mc{N}^{nr}$ a negative number as presented in Table~\ref{tab:cross}.
\bigskip

\begin{figure*}[]
\captionsetup[subfigure]{labelformat=empty}
{\includegraphics[width=0.8\textwidth]{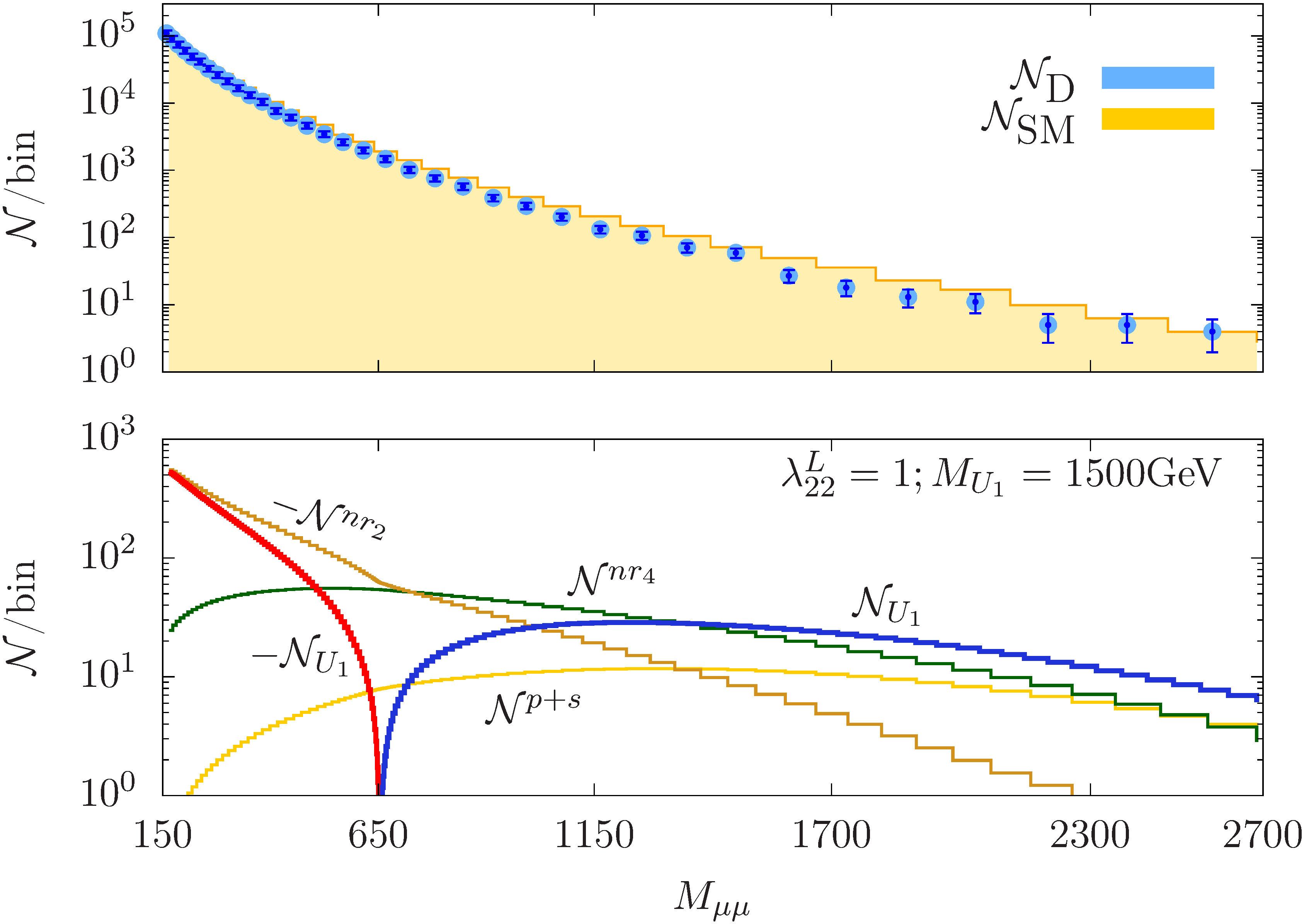}\label{fig:SM_BSM}}
\caption{The observed $M_{\m\m}$ distribution and the corresponding SM contributions from Ref.~\cite{Sirunyan:2021khd}. The errors are obtained using Eq.~\eqref{eq:errors}. (Lower panel)
The different signal components for a typical choice of parameters, $M_{U_1}=1.5$ TeV and $\lm_{22}^L =1$ ({\protect\hyperlink{sce:rk1a}{Scenario RK1A}}). The number of $U_1$ signal events are denoted by $\mc N_{U_1}$ [Eq.~\eqref{eq:NT}] and $\mc N^{p+s} = \mc N^p+\mc N^s$ denotes the events from pair and single production processes. Among the nonresonant contributions, $\mc N^{\rm nr_2}$ denotes the SM-BSM interference and $\mc N^{\rm nr_4}$ is the pure BSM nonresonant part.}  
\label{fig:SM_BSM}
\end{figure*}
\noindent 
For an illustration, we show the observed $\m\m$ data~\cite{Sirunyan:2021khd} and the corresponding SM contributions in Fig.~\ref{fig:SM_BSM}. In the lower panel we show the 
different signal components for $M_{U_1}=1.5$ TeV and $\lm_{22}^L =1$ (\hyperlink{sce:rk1a}{Scenario RK1A}).

\section{Limits estimation: $\chi^2$ tests}\label{sec:appendixB}
\noindent 
To obtain the limits on the parameter space in various $U_1$ scenarios, we recast the LHC $\ta\ta$ and $\m\m$ search data~\cite{Aad:2020zxo,Sirunyan:2021khd}. In particular, we perform $\chi^2$ tests to estimate the limits on parameters from the transverse (invariant) mass distribution of the $\ta\ta$ ($\m\m$) data. The method is essentially the same as the one used in Ref.~\cite{Mandal:2018kau} for $S_1$ LQ. Here we briefly outline the steps.
\begin{enumerate}
\item 
For each distribution, the test statistic can be defined as 
	\begin{equation}
        \chi^{2} = \sum_{i}\left(\frac{\mc N_{\rm T}^i(M_{U_1},\lm)-\mc N_{\rm D}^i}{\Delta \mc N^i}\right)^2
    \end{equation}
where the sum runs over the corresponding bins. Here, $\mc N_{\rm T}^i(M_{U_1},\lm)$ stands for expected (theory) events, and $\mc N_{\rm D}^i$ is the number of observed events in the $i^{th}$ bin. The number of theory events in the $i^{th}$ bin can be expressed
    \begin{align}
        \mc N_{\rm T}^i (M_{U_1},\lm)&= \mc N_{U_1}^i (M_{U_1},\lm)+ \mc N_{\rm SM}^i\nn\\
        & = \big[ \mc N^p(M_{U_1},\lm) +\mc  N^s(M_{U_1},\lm) + \mc N^{nr} (M_{U_1},\lm)\big] + \mc N_{\rm SM}^i.\label{eq:NT}
    \end{align}
Here, $\mc N_{U_1}^i$ and $\mc N_{\rm SM}^i$ are the total signal events from $U_1$ and the SM background in the $i^{th}$ bin, respectively. The total signal events are composed of $\mc N^p$, $\mc N^s$, and $\mc N^{nr}$ from Eqs.~\eqref{eq:nPair}, \eqref{eq:nSing}, and \eqref{eq:nNR}, respectively. 
 The details on how to calculate $\mc N_{U_1}^i$ for different scenarios is sketched in Appendix~\ref{sec:appendixA}. For the error $\Delta \mc N^i$, we use
    \begin{equation}
        \Delta \mc N^i = \sqrt{\left( \Delta \mc N^i_{stat}\right)^2 + \left( \Delta \mc N^i_{syst} \right)^2}\label{eq:errors}
    \end{equation} 
where $\Delta \mc N^i_{stat} = \sqrt{\mc N_D^i}$ and we assume a uniform $10$\% systematic error, i.e., $\Delta \mc N^i_{syst} = \delta^i \times \mc N_D^i$ with $\delta^i = 0.1$.
    
 \item
 In every scenario, for some discrete benchmark values of $M_{U_1}=M_{U_1}^b$ we compute the minimum of $\chi^2$ as $\chi^2_{min}(M_{U_1}^b)$ by varying the couplings $\lm$. 
 
 \item 
 In one-coupling scenarios (like \hyperlink{sce:rd1a}{Scenario RD1A}, \hyperlink{sce:rk1a}{Scenario RK1A}, etc.), we obtain the $1\sg$ and $2\sg$ confidence level upper limits on the coupling at $M_{U_1}=M_{U_1}^b$ from the values of $\lm$ for which $\Delta \chi^2(M_{U_1}^b,\lm)=\chi^2(M_{U_1}^b,\lm)-\chi^2_{min}(M_{U_1}^b)=1$ and $4$, respectively.
 
 In two-coupling scenarios (like \hyperlink{sce:rd2a}{Scenario RD2A}, \hyperlink{sce:rk2a}{Scenario RK2A}, etc.), we do the same, except we obtain the $1\sg$ and $2\sg$ limits (contours) from the $2$-variable limits on $\Delta \chi^2$; i.e., we solve 
 $\Delta \chi^2(M_{U_1}^b,\lm_1,\lm_2)=\chi^2(M_{U_1}^b,\lm_1,\lm_2)-\chi^2_{min}(M_{U_1}^b)=2.30$ and $6.17$, respectively.
 
 Similarly, we can obtain the limits for the scenarios with $n (\geq 2)$ free couplings by using the $n$-variable ranges for $\Delta\chi^2$.
\item 
 We obtain the limits for arbitrary values of $M_{U_1}$ by interpolating the limits for the benchmark masses.
    
\end{enumerate}

\bibliography{Leptoquark}{}
\bibliographystyle{JHEPCust}

\end{document}